\documentstyle[12pt]{article} 
\input epsf  
\newcommand{\Buildrel}[2]{\mathrel{\mathop{\kern 0pt #1}\limits_{#2}}}
\renewcommand{\theequation}{\arabic{section}.\arabic{equation}}
\newcounter{saveeqn}

\newcommand{\appeqn}{\setcounter{saveeqn}{1}%
\renewcommand{\theequation}{\Alph{section}.\arabic{equation}}}
\newcommand{\be}{\begin{equation}}
\newcommand{\ee}{\end{equation}}
\newcommand{\bea}{\begin{eqnarray}}
\newcommand{\eea}{\end{eqnarray}}
\newcommand{\nn}{\nonumber}   
\newcommand{\ba}{\bea \begin{array}}
\newcommand{\ea}{\end{array} \eea}
\renewcommand{\(}{\left(}
\renewcommand{\)}{\right)}
\newcommand{\T}[1]{\mbox{$\langle\,T^{#1}\,\rangle$}}
\newcommand{\tn}{\mbox{$\langle\,T^{11}_{\,n'}\,\rangle$}}
\newcommand{\sk}{\sum_{k=1}^{N-1}}
\newcommand{\sink}{\sin(\frac{\pi k}{2N})}
\newcommand{\p}{\partial}

\begin{document}

\title{Casimir Effect on a Finite Lattice}
\author{A. Actor\\
{\em Department of Physics, The Pennsylvania State University}\\
{\em Fogelsville, PA 18051, USA}\\
{\em E-mail: aaa2@psu.edu}
\and I. Bender\\
{\em Institut f\"ur Hochenergiephysik, Universit\"at Heidelberg}\\ 
{\em Philosophenweg 16, D-69120 Heidelberg, Germany}\\
{\em E-mail: bender@thphys.uni-heidelberg.de}
\and J. Reingruber\\
{\em Institut f\"ur Theoretische Physik, Technische Universit\"at 
M\"unchen}\\
{\em James-Franck-Strasse, D-85748 Garching, Germany}\\
{\em E-mail: reingrub@physik.tu-muenchen.de}}
\bigskip
\date{\today}
\maketitle
\thispagestyle{empty}
\pagenumbering{roman}
\begin{abstract}
\thispagestyle{empty}
\noindent Lattice quantum field theory is a well established branch of modern quantum field
theory (QFT). However, it has only peripherally been used for the 
investigation of Casimir systems -- i.e. for systems in which quantum fields are 
distorted by their interaction with classical background objects. This article
presents a Hamiltonian lattice formulation of static Casimir systems at a level of generality 
appropriate for an introductory investigation. Background structure
 -- represented by a lattice potential $V(x)$ -- is introduced along one spatial
 direction with translation invariance in all other spatial directions. It is 
 simple to extend this formulation to include arbitrary background structure in more 
than one spatial direction. Following some general analysis two specific finite 1D 
lattice QFT systems are analyzed in detail. The first has three Dirichlet boundaries 
at the lattice sites $x\,=\,0,\,l$ and $L$ ($L\,>\,l\,>\,0$) with vanishing lattice
potential $V(x)$ everywhere in between. The vacuum energy and vacuum stress 
tensor $T^{\mu\nu}$ for this system are calculated in $0\,<\,x\,<\,L$. Very careful 
attention must be and is given to renormalization in the (continuum) limit of 
vanishing lattice constant. Globally and locally this lattice system is seen to
closely mimic the corresponding $1D$ continuum system -- as one would hope. Then
we introduce a lattice potential $V(x)\,=\,c/(x-x_0)^2$ centered at $x\,=\,x_0\,<0$ 
to the left of the boundary at $x\,=\,0$ and extending {\em through} this
boundary and the middle Dirichlet boundary at $x\,=\,l$ out to the right-hand
boundary  $x\,=\,L\,>\,l$ and beyond. The vacuum energy and $T^{\mu\nu}$ are
calculated for this far more complicated system in the region $0\,<\,x\,<\,L$, again
with very good  results. The internal consistency of the lattice version of this system
is carefully examined. Our conclusion is that finite-lattice 
formulation provides a
powerful and effective tool, capable of solving completely many Casimir systems
which could not possibly be handled using continuum methods. This is precisely
our reason for introducing it. Future investigations (in one and more dimensions 
and in
dynamical as well as static contexts) will display more fully the power of this 
method.
\thispagestyle{empty}
\end{abstract}
\vfill\eject
\tableofcontents
\vfil\eject
\pagestyle{plain}
\pagenumbering{arabic}
\setcounter{page}{1}
\section{Introduction}
\label{sec.intro}
Let us here define Casimir QFT (quantum field theory) to be the study of quantum fields in flat space,
fields which coexist with and interact with classical background objects occupying 
the same space (in the simplest case Dirichlet or Neumann boundaries). 
This interaction distorts the quantum fields and all local observables connected with 
them away from the 
{\em spatial uniformity} (or translation invariance) that would characterize
these fields in {\em empty} flat space. This nonuniformity in flat space caused
by background objects is generally very pronounced, becoming even infinite as 
Dirichlet or Neumann boundary surfaces are approached. The background objects 
themselves experience measurable back forces --  a subtle example of
Newton's third law. These back forces are called {\em Casimir forces} and
their presence is often identified with the Casimir effect. However the deeper essence
of the Casimir effect is the distortion of the quantum fields caused 
by the background objects. Viewed this way the Casimir effect is obviously an
extremely fundamental and general phenomenon in quantum field theory. Any nonuniform
classical background whatsoever will produce a Casimir effect.\hfil\break
To the authors' knowledge all published work in Casimir QFT employs continuum
QFT. In certain other branches of QFT, of course, lattice formulations have become 
commonplace. Quite well-developed methods exist for doing QFT on a spatial or
spacetime lattice (see e.g. the book by Rothe \cite{Heinz}). Generally in lattice-QFT
applications spatial uniformity is assumed. Either one chooses the spatial lattice
to be infinite, or one imposes periodicity on a finite spatial lattice. There are
no background objects on the lattice. Rather, one is using the lattice to learn about
nonperturbative properties of interacting fields which are difficult if not impossible
to investigate in continuum theory.\hfil\break
Some years ago  two of the authors undertook the reformulation of Casimir 
QFT in lattice language. From the outset we have had two main objectives.\hfil\break
First, to gain a sense of how well lattice QFT works when applied to Casimir problems. 
By investigating a number of systems whose continuum versions can be solved explicitly
we found that lattice QFT is able to mimic and reproduce in a quite detailed way
both global and local features of continuum Casimir systems. While not unexpected this
did need to be established quantitatively before we could proceed to our second and
more important objective.\hfil\break
That objective is, to use lattice QFT to solve Casimir
problems which cannot be dealt with (adequately, or even at all) using continuum QFT.
It is unfortunate that practically all Casimir systems of physical interest belong
in this latter difficult-to-access category.\hfil\break
On a lattice one can work both analytically and numerically. Analytic methods tend 
to rather closely parallel continuum theory and, while interesting, do not necessarily 
offer greatly increased computational power, especially on an infinite lattice. Numerical
methods used on a finite lattice can be, however, very powerful. One can introduce 
static background objects on the finite spatial lattice and make the quantum field
$\phi(x)$ interact with these objects, the interaction being described by a potential in a 
functional Schr\"odinger equation, whose stationary solutions are expressible
in terms of the eigenfunctions and the spectrum of an ordinary Schr\"odinger equation 
associated with the functional one. 
These spatial modes and the spectrum can be found numerically.
Given these basic ingredients one can then compute everything numerically -- the
vacuum energy and the stress-energy-momentum tensor, Casimir forces and so on. 
This can surely be done for a
vast array of systems which could not be solved in continuum theory.\hfil\break
Of course renormalization {\em must} be performed to make lattice observables 
physically meaningful. Everything one calculates on a finite lattice is finite.
However, lattice quantum variables like vacuum  energy and the vacuum expectation value 
of the stress-energy-momentum tensor $T^{\mu\nu}$ contain the
seeds of ultraviolet (UV) divergences, i.e. terms which grow without limit in the 
continuum limit $a\to 0$ ($a$ is the lattice constant or spacing between lattice
points). For finite and decreasing $a$ these essentially informationless terms
eventually overwhelm the physical signal: their systematic removal is utterly 
essential. Lattice quantities which vanish as $a\to 0$ take care of themselves, of
course. What remains is the \lq\lq physical signal" -- a contribution independent
of $a$ which therefore neither blows up nor vanishes as $a\to 0$. All of this is
the process of renormalization.\hfil\break
For a number of years we have pursued this numerical strategy, for both standard 
and unconventional Casimir systems. To date none of this lattice QFT work has been reported
(excepting the brief descriptions \cite{ref2,ref3}). We begin here the systematic
development of lattice QFT with background objects and
its application to nontrivial Casimir systems. Before describing the contents of
the present article we pause to mention some nonstandard aspects of
Casimir QFT.\hfil\break
Traditional Casimir systems have \lq\lq hard" boundary surfaces -- i.e. boundary
surfaces $\partial\cal{M}$ with precise spatial location on which the quantum field
$\phi(x)$ has to satisfy some boundary condition. A perhaps more physically realistic
boundary can be fashioned by attaching to a Dirichlet boundary $\partial\cal{M}$ 
(on which $\phi(x)$ has to vanish) a static potential $V(x)$ which grows to infinity 
as $x$ approaches $\partial\cal{M}$ and falls off to zero more or less rapidly away
from $\partial\cal{M}$. The interaction of this classical potential with the quantum
field $\phi(x)$ is for simplicity assumed to be described by an interaction term
quadratic in $\phi(x)$ : $(1/2)\, \phi^2(x)\, V(x)$. We refer to such a modification of a hard Dirichlet surface
as {\em semihardening} \cite{ref4,ref5}. The potential function $V(x)$ represents
(one can say) {\em surface texture} attached to $\partial\cal{M}$. The wave
equation $[\partial_0^2\,-\,\Delta\,+\,m^2]\,\phi\,=\,0$ in usual hard-boundary 
scalar-field Casimir QFT with appropriate boundary conditions on $\phi$ is 
replaced by
\begin{equation}
\label{eq.1.1}
[\partial_0^2\,-\Delta\,+\,m^2\,+V(x)]\,\phi(x)\,=\,0\quad .
\end{equation}
Here $V(x)$ should be viewed as a (rather huge) generalization of ordinary
boundary conditions. This potential can represent many kinds of background structure
which interact with $\phi$ and are (for $V(x)\,<\,\infty$) partially transparent to
$\phi$. The semihardening potentials above are just one kind of background spatial
structure among many which can be introduced into QFT (see e.g. refs. 
\cite{ref6, ref8, vierzehn,fuenfzehn}). For us, Casimir QFT 
comprises all possible background structures interacting with quantum 
fields -- a truly vast subject. Many workers, of course, use the term 
``Casimir effect'' mainly in connection with boundary surfaces (see e.g. 
the reviews \cite{ref8, sechzehn}). Diffuse background structures in 
QFT are then often called ``background fields''.\hfil\break
\indent As mentioned earlier the fundamental phenomenon in Casimir QFT is the distortion
of otherwise spatially uniform fields and observable quantities away from spatial  
uniformity by the background $V(x)$. To really 
understand a Casimir system one must work locally. There is no better physical 
quantity to compute than the vacuum expectation value of the stress-energy-momentum 
tensor (hereafter stress tensor)
of a Casimir system which we will denote by $\langle\,T^{\mu\nu}(x)\,\rangle$ in the 
following. This quantity reveals the spatial nonuniformity of the system, its
nonuniform vacuum energy density and its Casimir forces. 
$\langle\,T^{\mu\nu}(x)\,\rangle$ can
be computed from the spatial modes and spectrum of $\phi(x)$, both in continuum theory
and on a  lattice. When working numerically on a finite lattice (using a lattice
potential $V(x)$) one can obtain $\langle\,T^{\mu\nu}(x)\,\rangle$ numerically for 
practically any kind of background structure.\hfil\break

In sec. 2 we begin our analysis of Casimir systems of the following type: A space 
of dimension $d$ is chosen with free (or periodic) \lq\lq boundary conditions"
in each of the directions $x_2,\,\cdots\,,\, x_d$. Along the $x_1$ axis we place 
parallel Dirichlet boundaries at $x_1\,=\,0$ and $x_1\,=\,l$. In sec. 2 we do not 
concern ourselves with the regions $x_1\,<\,0$ and $x_1\,>\,l$ external to the
interval $0\,\le\,x_1\,\le\,l$ of interest. These external regions can be 
\lq\lq attached" later on and given suitable global structure (e.g. periodicity in
$x_1$ over some larger interval) to complete the specification of the physical system.
We postpone doing this without loss of generality. In the interval $0\,\le\,x_1\,
\le\,l$ we introduce an arbitrary background potential $V(x_1)$. The formulation
of lattice scalar QFT on this spatial manifold is reviewed in detail 
in sec. 2, where detailed
lattice formulae for $\langle T^{\mu\nu}\rangle$ are given and the ground is prepared 
for our very specific lattice analyses to follow in secs. 3, 4.\hfil\break

In sec. 3  we go to one dimension $d=1$, set the background potential $V(x)\,=\,0$
and study scalar lattice QFT on a $1D$ lattice $0\,\le\,x\,\le\,l$ with Dirichlet
walls at $x\,=\,0,\,l$ and no structure in between. Choosing lattice constant
$a$ means $l=N a\, ,\,x=n a$ where $n=0,\,1,\cdots,\,N$ labels the lattice sites.
The continuum limit $a\to0$ (with $l$ fixed of course) means $N=l/a\to\infty$. As
described above we must identify and subtract those terms in the
lattice vacuum energy and $\langle T^{\mu\nu}\rangle$ of this system which diverge 
for $a\to 0$. The remaining $a$-independent 
parts of these functions are then to be brought into comparison with continuum
theory. In sec. 3 we present detailed analysis of this continuum limit on the $1D$ 
lattice. To enable us to make definite statements about the Casimir force on
the boundary $x=l$ we introduce an \lq\lq external" region $l\le x\le L$ with $L\gg l$
which can be analyzed on the lattice exactly as we do the \lq\lq internal" region
$0\le x\le l$. We then have a more complete physical system than just the internal
region $0\le x\le l$. To make it a really complete physical system the region
$x<0$ needs to be included and the global structure of the $x$ axis must be specified.
However there is no need to do all this here. Our goal is to show how well
hard-boundary renormalized lattice QFT mimics continuum QFT. That goal is nicely
achieved using the two adjacent $1D$ lattices $0\le x\le l,\,l\le x\le L$.
\hfil\break

Sec. 4 tackles a more difficult problem. Keeping spatial dimension $d=1$ we introduce
the Bessel semihardening potential $V(x)\,=\,c/(x-x_0)^2$ centered at $x=x_0<0$, allowing
this potential to extend arbitrarily far into $x>0$. Immersed in $V(x)$ we position
 Dirichlet point boundaries at $x=0,\,l,\,L$ as before. One can find the lattice
 vacuum energy $E^{vac}$ and $\langle T^{\mu\nu}\rangle$ numerically for this system with arbitrary 
background potential $V(x)$. In sec. 4.1 we show how to renormalize $E^{vac}$ and
$\langle T^{\mu\nu}\rangle$ for arbitrary $V(x)$. Then we adopt the Bessel potential above and
begin our principal numerical analysis. In a sense the numerical work in sec. 4.2
is the central part of this article. Using $E^{vac}_{ren}$ we calculate
{\em globally} the Casimir force on the middle boundary $x=l$ due to the other
Dirichlet boundaries and the potential $V(x)$. Using $\langle T^{\mu\nu}\rangle_{ren}$
 we calculate
{\em locally} the same Casimir force. Very precise agreement is found. Some
other things are computed as well, but let us limit our description of sec. 4 to 
the comment just made. This agreement between global and local calculations of a 
physical Casimir force in a quite nontrivial background potential gives us great 
confidence in the methods we are using. We have no doubt that equally good results
can be gotten for practically any background $V(x)$. Indeed we have had comparable success
with other potentials.\hfil\break

Concluding remarks are made in sec. 5. There we indicate the directions in 
lattice QFT theory we presently are pursuing. Three appendices dealing with 
technical matters enable us to present a substantially condensed narrative.
\hfil\break
\section{Scalar field on a finite lattice with background}
\label{sec.cont}
\setcounter{equation}{0}
To construct our Casimir system we begin with flat d-dimensional space and give it 
background structure along just one direction, say $x_1$. The other directions 
$\vec{x}_\perp\,=\,(x_2,\,\cdots,\,x_d)$ are free but -- in
preparation for the finite lattice  -- we make each of these 
coordinates periodic in $-L\le x_j\le L$ with period $2 L$. Periodicity is, of course, 
closely akin to \lq\lq free boundary conditions", neither of these being a true 
boundary condition at all. Thus $\vec{x}_\perp$ specifies position on a spatial 
$(d-1)$-dimensional torus $T^{d-1}$ representing the free directions and 
$x=(x_1,\,\vec{x}_\perp)$ is 
position in space. A real scalar field $\phi(x)$ with mass $m$ is now defined on 
this space. $\phi(x)$ interacts with and is made 
nonuniform by and exerts a back force on the background structure
along the $x_1$ direction. This interaction is what we wish to study 
in detail.\hfil\break
For completeness let us mention something we do not wish to study more 
than necessary: the so-called ``topological Casimir effect''. When a 
quantum field $\phi(x)$ is defined on a topologically nontrivial 
space, $\phi(x)$ is forced to be (nontrivially) different than it would 
be in free infinite space. This is itself a kind of Casimir effect. 
The vacuum energy density of $\phi(x)$ is shifted, vacuum stresses or 
Casimir forces are exerted by the modified field directly on the 
spatial manifold causing the modification etc (see e.g. the reviews 
\cite{ref8, sechzehn}). When, as here, one defines one's field on a 
(d-1)-dimensional torus $T^{d-1}$ (the periodic directions 
$x_{2},\cdots\, ,\, x_{d}$) there arises a topological Casimir effect 
parametrized by the finite circumferences of $T^{d-1}$. However, no 
``objects'' have been introduced anywhere; periodic boundary 
conditions do not involve real boundaries. The $T^{d-1}$ topological 
Casimir effect smoothly disappears as the torus circumferences become 
infinite.\hfil\break
Topological Casimir effects caused by global spatial structure {\em 
coexist} with the field distortion caused by locally nonuniform 
spatial backgrounds, or objects in space. One should, however, make 
the effort to keep these two aspects of QFT distinct in one's 
thinking. This seems particularly important when one replaces 
continuum QFT by lattice QFT. For numerical work the lattice has to 
be finite, inevitably introducing the topological effect under 
discussion, which lattice workers and others usually call 
``finite-size effects''. Unless one wishes specifically to study 
them, finite-size effects tend for obvious reasons to be regarded as 
a nuisance.\hfil\break
Many readers may be familiar with finite-size scaling theory (see e.g. 
refs \cite{siebzehn,achtzehn,neunzehn}) in 
the general area of statistical systems. This and other related work 
one finite-size effects in QFT has rather limited overlap with and 
relevance for the kind of investigation we are concerned with here. 
Our focus is on the distorting effect background {\em objects} have on 
quantum fields. Finite-size scaling theory usually deals with bulk 
effects and large scale collective phenomena near critical points, and 
the effect of small system size on such phenomena. Phase transitions 
are involved, and temperature plays an essential role. In Casimir QFT 
temperature is something of a sideshow (although not completely without 
interest), and phase transitions normally play no role whatsoever. 
True, the effects of surfaces and corners in statistical models are 
sometimes taken into account (see e.g. \cite{siebzehn}), and this 
overlaps with our work more substantially. However, an attempt to 
pursue this would take us far outside the scope of the present work, 
and necessitate substantially increased length in an already long paper.
\hfil\break
Resuming now our mathematical discussion, we position along $x_{1}$ 
two planar Dirichlet boundaries at $x_{1}=0$ and $x_{1}=l$. Between 
these boundaries we subject the quantum field $\phi(x)$ to a 
background potential
$V(x_1)\ge 0$. Lacking external regions $x_1<0$ and $x_1>l$ this Casimir system is 
physically incomplete. However it is easy to attach these external regions. As long as 
one does not try to discuss Casimir forces on the boundaries $x_1=0,\,l$ there is
no need to be concerned about anything outside $0< x< l$. Our purpose in this section 
will be the detailed formulation of the lattice quantum theory in $0<x < l$ in the 
presence of arbitrary background potential $V(x_1)$. We begin with a 
brief summary of the continuum QFT we wish to put on a spatial lattice. \hfil\break
\subsection{Continuum system}
\label{sec.quant}
The classical  Lagrangian of the real scalar field interacting with a background, 
represented by a classical potential $V(x_1)$ becoming infinite at $x_1=0,\,l$, is
\begin{eqnarray}
\label{eq2.1}
L=\int d^dx\,{\cal L}&=&\int d^{d}x\, \frac{1}{2}\((\p^{\mu}\phi)(\p_{\mu}\phi)-
(m^2+V(x_1))\phi^{2}\)\\
&=&\int d^{d}x\, \frac{1}{2}\(\dot{\phi}^{2}-\phi(m^2+V(x_1)-\triangle)\phi\)\,\nn
\quad .
\end{eqnarray}
Here in the partial integration used to reach the final equality the surface term 
vanishes because of the boundary conditions imposed on the classical field $\phi(x,t)$:
\begin{eqnarray}
\phi(x_1=0,\vec{x}_\bot,t)&=&\phi(x_1=l,\vec{x}_\bot,t)=0\quad (\mbox{due to }\;
V(0)=V(l)=\infty)\\
\phi(x_1,\vec{x}_\bot,t)&=&\phi(x_1,\vec{x}_\bot+2L\,\vec{e}_i,t)\,,\quad i=2,
\ldots,d \quad (\mbox{periodicity}),\nn
\end{eqnarray}
$\vec{e}_i$ being the unit vector in i-direction.
Using the canonically conjugate field
\be
\pi(x,t) \,\equiv\,{\delta L\over \delta\dot{\phi}(x,t)}\,=\,\dot{\phi}(x,t)
\ee
we obtain from $L$ by Legendre transformation the classical Hamiltonian
\bea
\label{eq2.2}
H=\int d^{d}x\,\pi(x,t)\dot{\phi}(x,t)-L\,=\,
\int d^{d}x\, \frac{1}{2}\(\pi^{2}+\phi(m^2+V(x_1)-\triangle)\phi\)\quad .
\eea
Canonical quantization means replacing the classical field $\phi(x,t)$ and its 
conjugate momentum $\pi(x,t)$ by hermitian operators $\Phi(x,t)$ and $\Pi(x,t)$
which obey the equal-time commutation relations
\bea
[\Phi(x,t),\Phi(x',t)]=[\Pi(x,t),\Pi(x',t)]&=& 0\,,\label{eq2.3}\\
\mbox{[} \Phi(x,t),\Pi(x',t) ] &=& i\,\delta (x-x')\,.\label{eq2.4}
\eea
Following standard procedure one can realize these commutation 
relations by expanding the field operator in terms of creation and 
annihilation operators $\hat{a}^{\dagger}_{k}$ and $\hat{a}_{k}$ which 
satisfy $[\hat{a}_{k},\hat{a}^{\dagger}_{k'}]\,=\,\delta_{k,k'}$ and 
$[\hat{a}_{k},\hat{a}_{k'}]\,=\,[\hat{a}^{\dagger}_{k},\hat{a}^{\dagger}_{k'}]\,=\,0$ ;
\be\label{eq25n}
\hat{\phi}(x,t)\,=\,\sum_{k}\,{1\over \sqrt{2 
\epsilon_{k}}}\,\left[\hat{a}_{k}\,e^{-i\epsilon_{k} t}\, u_{k}(x)\,+\,
\hat{a}^{\dagger}_{k}\,e^{i\epsilon_{k} t}\, u_{k}^{\ast}(x)\right]\quad .
\ee
Here the $u_{k}(x)$ and $\epsilon_{k}$ are eigenfunctions and the 
associated eigenvalues of the operator 
$[-\Delta\,+\,m^{2}\,+\,V(x_{1})]$\, , i.e.
\be
\label{eq2.7}
[m^2+V(x_1)-\triangle]\,u^k(x)\,=\,\varepsilon_k^2\,u^k(x)\quad .
\ee
Separating variables 
\label{eq2.6}
\bea
u^k(x)=v^{k_1}(x_1)\,w^{k_2}(x_2)\cdots w^{k_d}(x_d)&=&v^{k_1}(x_1)\,w^{\vec{k}_\bot}
(\vec{x}_\bot)\quad ,\\
k=(k_1,k_2,\ldots,k_d)&=&(k_1,\vec{k}_\bot)\quad ,\nn
\eea 
leads to
\bea
\(V(x_1)-\frac{\p^2}{(\p x_1)^2}\)\,v^{k_1}(x_1)&=&\rho_{k_1}^2\,v^{k_1}(x_1)\; ,\label{eq2.8}\\
-\frac{\p^2}{(\p x_{i})^2}\,w^{k_{i}}(x_{i})&=&\omega_{k_i}^2\,w^{k_{i}}(x_{i})\;,
\quad i=2,\ldots, d\;\; \mbox{and} \label{eq2.9}\\
\varepsilon^2_k&=&m^2+\rho_{k_1}^2+\sum_{i=2}^{d}\omega_{k_i}^2\quad ,\label{eq2.10}
\eea
where the boundary conditions on $v^{k_1}$ and $w^{k_i}$ are
\bea 
v^{k_1}(0)&=&v^{k_1}(l)\,=\,0\;,\\
w^{k_{i}}(x_{i})&=&w^{k_{i}}(x_{i}+2L)\,,\quad i=2,\ldots,d\;.\nn
\eea
Obviously for the free directions
\bea
\label{eq2.11}
w^{k_{i}}(x_{i})=\sqrt{\frac{1}{2L}}\,\,e^{i\frac{\pi}{L}k_{i}x_{i}}\; ,
\eea
with eigenvalues
\bea 
\label{eq2.12}
\omega_{k_i}^2=\frac{\pi^2}{L^2}\,k_{i}^2,\quad k_{i}\in {\bf Z}\; .
\eea
The functions $v^{k_1}(x_1)$ of course depend on the potential $V(x_1)$. These 
functions are chosen to be real.\hfil\break
Altogether we have
\bea 
u^{k}(x) & = & v^{k_1}(x_1)\,w^{\vec{k}_\bot}(\vec{x}_\bot)=v^{k_1}(x_1)
\(\frac{1}{2L}\)^{\frac{d-1}{2}}e^{i\frac{\pi}{L}\vec{k}_\bot\vec{x}_\bot}\quad ,
\label{eq2.13}\\
\varepsilon^2_k & = & m^2+\rho_{k_1}^2+\sum_{i=2}^{d}\frac{\pi^2}{L^2}\,k_{i}^2\; .
\label{eq2.15}
\eea
The set of modes $\lbrace u^k(x)\rbrace$ is complete and orthonormal
\bea\label{eq2.16}
\sum_{k}u^{k}(x)\,{u^{k}}^\ast(x')=\bar{\delta}(x-x')\quad\mbox{and}\quad 
\int d^{d}x\,u^{k}(x)\,{u^{k'}}^\ast(x)=\delta_{k,k'}\; ,
\eea
where
\bea
\sum_{k}\ldots\,=\,\sum_{k_1=1}^{\infty}\sum_{k_{2}=-\infty}^{\infty}\cdots 
\sum_{k_{d}=-\infty}^{\infty}\cdots\quad ,\quad
\delta_{k,k'}\,=\,\delta_{k_1,k_1'}\delta_{k_2,k_2'}\cdots \delta_{k_{d},k_{d}'}\quad .
\eea
Due to periodicity in
$x_2,\,\cdots,\, x_d$ we need to introduce here the periodic $\delta$-function 
$\tilde{\delta}(x)$ along these directions, distinguished from the usual 
$\delta$-function for a noncompact direction by its tilde. The spatial 
$\delta$-function is then
\bea
\label{eq2.4a}
\bar{\delta}(x-x')&=&\delta(x_1-x_1')\tilde{\delta}(\vec{x}_\bot-\vec{x}_\bot')\quad .
\eea
Our attention now shifts to the vacuum stress (energy-momentum) tensor of the system.
The classical canonical stress tensor for a real scalar field is
\be
\label{eq2.33}
T^{\mu\nu}=\frac{1}{2}\((\p^{\mu}\phi)(\p^{\nu}\phi)+(\p^{\nu}\phi)(\p^{\mu}\phi)\)-
g^{\mu\nu}\,{\cal L}\quad ,
\ee
with ${\cal L}$ given by the first equality in eq. (\ref{eq2.1}). 
Again standard procedure (see e.g. \cite{ref10}) leads to the 
following mode-sum formulae for the vacuum expectation value of the 
operator $T^{\mu\nu}$:
\bea
\langle 0|T^{00}(x)|0\rangle&=&\sum_{k}\frac{\varepsilon_k\, v^{k_1}(x_1)^2 }
{2(2L)^{d-1}}\label{eq2.38}\\
&+&\sum_{k}\frac{1}{4\varepsilon_k\,(2L)^{d-1}}\frac{\p}{\p x_1}
\(v^{k_1}(x_1)\frac{\p v^{k_1}(x_1)}{\p x_1}\)\nn\\
&=&\sum_{k}\frac{(2\varepsilon_k^2-\rho_{k_1}^2+V(x_1))\, v^{k_1}(x_1)^2 }
{4\varepsilon_k\,(2L)^{d-1}}\nn \\
&+&\sum_{k}\frac{1}{4\varepsilon_k\,(2L)^{d-1}}
\(\frac{\p v^{k_1}(x_1)}{\p x_1}\)^2\;,\nn\\ \nn \\
\langle 0|T^{11}(x)|0\rangle&=&\sum_{k}\frac{(\rho_{k_1}^2-V(x_1))\, v^{k_1}(x_1)^2 }
{2\varepsilon_k\,(2L)^{d-1}}\nn\\
&+&\sum_{k}\frac{1}{4\varepsilon_k\,(2L)^{d-1}}\frac{\p}{\p x_1}
\(v^{k_1}(x_1)\frac{\p v^{k_1}(x_1)}{\p x_1}\)\label{eq2.39}\\
&=&\sum_{k}\frac{(\rho_{k_1}^2-V(x_1))\, v^{k_1}(x_1)^2 }{4\varepsilon_k\,
(2L)^{d-1}}+\sum_{k}\frac{1}{4\varepsilon_k\,(2L)^{d-1}}\(\frac{\p v^{k_1}(x_1)}
{\p x_1}\)^2\nn\\
\mbox{and for $i=2,\,\ldots,\, d$}\nn\\
\langle 0|T^{ii}(x)|0\rangle&=&\sum_{k}\frac{\omega_{k_i}^2\, v^{k_1}(x_1)^2 }
{2\varepsilon_k\,(2L)^{d-1}}\label{eq2.40}\\
&-&\sum_{k}\frac{1}{4\varepsilon_k\,(2L)^{d-1}}\frac{\p}{\p x_1}\(v^{k_1}(x_1)
\frac{\p v^{k_1}(x_1)}{\p x_1}\)\nn\\
&=&\sum_{k}\frac{(\rho_{k_1}^2-V(x_1)+2\omega_{k_i}^2)\, v^{k_1}(x_1)^2 }
{4\varepsilon_k\,(2L)^{d-1}}\nn \\
&-&\sum_{k}\frac{1}{4\varepsilon_k\,(2L)^{d-1}}
\(\frac{\p v^{k_1}(x_1)}{\p x_1}\)^2 \nn\quad .
\eea 
The first equality in eq. (\ref{eq2.38}) shows that
\be
\int\, dx\,\langle 0|\, T^{00}\,|0\rangle\,=\,{1\over 2}\,\sum_k\,\epsilon_k
\ee
as one expects.
\subsection{Lattice system}
The Casimir system will now be redefined on a finite lattice, which is 
given
the same lattice constant $a$ in all 
spatial directions. Thus spatial position becomes $x=(x_1,\,\cdots,\,x_d)\,=
\,a\,(n_1,\,\cdots,\, n_d)$
where $n_1,\,\cdots,\, n_d$ all take integer values. The planar Dirichlet boundaries
at $x_1=0,\, l$ have lattice positions $n_1=0$ and $n_1=l/a\equiv N_1$. The  continuum
potential $V(x_1)$ becomes a lattice potential $V(n_1)\ge 0$. The free directions 
$\vec{x}_\perp\,=\,a\,\vec{n}_\perp$ are still periodic with period $2 L$:
\be 
-L\le x_i \le L\quad \longrightarrow\quad -\frac{L}{a}< n_i\le \frac{L}{a}=N_\bot\,,
\quad i=2,\ldots,d \, ,
\ee 
i.e. the lattice is periodic with period $2 N_\bot$ in the spatial directions
$i\,=\,2,\,\cdots\, ,\,d$. On the
lattice the Dirac $\delta$-function $\delta(x_1-x_1')$ is replaced by 
$\delta_{n_1,n_1'}/a$
with $\delta_{n_1,n_1'}$ the usual Kronecker $\delta$. Delta functions 
$\delta_{n_i n'_i}$ in the periodic directions $i>1$ are periodic of course.
\subsubsection{Quantization}
\label{sec.Quantization} 
We now quantize the lattice Casimir system carefully
 and in detail, using the 
Schr\"odinger picture. On the lattice the classical field $\phi(x,t)$ 
becomes
\be
\phi(x,t)=\phi(x_1,\vec{x}_\bot,t)\longrightarrow \phi_{n}(t)=\phi_{n_1,
\vec{n}_\bot}(t)\;.
\ee
subject to the boundary conditions
\bea
\phi_{0,\vec{n}_\bot}(t)&=&\phi_{N_1,\vec{n}_\bot}(t)=0\,,\\
\phi_{n_1,\vec{n}_\bot}(t)&=&\phi_{n_1,\vec{n}_\bot+2N_\bot\vec{e}_i}(t)\; ,
\quad i=2,\ldots,d\;.\nn
\eea
As before we ignore the external regions $n_1<0$ and $n_1>N_1=l/a$ beyond the 
Dirichlet planes. Since $\phi\equiv0$ on these planes the interior lattice $G$ on 
which $\phi(x)$ is a quantum field does {\em not} include these planes:
\be
G=\{n\in{\bf Z}^d|1\le n_1\le (N_1-1)\,;(-N_\bot+1)\le n_i \le N_\bot\,,
\quad i=2,\ldots d\,\}.
\ee
From eq. (\ref{eq2.1}) we obtain the classical Lagrangian
\bea\label{eq3.3}
L&=&\frac{1}{2}\left\{\sum_{n_1=0}^{N_1-1}+\sum_{n_1=1}^{N_1}\right\}
\sum_{n_2=-N_\bot+1}^{N_\bot}\!\!\cdots\!\!\sum_{n_d=-N_\bot+1}^{N_\bot}
\frac{a^d}{2}\(\dot{\phi}_{n}^2-\phi_{n}(m^2+V(n_1))\phi_{n}-(\nabla\phi)_{n}^2\) 
\nn \\
&=&\sum_{n\in G}\frac{a^d}{2}\(\dot{\phi}_{n}^2-\phi_{n}(m^2+V(n_1))\phi_{n}+
\phi_{n}(\triangle\phi)_{n}\)\;,
\eea
where
\bea
\sum_{n\in G}\ldots=\sum_{n_1=1}^{N_1-1}\sum_{n_2=-N_\bot+1}^{N_\bot}\cdots
\sum_{n_d=-N_\bot+1}^{N_\bot}\ldots
\eea
and $\triangle\,=\,\triangle_{n,n'}$ is the lattice Laplace operator (see Appendix A).
The specific $n_1$-summation prescription given in (\ref{eq3.3}) avoids introducing an
asymmetry in the $x_1$-direction by the asymmetry of the lattice derivative. 
The classical canonical momenta are
\be
\pi_n(t)=\frac{\p L}{\p \dot{\phi}_n(t)}=a^d\,\dot{\phi}_n(t)\,,
\quad \forall n\in G
\ee
and the classical Hamiltonian is
\bea\label{eq.3.4}
H=\sum_{n\in G}\pi_n(t)\dot{\phi}_n(t)-L=\sum_{n\in G}\, \frac{a^d}{2}
\(\frac{\pi_n^2}{a^{2d}}+\phi_n(m^2+V(n_1))\phi_n-\phi_n(\triangle \phi)_n\)\;.
\eea
Canonical quantization in the Schr\"odinger picture assigns to the classical field
$\phi_n$ and momentum $\pi_n$ time independent operators
\bea
\phi_n(t)\longrightarrow \Phi_n\,,\\
{\pi_n(t)\over a^d}\longrightarrow \Pi_n\,,\nn
\eea
which satisfy the commutation relations
\be
[\Phi_{n},\Phi_{n'}]=[\Pi_{n},\Pi_{n'}]=0 \;,
\ee
\be
\mbox{[}\Phi_{n},\Pi_{n'}]={i\over a^d}\,\delta_{n,n'} \;.
\ee
Note that in the continuum limit $a\to 0$
\bea
\frac{1}{a^d}\,\delta_{n,n'}&\to&\delta(x-x')\\ \label{eq3.6}
\Pi_n &\to&\Pi(x)\;\label{eq3.5}\; .
\eea
The lattice Hamilton operator is
\bea\
{\cal H}=\sum_{n\in G}\frac{a^d}{2}\(\Pi_n^2+\Phi_n(m^2+V(n_1))
\Phi_n-\Phi_n(\triangle\Phi)_n\)\; .
\eea
We go now to the field representation in which 
the field operators $\Phi_n$ are simply multiplicative: 
\be
\Phi_n\,\Psi[\{\phi_m\},t]\,=\,
\phi_n\,\Psi[\{\phi_m\},t]\, .
\ee
Here $\Psi[\{\phi_m\},t]$ is the lattice analogue of the
wave functional $\Psi[\phi,t]$ of continuum theory, i.e. a {\em function} of
the c-number variables $\phi_m$. The canonical-momentum operators $\Pi_n$ become 
the differential operators $-(i/a^d)\, \p/\p\phi_n$.
In this representation the Hamilton operator assumes the form
\bea\label{eq3.7}
{\cal H}&=&\sum_{n \in G}\(-\frac{1}{2a^d}\frac{\p^2}{(\p\phi_n)^2}+\frac{a^d}{2}\,
\phi_n(m^2+V(n_1))\phi_n-\phi_n(\triangle\phi)_n\)\nn\\
&=&\sum_{n\in G}\(-\frac{1}{2a^d}\frac{\p^2}{(\p\phi_n)^2}+\frac{a^d}{2}\,
\phi_n\sum_{n'\in G}{\cal O}_{n,n'}\phi_{n'}\)\; 
\eea
with (see eq. (\ref{A3}))
\bea\label{eq3.9}
{\cal O}_{n,n'}&=&(m^2+V(n_1))\,\delta_{n,n'}-\triangle_{n,n'}\nn\\
&=&(m^2+V(n_1))\,\delta_{n,n'}-\sum_{j=1}^{d}\frac{1}{a^2}\(\delta_{n+\vec{e}_j,n'}-
2\delta_{n,n'}+\delta_{n-\vec{e}_j,n'}\)\;.
\eea
This matrix is symmetric in $(n,\,n')$ and positive definite.
We need its eigenvectors and eigenvalues to diagonalize 
the Hamilton operator. The eigenvectors $\{u_n^k\}$ fulfilling the eigenvalue equation
\be \label{eq3.9a}
\sum_{n'\in G} {\cal O}_{n,n'} u^k_{n'}=\varepsilon^2_k u^k_n\,.
\ee
can be written as a direct product:
\be\label{eq3.10}
 u^k_n=v_{n_1}^{k_1}w^{k_2}_{n_2}\cdots w^{k_d}_{n_d}=v_{n_1}^{k_1}
 w^{\vec{k}_\bot}_{\vec{n}_\bot}\; ,
\ee
with
\[ k=(k_1,k_2,\ldots,k_d)=(k_1,\vec{k}_\bot)\quad .\]
The indices $k$ label the linearly independent vectors, of which there are exactly 
as many as there are points on the lattice -- namely $(N_1-1)\,(2\,N_\perp)^{d-1}$
which is also the dimenson of the real symmetric matrix ${\cal O}_{n,n'}$. One can therefore choose the individual $k$'s so they take the 
same values as the $n_i$ and consequently form the same lattice:
\be 
G=\{k\in{\bf Z}^d|1\le k_1\le (N_1-1)\,;(-N_\bot+1)\le k_i \le N_\bot\,,
\quad i=2,\ldots d\,\}.
\ee
Choosing the lattice as described above guarantees that the boundary conditions are
automatically fulfilled.\hfil\break
The separation formula (\ref{eq3.10}) leads to (compare eqs. (\ref{eq2.8})\,-\,
(\ref{eq2.10}))
\bea
\sum_{n'_1=1}^{N_1-1} \,[V(n_1)\delta_{n_1,n'_1}\,-\,\frac{1}{a^2}
(\delta_{n_1+1,n'_1}-
2\delta_{n_1,n'_1}+\delta_{n_1-1,n'_1}))]\,v^{k_1}_{n'_1}&=&\rho_{k_1}^2 
v^{k_1}_{n_1}\quad ,\quad\label{eq3.13}\\
\sum_{n'_i=-N_\bot+1}^{N_\bot}-\frac{1}{a^2}(\delta_{n_i+1,n'_i}-
2\delta_{n_i,n'_i}+
\delta_{n_i-1,n'_i})\,w^{k_i}_{n'_i}\,=\,\omega_{k_i}^2 w^{k_i}_{n_i}\; ,\;
i&=&2,\cdots,d \label{eq3.14}
\eea
where $\rho_{k_1}^2,\,\omega_{k_i}^2$ are related to the eigenvalues $\varepsilon_k^2$
of (\ref{eq3.9a}) by $\varepsilon_k^2=m^2+\rho_k^2+\sum_{i=2}^d\omega_{k_i}^2$.
While the $v_{n_1}^{k_1}$ can in general be obtained only numerically, the modes
for the free directions can again be gotten analytically:
\bea
w^{k_{i}}_{n_{i}}&=&\sqrt{\frac{1}{2N_\bot}}\,\,e^{i\frac{\pi}{N_\bot}k_{i}n_{i}}\,,
\\
\omega_{k_i}^2&=&\frac{4}{a^2}\,\sin^2 (\frac{\pi k_i}{2N_\bot})\;.\nn
\eea
In summary
\bea
u^k_n&=&v_{n_1}^{k_1}w^{\vec{k}_\bot}_{\vec{n}_\bot}=v_{n_1}^{k_1}
\(\frac{1}{2N_\bot}\)^{\frac{d-1}{2}}e^{i\frac{\pi}{N_\bot}\vec{k}_\bot\vec{n}_\bot}
\;,\label{eq3.16}\\
\varepsilon_k^2&=& m^2+\rho_{k_1}^2+\sum_{i=2}^{d}\frac{4}{a^2}\,
\sin^2 (\frac{\pi k_i}{2N_\bot})\;.\label{eq3.18}
\eea
These formulae are the lattice analogue of eqs. (\ref{eq2.13}), (\ref{eq2.15}) in
the continuum problem with one little difference: the exact lattice analog of the
orthogonality- and completeness relation (\ref{eq2.16}) would be
$\sum_{n\in G}\,a^d\,u_n^k\,(u_n^{k'})^\ast\,=\,\delta_{k,k'}\, ,\,
\sum_{k\in G}\,u_n^k\,(u_n'^k)^\ast\,=\,\delta_{n,n'}/a^d$. In the limit $a\to 0$
these expressions tend to (\ref{eq2.16}). But on the lattice it is more natural
to work with the relations
\bea
\sum_{n\in G} u^k_n (u^{k'}_n)^{\ast} &=& \delta_{k,k'}\;,\label{eq3.19}\\
\sum_{k\in G} u^k_n (u^k_{n'})^{\ast} &=& \delta_{n,n'} \label{eq.3.20}\; .
\eea
It is therefore the combination $u_n^k/\sqrt{a^d}$ which in the limit $a\to 0$ 
tends to $u^k(x)$.\hfil\break
To diagonalize the Hamilton operator we 
make the following expansion of $\phi_n$ in terms of the eigenvectors  
$u_n^k$, 
\bea
\phi_n=\sum_{k\in G} {u^k_n\over \sqrt{a^d}} \,\hat{\phi}_k \quad ,\label{eq3.22}
\eea
with the inversion
\bea
\hat{\phi}_k=\sum_{n\in G} \sqrt{a^d} (u^k_n)^{\ast}\, \phi_n\,=\,
\sum_{n\in G}\, a^d\({u_n^k\over \sqrt{a^d}}\)^\ast\,\phi_n   \;.\label{eq3.22a}
\eea
Because the $\phi_n$ are real we have
\be\label{eq3.23}
\hat{\phi}_k^{\ast}=\hat{\phi}_{k_1,\vec{k}_\bot}^{\ast}=\hat{\phi}_{k_1,-
\vec{k}_\bot}\;.
\ee
Eqs. (\ref{eq3.22}) and (\ref{eq3.22a}) imply for the derivatives
\bea
\frac{\p}{\p \phi_n}&=&\sum_{k\in G}\sqrt{a^d} (u^k_n)^{\ast}\frac{\p}
{\p \hat{\phi}_k}\quad ,\label{eq3.24}\\
\frac{\p}{\p \hat{\phi}_k}&=&\sum_{n\in G} {1\over \sqrt{a^d}} u^k_n \,\frac{\p}
{\p \phi_n}\quad .\label{eq3.25}
\eea
In terms of the $\hat{\phi}_k$ the Hamilton operator (\ref{eq3.7}) becomes
\bea\label{eq3.26}
{\cal H}=\frac{1}{2}\,\sum_{k\in G}\(-\frac{\p^2}{\p\hat{\phi}_k \p\hat{\phi}^{\ast}_k}+
\varepsilon_k^2\,\hat{\phi}_k\hat{\phi}^{\ast}_k \)\; ,
\eea
a finite sum over uncoupled harmonic oscillator Hamiltonians. The 
remaining steps are familiar. We introduce annihilation/creation 
operators
\bea
A_{k}&=&{1\over 
\sqrt{2\epsilon_{k}}}\,\(\epsilon_{k}\hat{\phi}_{k}\,+\,
{\p\over\p\hat{\phi}^{\ast}_{k}}\)\quad , \\
A^{\dagger}_{k}&=&{1\over 
\sqrt{2\epsilon_{k}}}\,\(\epsilon_{k}\hat{\phi}^{\ast}_{k}\,-\,
{\p\over\p\hat{\phi}_{k}}\)\quad ,
\eea
with commutation relations $[A_{k},A^{\dagger}_{k'}]\,=\,\delta_{k k'}\, ,\,
[A_{k},A_{k'}]\,=\,[A^{\dagger}_{k},A^{\dagger}_{k'}]\,=\,0$,
in terms of which the Hamilton operator (\ref{eq3.26}) becomes
\be
{\cal H}\,=\,\frac{1}{2}\,\sum_{k\in G} \varepsilon_k\(A^{\dagger}_{k}A_{k}+
A_{k}A^{\dagger}_{k}\)\,=\,\sum_{k\in G} \varepsilon_k\(A^{\dagger}_{k}A_{k}+
\frac{1}{2}\)\;.\nn
\ee
One recognizes that one is dealing with a finite set of uncoupled 
harmonic oscillators whose frequencies are the spectrum 
$\{\epsilon_{k}\}$. A Fock space can now be constructed, whose ground 
state $|0\rangle$ is of course defined by 
$A_{k}\,|0\rangle\,=\,0\,\forall k$.
The (finite) energy of this state is
 \be\label{eq3.27}
E_{0}=\frac{1}{2}\,\sum_{k\in G}\varepsilon_k\quad .
\ee
The representation of $\phi_n$ in terms of creation and annihilation operators is
\be
\phi_n=\sum_{k\in G}\frac{1}{\sqrt{2\varepsilon_k}}\(\frac{u^k_n}
{\sqrt{a^d}}A_{k}+\frac{(u^{k}_n)^{\ast}}{\sqrt{a^d}}A^{\dagger}_{k}\)\;.
\ee
\subsubsection{Vacuum stress tensor}
\label{sec.spanntens}
To compute the vacuum expectation of the stress tensor operator 
(\ref{eq2.33}) we recall that, in the Schr\"odinger picture and field 
representation,  $\Pi(x,t)\,=\,\dot{\phi}(x,t)\,=\,-i 
{\delta\over\delta\phi(x)}\, .$ Thus
\bea
T^{00}(x)&=&\frac{1}{2}\(-\frac{\delta^2}{(\delta\phi)^2}+(\nabla\phi)^2+
(m^2+V(x_1))\phi^{2}\)\nn\\
&=&\frac{1}{2}\(-\frac{\delta^2}{(\delta\phi)^2}+\phi(m^2+V(x_1)-\triangle)\phi\)+
\frac{1}{2}\nabla(\phi\nabla\phi)\;,\label{eq2.34}\\
T^{ii}(x)
&=&\frac{1}{2}\(-\frac{\delta^2}{(\delta\phi)^2}-\phi(m^2+V(x_1)-\triangle)\phi\)
\label{eq2.35}\\
&-&\frac{1}{2}\nabla(\phi\nabla\phi)+\(\frac{\p\phi}{\p x_{i}}\)^2\,,\nn\\
T^{0i}(x)&=&-\frac{i}{2}\(\frac{\p\phi}{\p x_{i}}\)\frac{\delta}{\delta \phi}-
\frac{i}{2}\frac{\delta}{\delta \phi}\(\frac{\p\phi}{\p x_{i}}\)\;,\nn\\
T^{ij}(x)&=&\(\frac{\p\phi}{\p x_{i}}\)\(\frac{\p\phi}{\p x_{j}}\)\;,\quad i\neq j 
\,.\nn
\eea
With the help of the correspondence (\ref{eq3.5}) and eqs. (\ref{A10}), 
(\ref{A11}) one recasts (\ref{eq2.34}), (\ref{eq2.35}) in the form
\bea
T^{00}_n&=&\frac{1}{2}\(-\frac{1}{a^{2d}}\frac{\p^2}{(\p \phi_n)^2}+
\phi_n\sum_{n'\in G}{\cal O}_{n,n'}\phi_{n'}\)+\sum_{j=1}^{d}
\frac{\phi_{n+\vec{e}_j}^2-2\phi_n^2+\phi_{n-\vec{e}_j}^2}{4a^2}\,,\quad\label{eq3.31}\\
T^{ii}_n&=&\frac{1}{2}\(-\frac{1}{a^{2d}}\frac{\p^2}{(\p \phi_n)^2}-
\phi_n\sum_{n'\in G}{\cal O}_{n,n'}\phi_{n'}\)-\sum_{j=1}^{d}
\frac{\phi_{n+\vec{e}_j}^2-2\phi_n^2+\phi_{n-\vec{e}_j}^2}{4a^2}\nn\\
& &-\phi_n\frac{\phi_{n+\vec{e}_i}-2\phi_n+\phi_{n-\vec{e}_i}}{a^2}+
\frac{\phi_{n+\vec{e}_i}^2-2\phi_n^2+\phi_{n-\vec{e}_i}^2}{2a^2}\label{eq3.32}\,\,.
\eea 
The next step is to take the vacuum expectation values of the 
operators (\ref{eq3.23}), (\ref{eq3.24}). For this purpose it is 
useful to note that
\bea
\hat{\phi}_{k}&=&\hat{\phi}_{k_1,\vec{k}_\bot}=\frac{1}{\sqrt{2\varepsilon_k}}
\(A_{k_1,\vec{k}_\bot}+A^{\dagger}_{k_1,-\vec{k}_\bot}\)\;,\label{eq2.28}\\ 
\frac{\p}{\p\hat{\phi}_{k}}&=&\frac{\p}{\p\hat{\phi}_{k_1,\vec{k}_\bot}}=
\sqrt{\frac{\varepsilon_k}{2}}\(A_{k_1,-\vec{k}_\bot}-A^{\dagger}_{k_1,
\vec{k}_\bot}\)\,.\label{eq.2.29}
\eea
One then verifies
\bea
\langle 0|\hat{\phi}_{k}\hat{\phi}_{k'}^{\ast}|0\rangle&=&\frac{1}{2\varepsilon_k}\,
\delta_{k,k'}\;,\\
\langle 0|\frac{\p^2}{\p\hat{\phi}_{k}\p\hat{\phi}_{k'}^{\ast}}|0\rangle&=& -
\frac{1}{2}\,\varepsilon_k\,\delta_{k,k'}\;,\nn\\
\langle 0|\frac{\p}{\p\hat{\phi}_{k}}\hat{\phi}_{k'}|0\rangle\,=\,-
\langle 0|\hat{\phi}_{k}\frac{\p}{\p\hat{\phi}_{k'}}|0\rangle&=&\frac{1}{2}\,
\delta_{k,k'}\; ,\nn
\eea
and
\bea
\langle 0|-\frac{\p^2}{a^d(\p \phi_n)^2}|0\rangle&=&\langle 0|a^d\phi_n
\sum_{n'\in G}{\cal O}_{n,n'}\phi_{n'}|0\rangle \\
&=&\frac{1}{2}\sum_{k\in G}\varepsilon_k\,|u^k_n|^2\,\, .\nn
\eea
Using (\ref{eq3.13}), (\ref{eq3.14}) and the completeness relations 
for the eigenvectors one finds
\bea
\langle 0|T^{00}_n|0\rangle&=&\sum_{k\in G}\frac{\varepsilon_k}{2a^d}|u^{k}_n|^2+
\sum_{k\in G}\sum_{j=1}^{d}\frac{1}{8 a^{d+2} \varepsilon_k}
\(|u^k_{n+\vec{e}_j}|^2-2|u^k_n|^2+|u^k_{n-\vec{e}_j}|^2\)\, ,\label{eq3.33}\\
\langle 0|T^{11}_n|0\rangle&=&\sum_{k}\frac{\rho_{k_1}^2-V(n_1)}{2a^d\varepsilon_k}
\,|u^{k}_n| ^2+\sum_{k\in G}\frac{1}{8 a^{d+2} \varepsilon_k}
\(|u^k_{n+\vec{e}_1}|^2-2|u^k_n|^2+|u^k_{n-\vec{e}_1}|^2\)\nn\\
& &\hspace{1cm}-\sum_{k\in G}\sum_{j=2}^{d}\frac{1}{8 a^{d+2} \varepsilon_k}
\(|u^k_{n+\vec{e}_j}|^2-2|u^k_n|^2+|u^k_{n-\vec{e}_j}|^2\)\label{eq3.34}
\eea
and for $i\,=\,2,\,\cdots,\, d$
\bea
\langle 0|T^{ii}_n|0\rangle&=&\sum_{k}\frac{\omega_{k_i}^2}{2a^d\varepsilon_k}
\,|u^{k}_n| ^2+\sum_{k\in G}\frac{1}{8 a^{d+2}\varepsilon_k}
\(|u^k_{n+\vec{e}_i}|^2-2|u^k_n|^2+|u^k_{n-\vec{e}_i}|^2\)\nn\\
& &\hspace{1cm}-\sum_{k\in G}\sum_{j=1\atop j\neq i}^{d}\frac{1}{8 a^{d+2} 
\varepsilon_k}\(|u^k_{n+\vec{e}_j}|^2-2|u^k_n|^2+|u^k_{n-\vec{e}_j}|^2\)\;\;.
\label{eq3.35}
\eea
Finally from eq. (\ref{eq3.16})
\be
|u^k_n|^2=\frac{(v^{k_1}_{n_1})^2}{(2N_\bot)^{d-1}}
\ee 
so the preceding formulae can be simplified. Thus with $L\,=\,a\,N_\perp$
\bea
\langle 0|T^{00}_n|0\rangle&=&\sum_{k\in G}\frac{\varepsilon_k\,(v^{k_1}_{n_1})^2}
{2 a\,(2L)^{d-1}}+\sum_{k\in G}\frac{(v^{k_1}_{n_1+1})^2-2(v^{k_1}_{n_1})^2+
(v^{k_1}_{n_1-1})^2}{8 a^3 \varepsilon_k (2L)^{d-1}}\,\, ,\label{eq3.36}\\
\langle 0|T^{11}_n|0\rangle&=&\sum_{k\in G}\frac{(\rho_{k_1}^2-V(n_1))
(v^{k_1}_{n_1})^2}{2a\varepsilon_k\,(2L)^{d-1}}\nn\\
&+&\sum_{k\in G}\frac{(v^{k_1}_{n_1+1})^2-2(v^{k_1}_{n_1})^2+(v^{k_1}_{n_1-1})^2}
{8 a^3 \varepsilon_k (2L)^{d-1}}\,\, ,\label{eq3.37}\\
\langle 0|T^{ii}_n|0\rangle&=&\sum_{k\in G}\frac{\omega_{k_i}^2(v^{k_1}_{n_1})^2}
{2a\varepsilon_k\,(2L)^{d-1}}\nn\\
&-&\sum_{k\in G}\frac{(v^{k_1}_{n_1+1})^2-2(v^{k_1}_{n_1})^2+(v^{k_1}_{n_1-1})^2}
{8 a^3 \varepsilon_k (2L)^{d-1}}\,\, ,\quad i\ge2 \, .\label{eq3.38}
\eea
Eqs. (\ref{eq3.36})\,-\,(\ref{eq3.38}) are the lattice analogues of
eqs. (\ref{eq2.38})\,-\,(\ref{eq2.40}) in the continuum problem. One can transform
the latter directly to the lattice with the help of eqs. (\ref{A10}), 
(\ref{A11}) and the
correspondence
\be 
u^k(x) \to \frac{u^k_n}{\sqrt{a^d}}\quad ,\quad \(v^{k_1}(x_1)\)^2\to 
\frac{(v^{k_1}_{n_1})^2}{a}\;.
\ee 
Eqs. (\ref{eq3.36})\,-\,(\ref{eq3.38}) are valid on the lattice $n\in G$. However
they do not hold for the Dirichlet boundary surfaces $n_1=0,\, N_1$ which have to
be dealt with separately. On these surfaces one must use the unsymmetric derivative
formula discussed in Appendix A which ensures a smooth continuation of the lattice 
mathematics onto the Dirichlet boundary surfaces.\hfil\break
The calculation of $\langle T^{\mu\nu}\rangle$ on the Dirichlet surfaces using the 
one-sided derivative is equivalent (see eqs. (\ref{A8}), (\ref{A9})) to introducing in 
eqs. (\ref{eq3.36})\, -\,(\ref{eq3.38}) the additional definitions
\bea
(v^{k_1}_{-1})^2&:=&(v^{k_1}_{1})^2\,\,,\label{eq3.39}\\
(v^{k_1}_{N_1+1})^2&:=&(v^{k_1}_{N_1-1})^2\,\,.\label{eq3.40}
\eea
Thus on the Dirichlet boundary surfaces
\bea
\langle 0|T^{00}_{0,\vec{n}_\bot}|0\rangle&=&\langle 0|T^{11}_{0,\vec{n}_\bot}
|0\rangle =\sum_{k\in G}\frac{(v^{k_1}_{1})^2}{4 a^3 \varepsilon_k (2L)^{d-1}}=
-\langle 0|T^{ii}_{0,\vec{n}_\bot}|0\rangle \,\,,\\
\langle 0|T^{00}_{N_1,\vec{n}_\bot}|0\rangle&=&\langle 0|T^{11}_{N_1,\vec{n}_\bot}
|0\rangle=\sum_{k\in G}\frac{(v^{k_1}_{N_1-1})^2}{4 a^3 \varepsilon_k (2L)^{d-1}}=
-\langle 0|T^{ii}_{N_1,\vec{n}_\bot}|0\rangle \,\,.\nn
\eea  
We see that the unregularized $\langle T^{00}\rangle$ and $\langle T^{11}\rangle$ are
the same on the boundary surfaces. However, the renormalizations of 
$\langle T^{00}\rangle$ and $\langle T^{11}\rangle$ turn out to be different 
(secs. 3 and 4). Thus the renormalized tensor components $\langle T^{00}\rangle_{ren}$
and $\langle T^{11}\rangle_{ren}$ on the boundary surfaces end up being different.\hfil
\break
Using the lattice definition  (\ref{A16}) of "spatial integration" along $x_1$ we find
\bea\label{eq3.41}
\frac{1}{2}\(\sum_{n_1=1}^{N_1}+\sum_{n_1=0}^{N_1-1}\)\sum_{\vec{n}_\bot}\langle 
0|T^{00}_n|0\rangle\,a^d=\sum_{k\in G}\frac{1}{2}\varepsilon_k=E_0\,\, .
\eea
This relation must hold, of course, for $\langle T^{00}_n\rangle$ to be
the unrenormalized lattice vacuum energy density.
\subsubsection{The case $V(n_1)\,=\,0$}
\label{sec.nullpot}
When the lattice potential $V(n_1)\,=\,0$ between the Dirichlet walls $n_1=0,\,N_1$ one 
can easily solve the lattice Schr\"odinger mode equation (\ref{eq3.13}). The 
eigenvectors and eigenvalues are
\bea
v^{k_1}_{n_1}&=&\sqrt{ \frac{2}{N_1}}\, \sin (\frac{\pi}{N_1}k_1 n_1)\;,
\label{eq3.42}\\
\rho_{k_1}^2 &=& \frac{4}{a^2}\sin^2(\frac{\pi k_1}{2N_1})\;,\quad k_1=1,2,3,
\ldots\,,\, N_1\quad .\label{eq3.43}
\eea
Thus
\bea
u^k_n&=&\sqrt{\frac{2}{N_1}}\,\sin(\frac{\pi k_1}{N_1}n_1)\(\frac{1}
{2N_\bot}\)^{\frac{d-1}{2}}e^{i\frac{\pi}{N_\bot}\vec{k}_\bot\vec{n}_\bot}\,,\nn\\
\varepsilon^2_k&=&m^2+\frac{4}{a^2}\sin^2(\frac{\pi k_1}{2N_1})+\sum_{i=2}^{d}
\frac{4}{a^2}\sin^2(\frac{\pi k_i}{2N_\bot})\;.\label{eq3.45}
\eea
The lattice vacuum energy is
\be\label{eq3.46}
E_0=\frac{1}{2}\,\sum_{k\in G}\varepsilon_k=\frac{1}{2}\,\sum_{k\in G}
\sqrt{m^2+\frac{4}{a^2}\sin^2(\frac{\pi k_1}{2N_1})+\sum_{i=2}^{d}\frac{4}{a^2}
\sin^2(\frac{\pi k_i}{2N_\bot})}\,\, .
\ee
In computing the vacuum stress tensor we make use of the identities
\bea
\sin^2(\alpha \,(n+1))-2\sin^2(\alpha\, n)+\sin^2(\alpha \,(n-1))&=&2\sin^2(\alpha)
\cos(2\alpha \,n) \quad,\quad\\
\sin^2(\alpha )&=&4\sin^2\(\frac{\alpha}{2}\)\(1-\sin^2\(\frac{\alpha}{2}\)\)\,.\nn
\eea
Thus
\bea\label{eq3.47}
& &\hspace{-1cm}\sin^2(\frac{\pi k_1}{N_1}(n_1+1))-2\sin^2(\frac{\pi k_1}{N_1}n_1)+
\sin^2(\frac{\pi k_1}{N_1}(n_1-1))=\quad\quad\\
& &\hspace{2cm}2a^2\rho_{k_1}^2\cos\(\frac{2\pi k_1 n_1}{N_1}\)-2a^2\rho^2_{k_1}
\sin^2\(\frac{\pi k_1}{2N_1}\)\cos\(\frac{2\pi k_1n_1}{N_1}\)\;.\nn
\eea
Then from eqs. (\ref{eq3.36})\,-\,(\ref{eq3.38}) and with $l=N_1/a$
\bea
\langle 0|T^{00}_n|0\rangle&=&\frac{1}{l\,(2L)^{d-1}}\sum_{k\in G}\frac{1}{2}\,
\varepsilon_k-\sum_{k\in G} \frac{(\varepsilon_k^2-\rho_{k_1}^2)}{2l(2L)^{d-1}\,
\varepsilon_k}\cos\(\frac{2\pi k_1 n_1}{N_1}\)\nn\\
& &-\sum_{k\in G}\frac{\rho_{k_1}^2}{2l(2L)^{d-1}\,\varepsilon_k}\sin^2\(
\frac{\pi k_1}{2N_1}\)\cos\(\frac{2\pi k_1n_1}{N_1}\)\,,\label{eq3.48}\\
\langle 0|T^{11}_n|0\rangle&=&\sum_{k\in G}\frac{\rho_{k_1}^2}{2l\,(2L)^{d-1}
\varepsilon_k}\nn\\
&-&\sum_{k\in G}\frac{\rho_{k_1}^2}{2l(2L)^{d-1}\,\varepsilon_k}\sin^2\(\frac{\pi k_1}
{2N_1}\)\cos\(\frac{2\pi k_1n_1}{N_1}\)\,,\label{eq3.49} \\
\langle 0|T^{ii}_n|0\rangle&=&\frac{1}{l\,(2L)^{d-1}}\sum_{k\in G}\frac{1}{2}\,
\omega_{k_i}^2-\sum_{k\in G}\frac{\rho_{k_1}^2+\omega_{k_i}^2}{2l(2L)^{d-1}\,
\varepsilon_k}\cos\(\frac{2\pi k_1 n_1}{N_1}\)\nn\\
& &+\sum_{k\in G}\frac{\rho_{k_1}^2}{2l(2L)^{d-1}\,\varepsilon_k}\sin^2\(\frac{\pi k_1}
{2N_1}\)\cos\(\frac{2\pi k_1n_1}{N_1}\)\,,\quad i\geq 2\,.\label{eq3.50}
\eea
Comparing eqs. (\ref{eq3.48})\,-\,(\ref{eq3.50}) and the continuum formulae
(\ref{eq2.38})\,-\,(\ref{eq2.40}) with background potential $V(x_{1})=0$  and 
$v^{k_{1}}(x_{1})=\sqrt{2/l}\,\sin(\pi k_{1} x_{1}/l),\, 
k_{1}=1,2,3,\cdots$ one finds in the former the presence of a 
{\em lattice artefact} term
\be\label{eq3.51}
A\,\equiv\,\sum_{k\in G}\frac{2\sin^4(\frac{\pi k_1}{2N_1})\cos(\frac{2\pi k_1n_1}{N_1})}
{a^2l(2L)^{d-1}\,\varepsilon_k}\,\, .
\ee
This term has no continuum counterpart and it diverges for $a\to 0$. In Appendix B we
investigate the 1D version of $A$ in considerable detail and conclude that it is 
essentially devoid of physical information. 
\section{One dimensional lattice without background potential}
\label{sec.ohneback}
\setcounter{equation}{0}
In this and the following section we specialize to one-dimensional lattice systems 
with no additional free directions $x_2,\,\cdots,\,x_d$. We drop the subscript 1
and first consider the case of vanishing potential $V(x)$. In sec. 4
we will introduce a specific nonzero potential along the lattice. In both sections 
we wish to deal with Casimir forces. This means we must also consider one or more regions 
{\em external} to the interval $0<x<l$ or $0<n<N=l/a$. This is easily done. We 
can attach external regions which are exactly like the internal region $0<x<l$
(except for their length) with $V(x)=0$ and a Dirichlet boundary at either end. When
we calculate the lattice vacuum energy $E_{vac}(l)$ in $0<x<l$ the same calculation
gives us the lattice vacuum energy $E_{vac}(L-l)$ in the adjacent interval $l<x<L$.
Likewise one calculation gives us the lattice $T^{\mu\nu}$ in both intervals. Our
principal goals here are to understand in detail how to renormalize $E_{vac}$ and
$T^{\mu\nu}$ in the continuum limit $a\to 0$. Then we can calculate the Casimir
force on the common Dirichlet boundary $x=l$ in two ways: globally using $E_{vac}$
and locally using $T^{11}$. Comparison with the continuum version of this system will 
be important. We shall find that in every respect the lattice theory closely
approximates the continuum system as expected. It will be convenient to discuss
the cases of zero and nonzero field mass separately.\hfil\break
\subsection{Renormalized vacuum energy}
\label{sec.renormalizedVE}
\subsubsection{Zero mass}
\label{sec.m0}
We find from eq. (\ref{eq3.46}) the lattice vacuum energy in the internal region 
$0\le x\le l$:
\be\label{eq4.1}
E_{reg}(l,a)=\frac{1}{2} \sk \rho_{k}=\frac{1}{a} \sk \sin{(\frac{\pi k}{2N})}
\ee
where $N=l/a$. The label \lq\lq {\it reg}" here (and subsequently) is used to denote lattice
quantities. The finite sum (\ref{eq4.1}) can be evaluated using \cite{ref7}
\be
\sk \sin(kx)=\sin\(\frac{Nx}{2}\)\sin\(\frac{(N-1)x}{2}\)\csc\(\frac{x}{2}\)
\ee 
with the result
\bea\label{eq4.2} 
E_{reg}(l,a)=\frac{1}{2a}\( \frac{1}{\tan(\frac{\pi a}{4l})}-1 \) \, . 
\eea
Expanding in powers of $a$ yields
\be\label{eq4.2a}
E_{reg}(l,a)=\frac{2l}{\pi a^2}-\frac{1}{2a}-\frac{\pi}{24l}+{\cal O}(a^2)\, .
\ee
Subtracting all terms in eq. (\ref{eq4.2}) which diverge in the continuum limit 
$a\to 0$ we recover the continuum Casimir energy for a real 
1D massless scalar field confined between Dirichlet boundaries \cite{ref8,ref9,ref10},
\bea\label{eq4.4}
E(l)=\lim_{a\to 0}\(E_{reg}(l,a)-\frac{2l}{\pi a^2}+\frac{1}{2a}\)=-
\frac{\pi}{24\,l}\;
\;.
\eea
To directly derive $E(l)$ from the lattice energy (\ref{eq4.1}) one can also do the 
following:
\bea
\label{eq4.5}
\lim_{N\to\infty}\left[ \sk \(\frac{1}{N\sink}\)^s\right]&=&
\(\frac{2}{\pi}\)^s\sum_{k=1}^{\infty}\(\frac{1}{k}\)^s\nn \\ 
&=& \(\frac{2}{\pi}\)^s\zeta(s),\qquad \rm{Re}(s)>1 \;
\eea  
where $\zeta(s)$ is the Riemann $\zeta$ function \cite{ref11}. Because $\zeta(s)$
can be continued throughout the $s$-plane (and is a meromorphic function with its 
only pole at $s=1$) one can use the preceding formula to evaluate the $N\to\infty$ 
limit of eq. (\ref{eq4.1}):
\bea
\lim_{N\to\infty}\,E_{reg}(l,a)&=&\lim_{N\to\infty}{1\over a N}\,
\sum_{k=1}^{N-1}N\,\sin{\pi k\over 2 N} \\
&=&{1\over l} {\pi\over 2}\,\zeta(-1)\,=\,-{\pi\over 24 l}\nn \quad .
\eea
To have a more complete physical system we add the external region $l\le x\le L$.
Exactly the same calculations with $l$ replaced by $L-l$ give the lattice and 
continuum vacuum energies in this region; these are $E_{reg}(L-l,a)$ and
$E(L-l)$ respectively. The total lattice and continuum vacuum energies in 
$0\le x\le L$ are then
\bea
E(l,L,a)&\equiv&E_{reg}(l,a)+E_{reg}(L-l,a) \quad ,\\
E(l,L)&\equiv& E(l)+E(L-l)\quad .\nn
\eea
Similarly an external region $-L'\le x\le 0$ could be attached to the other side 
of the interval $[0,l]$. Additional external regions beyond these could be attached.
Global structure on $x$ could also be imposed; e.g. periodicity on some large interval,
or noncompactness for $x\to \pm\infty$. We shall do none of these things here. As stated,
our sole purpose will be to probe the closeness of the lattice and continuum 
descriptions in $0\le x\le L$.\hfil\break
Consider for example the global Casimir force on the Dirichlet boundary at $x=l$.
The continuum definition of this force (in the direction of positive $x$) is
\bea\label{eq4.6}
F_{Cas}(l,L)&=&-{\partial\over \partial l}\,E(l,L)\nn \\
&=&-{\pi\over 24\, l^2}\,+\,{\pi\over 24\, (L-l)^2}\quad .
\eea
On the lattice one would define
\be\label{eq4.7}
F_{Cas}(l,L,a)=-\frac{E_{reg}(l+a,a)+E_{reg}(L-(l+a),a)-E_{reg}(l,a)-E_{reg}(L-l,a)}
{a}\,.
\ee
In the limit $a\to 0$ and using eq. (\ref{eq4.2}) one verifies that the same Casimir
force (\ref{eq4.6}) is recovered without the need to perform explicit subtractions.
\subsubsection{Nonzero mass}
\label{sec.m>0}
For nonzero mass the lattice vacuum energy in the internal region $0\le x\le l$
\bea\label{eq4.8}
E_{reg}(m,l,a)&=&\frac{1}{2}\sk\sqrt{m^2+\frac{4}{a^2}\sin^2(\frac{\pi k}{2N})}\nn\\
&=&\frac{1}{2l}\sk\sqrt{(ml)^2+4 N^2\sin^2(\frac{\pi k}{2N})} 
\eea
evidently cannot be evaluated in closed form. It can, however, usefully be expanded 
in powers of $m l$:
\bea\label{eq4.9}
E_{reg}(m,l,a)&=&\frac{N}{l}\sk \sink +\frac{1}{8} m^2 l \sk \frac{1}{N\sink}\nn\\
& &+\frac{1}{l}\sum_{\nu =2}^{\infty}c_{\nu}\(\frac{ml}{2}\)^{2\nu}\sk \(\frac{1}
{N\sink}\)^{2\nu -1} \, ,\\
c_\nu&=&{{1/2}\choose{\nu}}\,=\,{\Gamma(3/2)\over \nu!\,\Gamma(3/2-\nu)}\quad .\nn
\eea
The first term in eq. (\ref{eq4.9}) is just the $m=0$ energy (\ref{eq4.1}) which
we already know how to renormalize. All of the terms in eq. (\ref{eq4.9}) labeled
 by $\nu\ge 2$ are convergent in the continuum limit. Each of them can, in this limit, 
 be expressed in terms of the Riemann $\zeta$ function using eq. (\ref{eq4.5}) with
 $s$ values $s=2\nu-1=3,\, 5,\cdots$ for which this latter series converges. Hence 
they require no renormalization. However the second term in eq. (\ref{eq4.9}) does
require renormalization. This is evident from eq. (\ref{eq4.5}) where evaluation
of the term in question corresponds to using $\zeta(s)$ at its pole $s=1$. 
More directly, the second term in eq. (\ref{eq4.9}) diverges logarithmically for 
$N\to\infty$. Indeed for $N\gg k$
\be
N\sink \approx \frac{\pi k}{2}
\ee 
and it is well known that 
\bea
\lim_{N\to\infty}\(\sum_{k=1}^{N}\frac{1}{k}-\ln{N}\) \equiv 
\gamma = 0.5772\ldots 
\eea
is a definition of Euler's constant $\gamma$. Analogously we define here
\bea\label{eq4.10}
\sk \frac{1}{N\sink} &=& \frac{2}{\pi}\ln{N}+ c(N)\, , \\
\lim_{N\to\infty}(c(N))&=& 0.52125\ldots \nn
\eea
which specifies another sequence $\{c(N)\}$ whose limiting value for $N\to \infty$ can
be obtained numerically with the result shown in eq. (\ref{eq4.10}). Thus the second 
term on the right of eq. (\ref{eq4.9}) behaves as 
\be\label{eq4.10a}
-{1\over 4 \pi} m^2 l\, \ln({a\over l})\,+\,{m^2 l\over 8}\, c(N) \quad , 
\ee 
i.e. there is an additional logarithmic divergence which has
to be subtracted. By subtracting from $E_{reg}(m,l,a)$ the first term of (\ref{eq4.10a})
as it stands, we would incorrectly discard the $l$-dependence of this term. We must therefore 
split this term 
into the finite term ${1\over 4 \pi}\, m^2 l\, \ln( \mu l)$ and a term 
$-{1\over 4 \pi}\, m^2 l \,\ln( \mu a)$ diverging for $a\to 0$. Here $\mu$ is an arbitrary constant with dimension 
[mass] which has to be introduced to make the arguments of the logarithms
dimensionless. We denote by $E(m,l,a,\mu)$ the expression obtained by subtracting
this latter term, in addition to the divergent terms found in (\ref{eq4.2}), from
$E_{reg}(m,l,a)$, i.e.
\bea\label{eq4.11}
E(m,l,a,\mu)&\equiv&E_{reg}(m,l,a)-\frac{2l}{\pi a^2}+\frac{1}{2a}+\frac{m^2 l}{4\pi}
\ln(\mu a)\quad .
\eea
$E(m,l,a,\mu)$ converges in the limit $a\to 0$. The renormalized continuum vacuum 
energy in the internal region $0\le x\le l$ can therefore be defined to be
\be
E(m,l,\mu):=\lim_{a\to 0}E(m,l,a,\mu)\,.
\ee  
All the preceding global formulae  can be transferred to the external region
$l\le x\le L$ by simply replacing $l$ with $L-l$. The total vacuum energy for the 
system comprising the region $0\le x \le L$ with Dirichlet boundaries at $x=0,\, l$ and
$L$ is $E(m,l,a,\mu)\,+\,E(m,L-l,a,\mu)$. The Casimir force on the boundary 
point at $x=l$ is then
\bea\label{eq4.12}
F_{Cas}(m,l,L)\,=\,-\lim_{a\to 0}\,{1\over a}\,[E(m,l+a,a,\mu)&+&E(m,L-l-a,a,\mu)\\
&-& E(m,l,a,\mu)\,-\,E(m,L-l,a,\mu)]\quad .\nn
\eea
One can easily convince oneself that the divergences which have to be subtracted in order
to make the renormalized continuum vacuum energy $E(m,l,\mu)$ finite, cancel in the 
difference on the right of eq. (\ref{eq4.12}) in the limit $a\to 0$. 
I. e. in the lattice calculation of $F_{Cas}$
these terms do not need to be subtracted from $E_{reg}$ and
the measurable Casimir force, unlike the vacuum energy, does {\em not} depend
on the arbitrary parameter $\mu$. 
\hfil\break

\noindent\underline{Physically fixing $\mu$}\hfil\break
The dependence of the $m>0$ renormalized continuum vacuum energy $E(m,l,\mu)$ on
the arbitrary mass parameter $\mu$ seems unsatisfying. This does not, of course, mean that 
it is wrong. A vacuum energy in QFT evidently need not be absolutely calculable.
(However {\em shifts} in vacuum energies -- i.e. Casimir energies and their 
associated Casimir forces -- should be absolutely calculable as is the case here.)
We now propose a way to select a particular value of $\mu$ which, while not dictated
by general theory, is physically motivated.\hfil\break
To this end let us make the {\em ansatz}
\be
\mu=\kappa m
\ee
where $\kappa$ is a dimensionless proportionality constant which will be determined by
the requirement
\bea\label{eq4.13}
\lim_{l\to \infty}\(-\frac{\p E(m,l,\kappa m)}{\p l}\)=0
\eea
This means that the contribution to  the Casimir force on the boundary $x=l$ is
determined by $-\,{\p\over\p l}\,E(m,l,\mu)$ alone if $L$ is chosen to be $\infty$.
One can verify that for $\mu=m$
\bea 
\lim_{l\to \infty}E(m,l,\mu=m) =-\frac{m^2l}{4\pi}\ln(\chi)-\frac{m}{4}\,,
\label{eq4.14}
\eea
where $\chi$ is a {\em specific} constant which can be determined numerically
with the result $\ln\chi=-2.579436$. Using this $\chi$ we define a new renormalized 
continuum vacuum energy by
\bea\label{eq4.15}
E(m,l)&\equiv&E(m,l,\mu=m)\,+\,{m^2 l\over 4 \pi}\,\ln\chi\nn\\
&=&\lim_{a\to 0}\(E_{reg}(m,l,a)-\frac{2l}{\pi a^2}+\frac{1}{2a}+\frac{m^2l}{4\pi}
\ln(\chi ma)\)\,,
\eea
which has the large-$l$ behavior
\bea\label{eq4.16}
\lim_{l\to \infty} E(m,l)=-\frac{m}{4}\,.
\eea
Hence, if we identify the parameter $\kappa$ introduced above with the numerically
determined constant $\chi$, this vacuum energy $E(m,l)$ achieves the goal set in 
eq. (\ref{eq4.13}). Let us give a bit
more detail. From eqs. (\ref{eq4.4}), (\ref{eq4.8})\,-\,(\ref{eq4.10})
\bea\label{eq4.17}
& &\hspace{-1.5cm}E(m,l)\,=\,-\frac{\pi}{24l}+\frac{1}{4\pi}m^2l\ln(\tilde{\chi} ml)\nn\\
&+&\frac{1}{2l}\lim_{N\to\infty}\left[ \sk \( \sqrt{(ml)^2+4N^2\sin^2
{(\frac{\pi k}{2N})}}-2N\sink-\frac{(ml)^2}{4N\sink}\)\right]\nn\\
&=&-\frac{\pi}{24l}\,+\,\frac{1}{4\pi}m^2l\ln(\tilde{\chi} ml)+\frac{1}{2l}
\sum_{k=1}^{\infty}\( \sqrt{(ml)^2+(\pi k)^2}-\pi k-\frac{(ml)^2}{2\pi k}\)\, 
\eea 
where $\ln(\tilde{\chi})$ differs from $\ln(\chi)$ by $\pi\,c(\infty)/2$, $c(\infty)$
being the constant introduced in (\ref{eq4.10}):
\bea\label{eq4.18}
\ln(\tilde{\chi})=\ln(\chi)+\frac{\pi}{2}\,0.52125\ldots=-1.76066\ldots\,.
\eea 
In fig. 1 we plot $E(m,l)$ versus $l$ for three different masses. An 
arbitrary length unit is involved in Fig. 1. The relevant scaling property of $E(m,l)$ is obvious in 
(\ref{eq4.17}). In natural units ($\hbar=c=1$) the only dimensional 
quantities available are $l$ and $m$, with $m\,l$ dimensionless. 
Thus $l$ sets the scale in (\ref{eq4.17}). $E(m,l)$ is negative,
increasingly so for increasing $m$ and fixed $l$. For fixed $m$ and increasing $l$
 we see that $E(m,l)$ increases, corresponding to an attractive Casimir force $F_{Cas}$
 between the boundaries at $x=0$ and $x=l$, where due to our fixing of $\mu$ no account
 has to be taken of the external regions $x<0,\,x>l$ provided they are chosen to extend
 to infinity. 
\begin{figure}
\centerline{%
\epsfxsize=5in
 \epsfbox{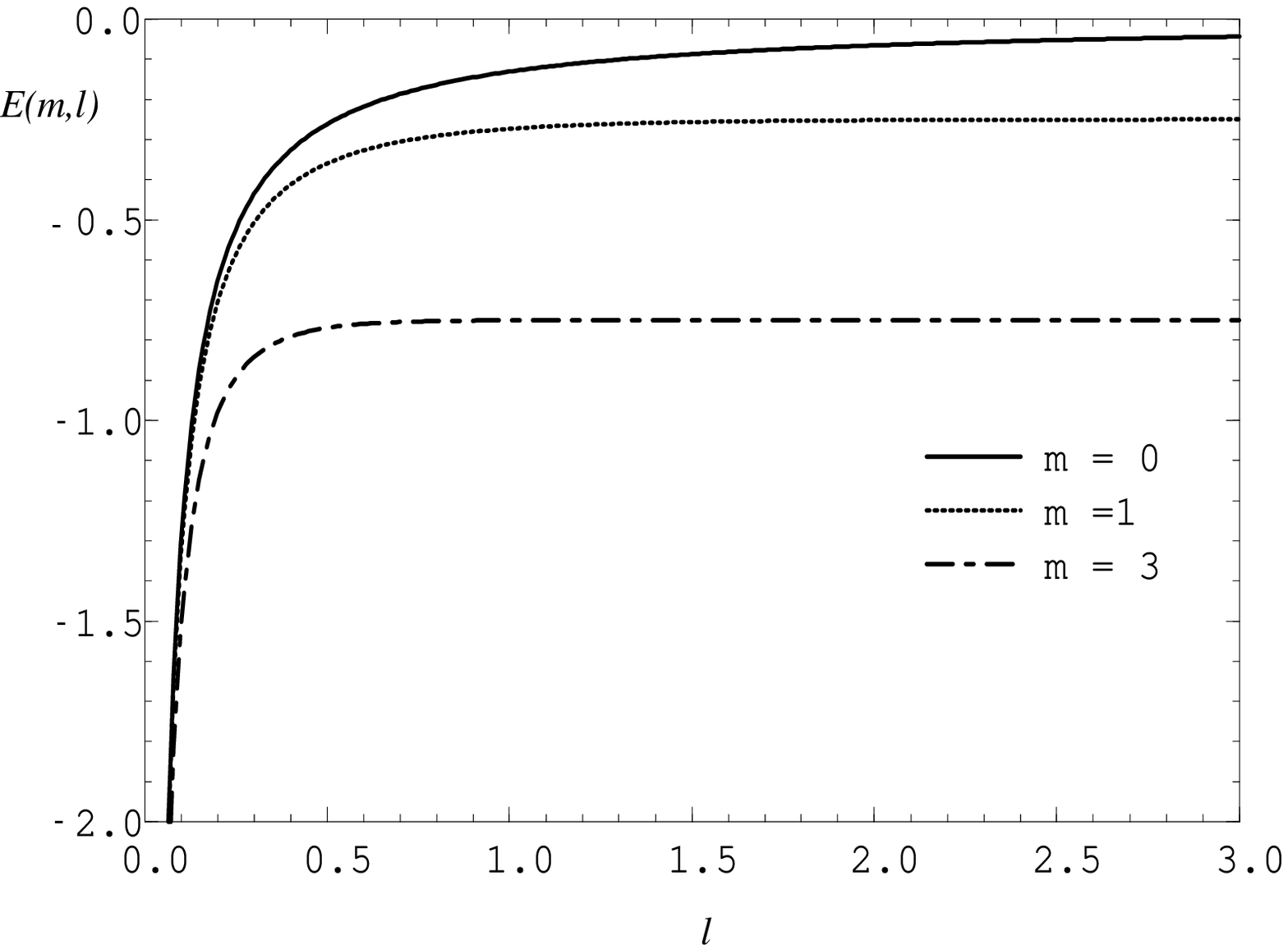}}
\caption[fig. 1]{\parbox[t]{12cm}{$E(m,l)$ from (\ref{eq4.15}) with  
$\chi=0.0758166$. $E(m,l)$ is summed to $k=1000$.\\
In this and the following figures an arbitrary length unit is used.}}
\end{figure} 
In this case the Casimir force on the boundaries at $x=0,\,l$ becomes
\bea\label{eq4.20}
F_{Cas}(m,l)&=&\lim_{a\to 0}\(\frac{2}{a^2}\,\sk \frac{\sin^2(\frac{\pi k}{2N})}
{\sqrt{(ml)^2+4N^2\sin^2(\frac{\pi k}{2N})}}\right.\nn\\
&-&\left.\frac{2}{\pi a^2}+\frac{1}{2al}-
\frac{m^2}{4\pi}\ln(\chi ma)-\frac{m^2}{4\pi}\)\\
&=&-\frac{\pi}{24l^2}-\frac{m^2}{4\pi}(\ln{(\tilde{\chi} ml)}+1)\nn\\
&+&\frac{1}{2l^2}
\sum_{k=1}^{\infty} \(\frac{(\pi k)^2}{\sqrt{(ml)^2+(\pi k)^2}}-\pi k + 
\frac{(ml)^2}{2\pi k}\)\nn\,.
\eea

\noindent \underline{Comparison with analytic results}\hfil\break
Many authors have calculated the vacuum energy shift of the massless and massive 1D 
scalar field
confined between Dirichlet points. An early and useful analysis in arbitrary
dimension was given by Ambj$\o$rn and Wolfram \cite{ref9}. We simply quote their result 
for d=1:
\bea
E_{AW}(m=0,l)&=&-\frac{\pi}{24l}\,,\\
E_{AW}(m,l)&=&-\frac{m}{2\pi}\sum_{n=1}^{\infty}\frac{K_1(2mln)}{n}\,,\nn\\
\lim_{l\to \infty}E_{AW}(m,l)&=&0 \,,\nn
\eea 
where $K_1(z)$ is a modified Bessel function. Note that no arbitrary mass parameter
$\mu$ appears in $E_{AW}(m,l)$ and that $\lim_{l\to\infty} E_{AW}(m,l)$=0, i.e.
 the situation is essentially the same as above after fixing $\mu$ by the requirement
 ${\p\over \p l}\,E(m,l)=0$, the only difference being that our $E(m,l)$ tends to 
 $-m/4$ for $l\to\infty$ whereas $\lim_{l\to\infty}\,E_{AW}(m,l)$\,=\,0. Let us form the ratio
\bea\label{eq4.22}
e_{AW}(ml)\,\equiv\,\frac{E_{AW}(m,l)}{E_{AW}(m=0,l)}=\frac{12ml}{\pi^2}\sum_{n=1}^{\infty}
\frac{K_1(2mln)}{n}
\eea
and the corresponding ratio using our result $E(m,l)$:
\bea\label{eq4.23}
\hspace{-.5cm}e(ml)&\equiv&\frac{E(m,l)+\frac{m}{4}}{E(m=0,l)}
\,=\,1-\frac{6ml}{\pi}-\frac{6(ml)^2}{\pi^2}\ln(\tilde{\chi} ml)\nn\\
&-&\frac{12}{\pi}\sum_{k=1}^{\infty}\( \sqrt{(ml)^2+(\pi k)^2}-\pi k-\frac{(ml)^2}
{2\pi k}\)\; .
\eea
Numerically very close agreement of these ratios up to ${\cal O}(10^{-5})$ - 
${\cal O}(10^{-6})$ is found. Their difference increases for 
large $m l$ as $(ml)^2$;  this arises from our imperfect knowledge of the number
$\ln \tilde{\chi}$. 
\subsection{Renormalized vacuum stress tensor}
\label{sec.renorm}
\subsubsection{$\langle T^{11}\rangle$}
\label{T11}
From eq. (\ref{eq3.49}) we find for the internal region
\be\label{eq4.24}
\T{11}_{reg}(m,l,n,a)=\frac{2}{a^2}\sk\frac{\sin^2(\frac{\pi k}{2N})}{\sqrt{(ml)^2+
4N^2\sin^2(\frac{\pi k}{2N})}}\,+\,A(m,l,n,a)
\ee
where the lattice artefact term
\be 
A(m,l,n,a)\,\equiv\,-\frac{2}{a^2}\sk\frac{\sin^4(\frac{\pi k}{2N})
\cos(\frac{2\pi k n}{N})}{\sqrt{(ml)^2+4N^2\sin^2(\frac{\pi k}{2N})}}
\ee 
has already been mentioned. This term has no counterpart in the continuum theory and
it is unphysical. As discussed in Appendix B it makes no contribution to the Casimir 
force. The artefact $A$ diverges as $a\to 0$ and must be subtracted -- but not yet for 
there is some interplay (or more precisely a cancellation) between the leading
divergence in $A$ and a nonleading divergence in the first sum in eq. (\ref{eq4.24}).\hfil\break
For zero mass $m=0$ one can express eq. (\ref{eq4.24}) using eqs. (\ref{eq4.2}),
( \ref{B11})
in the form
\bea\label{eq4.25}
\T{11}_{reg}(l,n,a)=\frac{2}{\pi a^2}-\frac{\pi}{24l^2}+f(l,n,a)+{\cal O}(a^2)\,,
\eea
where the lattice function $f(l,n,a)$ is defined by eq. (\ref{B12}). Because $f\to 0$ as 
$a\to 0$ in the internal region $0< x< l$ the renormalized continuum tensor for $m=0$ 
in this region is
\bea
\T{11}\,=\,-\frac{\pi}{24\,l^2}=-\frac{\p E(l)}{\p l}
\eea
as it should be (recall eq. (\ref{eq4.6})). For any $m\ge 0$ the continuum 
$\langle T^{11}\rangle$ is {\em constant} between parallel Dirichlet planes in
arbitrary dimension (see e.g. ref. \cite{ref10}). This constant value of
$\langle T^{11}\rangle$ equals the Casimir force/area on the enclosing 
Dirichlet boundaries when no force acts from outside on these boundaries (because e.g.
 the external region is infinite).\hfil\break
 At the lattice level note that the physical first term in eq. (\ref{eq4.24}) is
 indeed independent of lattice position $n$. 
 Thus the lattice formulation 
 has already scored a success in reproducing this known continuum feature:
 \bea\label{eq4.26}
\T{11}_{phys}\,=\,\T{11}(m,l,a)=\frac{2}{a^2}\sk\frac{\sin^2(
\frac{\pi k}{2N})}{\sqrt{(ml)^2+
4N^2\sin^2(\frac{\pi k}{2N})}}\,.
\eea
On the other hand the lattice artefact second term in eq. (\ref{eq4.24})
{\em does} depend on $n$ -- another indication of its unphysical nature.
\hfil\break
It is not difficult to see that 
$\langle T^{11}\rangle(m,l,a)$ coincides with the nonrenormalized Casi\-mir force
obtained from the nonrenormalized global vacuum  energy. Thus 
on the lattice one obtains the same Casimir force using global and local methods,
just as one does in the continuum formulation. Because we know the $a\to 0$ 
divergent structure of eq. (\ref{eq4.26}) from the global problem we can write down
immediately the renormalized result
\bea\label{eq4.27}
\T{11}(m,l,\mu)&=&\lim_{a\to 0}\( \frac{2}{a^2}\,\sk \frac{\sin^2(\frac{\pi k}{2N})}
{\sqrt{(ml)^2+4N^2\sin^2(\frac{\pi k}{2N})}}\right.\\
&-&\left.\frac{2}{\pi a^2}+\frac{1}{2al}-\frac{m^2}{4\pi}\ln(\mu a) \)\, .\nn
\eea
Of course the arbitrary parameter $\mu$ is again present.\hfil\break

\noindent \underline{Physically fixing $\mu$}\hfil\break
The condition chosen here to fix $\mu$ is
\bea
\T{11}(m,l)=-\frac{\p E(m,l)}{\p l}
\eea
where $E(m,l)$ is the fixed-$\mu$ vacuum energy (\ref{eq4.17}). Equivalently
\be 
\lim\limits_{l \to \infty}\T{11}(m,l)=0\quad .
\ee 
It is important to emphasize that in general
\bea
\T{11}(m,l,\mu)\neq -\frac{\p E(m,l,\mu)}{\p l}\,,
\eea
assuming one uses the same $\mu$ on both sides. Directly from eq. (\ref{eq4.18}) we 
obtain the fixed-$\mu$ vacuum tensor
\bea\label{eq4.28}
& &\hspace{-1cm}\T{11}(m,l)\,=\,-\frac{\p E(m,l)}{\p l}\\
& &=\lim_{a\to 0}\(\frac{2}{a^2} \sk \frac{\sin^2(\frac{\pi k}{2N})}{\sqrt{(ml)^2+
4N^2\sin^2(\frac{\pi k}{2N})}}-\frac{2}{\pi a^2}+\frac{1}{2al}-
\frac{m^2}{4\pi}\ln(\tau ma) \)\, ,\nn
\eea
where\bea\label{eq4.29}
\ln(\tau)= \ln(\chi)+1=-1.579438 \ldots \,\,\,.
\eea
Here fixing $\mu$ has led to $\mu=\tau m$. In the global vacuum energy it was
$\mu=\chi m$.

With the $V(n)\ne 0$ system of sec. 4 in mind let us say something more about the 
renormalization of eq. (\ref{eq4.24}). Subtract the lattice artefact term $A$ (using
eq. (\ref{B16}) for it) in addition to the other subtractions in eq. (\ref{eq4.27}):
\bea\label{eq4.30}
\T{11}(m,l,\mu)&=&\lim_{a \to 0}\(\T{11}_{reg}(m,l,n,a)-\frac{2}{\pi a^2}-
\frac{m^2}{4\pi}\ln(\mu a) \right. \nn\\
&&\hspace{4cm}\left.-f(l,n,a)-\frac{m^2}{8}h(\frac{l}{a},n)\)\,,
\eea
or with $\mu=\tau m$
\bea
\T{11}(m,l)&=&\lim_{a \to 0}\(\T{11}_{reg}(m,l,n,a)-\frac{2}{\pi a^2}-\frac{m^2}
{4\pi}\ln(\tau m a) \right. \nn\\
&&\hspace{4cm}\left.-f(l,n,a)-\frac{m^2}{8}h(\frac{l}{a},n)\)\label{eq4.31}\, ,
\eea
where $f,\,h$ are defined in Appendix B. As indicated both of these functions depend
on position $x=n a$. However, in the continuum limit they both vanish everywhere
except {\em on} the boundaries $n=0,\,N$ where they take values independent of
$l$: $f\to const/a^2$ while $h\to 2/\pi$. Subtracting this position dependence in
 eq. (\ref{eq4.31}) yields a $\langle T^{11}\rangle$ independent of lattice position.
 To show this we rewrite eq. (\ref{eq4.31}) in the form
 \bea\label{eq4.32}
\T{11}(m,l)&=&\lim_{a\to 0}\sum_{k=1}^{N-1}\left\{ \frac{2N^2
\sin^2(\frac{\pi k}{2N})}{l^2\sqrt{(ml)^2+4N^2\sin^2(\frac{\pi k}{2N})}}
-\frac{N}{l^2}\sink+\frac{m^2}{8N\sink}\nn\right.\\
& &-\left.\frac{2N^2\sin^4(\frac{\pi k}{2N})\cos(\frac{2\pi k x}{l})}{l^2
\sqrt{(ml)^2+4N^2\sin^2(\frac{\pi k}{2N})}}+\frac{N}{l^2}\sin^3(\frac{\pi k}{2N})
\cos(\frac{2\pi k x}{l})\nn\right.\\
& &\left.-\frac{m^2}{8N}\sink \cos(\frac{2\pi k x}{l})\right\}-
\frac{\pi}{24l^2}-\frac{m^2}{4\pi}\ln(\tilde{\tau}ml)\\
&=&-\frac{\pi}{24l^2}-\frac{m^2}{4\pi}\ln{(\tilde{\tau} ml)} + \frac{1}{2l^2} 
\sum_{k=1}^{\infty} \(\frac{(\pi k)^2}{\sqrt{(ml)^2+(\pi k)^2}}-\pi k + 
\frac{(ml)^2}{2\pi k}\),\nn
\eea
with
\bea
\ln(\tilde{\tau})&=&\ln(\tau)+\frac{\pi}{2}\,0.52125 \ldots  \\
&=&\ln(\tilde{\chi})+1=-0.76066\ldots\,\, .\nn
\eea
The two terms following 
the first term in curly brackets are the first two terms of its expansion in $m^2$
multiplied by (-1). This removes all $a\to 0$
divergences and the resulting limiting value (second equality) follows 
straightforwardly. Expansion of the latter in powers of $m l$ yields
\bea
\T{11}(m,l)&=&-\frac{m^2}{4\pi}\ln(\tilde{\tau} ml)+\frac{\pi}{2l^2}
\sum_{\nu=0 \atop\nu\neq 1}^{\infty}d_{\nu}\(\frac{ml}{\pi}\)^{2\nu}\zeta(2\nu-1)
\quad,\mbox{with}\\
d_\nu&=&{{-1/2}\choose {\nu}}\,=\,(-1)^\nu \,{\Gamma(1/2+\nu)\over \nu!\,\sqrt{\pi}}
\quad .\nn
\eea
In fig. 2 we plot the ratio 
$\langle T^{11}\rangle (m,l)/\langle T^{11}\rangle (m=0,l)$ versus $m l$. The 
exponential fall-off of this function with increasing $m l$ is evident.\hfil\break
\begin{figure}
\centerline{%
\epsfxsize=4.5in
 \epsfbox{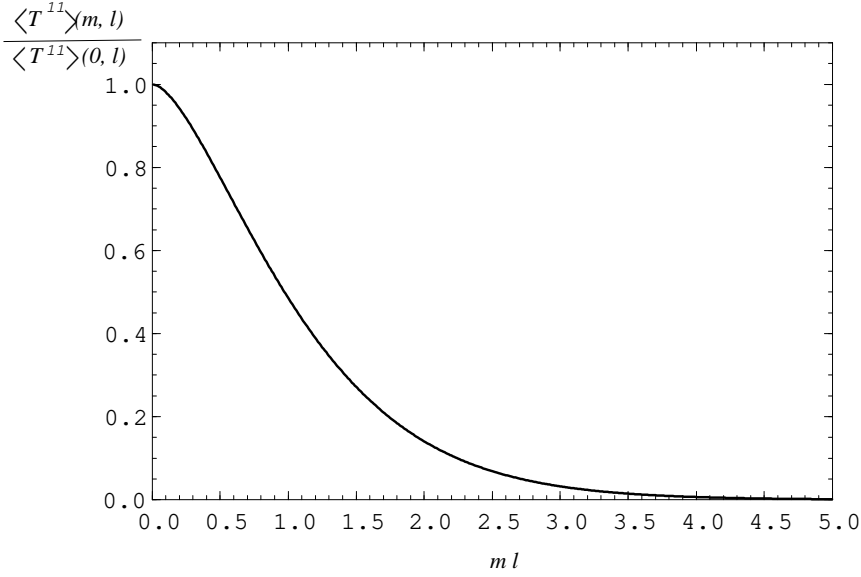}}
\caption[fig. 4] {\parbox[t]{12cm}{The ratio $\langle T^{11}\rangle(m,l)/\langle T^{11}\rangle(0,l)$
with $\ln\tilde{\tau}=-0.76066$ versus $m l$. Eq.~(\ref{eq4.32}) is summed to
$k=1000$.}}
\end{figure}
\subsubsection{$\langle T^{00}\rangle$}
\label{sec.T00a}
From eq. (\ref{eq3.48}) we have
\bea\label{eq4.33} 
\T{00}_{reg}(m,l,n,a)&=&\frac{E_{reg}(m,l,a)}{l}-\frac{m^2}{2}
\sk\frac{\cos(\frac{2\pi k n}{N})}{\sqrt{(ml)^2+4N^2\sin^2(\frac{\pi k}{2N})}}\nn\\
& &+A(m,l,n,a)\,.
\eea
The property
\be 
\sum_{n=1}^N\,\cos({2 \pi k n\over N})\,=\,0 \quad \mbox{for}\quad k\ne 0
\ee 
then yields immediately (compare eq. (\ref{eq3.41}))
\be\label{eq4.34}
a\,\sum_{n=1}^{N}\T{00}_{reg}(m,l,n,a)=E_{reg}(m,l,a)\,.
\ee
Let us now renormalize eq. (\ref{eq4.33}).\hfil\break
The second term in eq. (\ref{eq4.33}) remains finite in the limit $N\to \infty$ away
from the boundaries
\bea 
\lim_{N\to \infty}\(\sk\frac{\cos(\frac{2\pi k n}{N})}{\sqrt{(ml)^2+
4N^2\sin^2(\frac{\pi k}{2N})}}\)=\sum_{k=1}^{\infty}\frac{\cos(
\frac{2\pi k x}{l})}{\sqrt{(ml)^2+(\pi k)^2}}\,.
\eea
This series converges for $0< x< l$. However its behavior near the boundaries is 
$\ln x$ and $\ln (l-x)$ for $x\to 0$ and $x\to l$ respectively (where the series
diverges logarithmically). The renormalized energy density
\bea\label{eq4.35}
& &\hspace{-.5cm}T^{00}(m,l,x,\mu)\,=\,\frac{E(m,l,\mu)}{l}-\frac{m^2}{2}\sum_{k=1}^{\infty}
\frac{\cos(\frac{2\pi k x}{l})}{\sqrt{(ml)^2+(\pi k)^2}}\\
& &=\lim_{a\to 0}\(\T{00}_{reg}(m,l,\frac{x}{a},a)-\frac{2}{\pi a^2}+
\frac{m^2}{4\pi}\ln(\mu a)-f(l,\frac{x}{a},a)-\frac{m^2}{8}
h(\frac{l}{a},\frac{x}{a})\)\,,\nn 
\eea
therefore contains the expected boundary divergences. Fixing $\mu$ by $\mu=\chi m$ we 
obtain
\bea\label{eq4.36}
& &\hspace{-.5cm}T^{00}(m,l,x)\,=\,\frac{E(m,l)}{l}-\frac{m^2}{2}\sum_{k=1}^{\infty}
\frac{\cos(\frac{2\pi k x}{l})}{\sqrt{(ml)^2+(\pi k)^2}}\\
& &=\lim_{a\to 0}\(\T{00}_{reg}(m,l,\frac{x}{a},a)-\frac{2}{\pi a^2}+
\frac{m^2}{4\pi}\ln(\chi m a)-f(l,\frac{x}{a},a)-\frac{m^2}{8}
h(\frac{l}{a},\frac{x}{a})\).\nn 
\eea
Summing eq. (\ref{eq4.36}) over the lattice $0< x< l$ we find with the help of
eqs. (\ref{B7}), (\ref{B15})
\bea\label{eq4.37}
& &\hspace{-1.2cm}\sum_{n=1}^{N}a\,\(\T{00}_{reg}(m,l,n,a)-\frac{2}{\pi a^2}+
\frac{m^2}{4\pi}\ln(\chi m a)-f(l,n,a)-\frac{m^2}{8}h(\frac{l}{a},n)\)\nn\\
& &=E_{reg}(m,l,a)-\frac{2l}{\pi a^2}+\frac{1}{2a}+\frac{m^2l}{4\pi}\ln(\chi ma)\,,
\eea
which is the renormalized global vacuum energy (\ref{eq4.15}) as it should be.
This entitles us to assign parameter $\chi$ the same value it has in eq. (\ref{eq4.15}).
Here we do not have a condition to impose which fixes $\chi$ nor do we need one. The 
connection (\ref{eq4.37}) is quite sufficient.\hfil\break
\begin{figure}
\centerline{%
\epsfxsize=4in
 \epsfbox{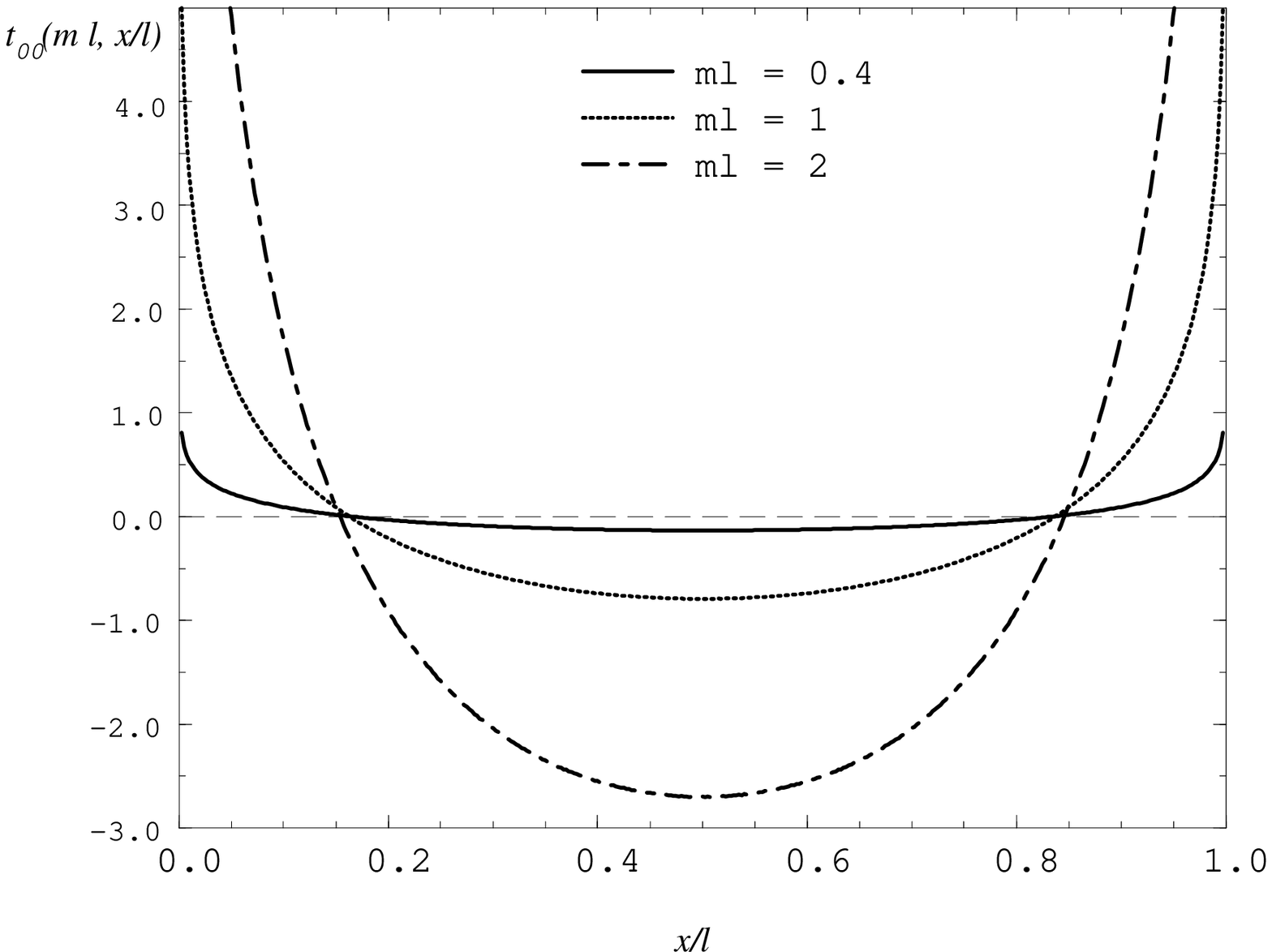}}
\caption[fig. 5] {\parbox[t]{12cm}{The function eq. (\ref{eq4.38}) summed to $k=1000$ over the interval 
\mbox{$0< x/l< 1$.}}}
\end{figure}
The energy density $\langle T^{00}\rangle$ is in contrast with the constant
$\langle T^{11}\rangle$ {\em strongly} position dependent, especially important
being logarithmic divergences at the boundaries. To display this we plot in fig. 3 the
quantity
\bea\label{eq4.38}
t^{00}(ml,\frac{x}{l})&\equiv&\frac{\T{00}(m,l,x)}{\T{00}(m=0,l)}-
\frac{E(m,l)}{E(m=0,l)}\\
&=&\frac{12\,(ml)^2}{\pi}\sum_{k=1}^{\infty}\frac{\cos(\frac{2\pi k x}{l})}
{\sqrt{(ml)^2+(\pi k)^2}}\nn
\eea
versus $x/l$ for three values of $m l$. The boundary divergences in 
$\langle T^{00}\rangle$ are inseparable from the boundaries themselves \cite{ref12}.
It is gratifying that the lattice formalism very closely reproduces \cite{ref2,ref3} this
crucially-important and prominent feature of continuum QFT.\hfil\break
\section{One dimensional lattice with Bessel potential background}
\label{sec.Bessel}
\setcounter{equation}{0}
Continuing on the 1D lattice with Dirichlet point boundaries at $x_1=0,\, x_2=l$ and
$x_3=L$ (in this section it will be convenient to make this change of notation) we now
introduce a background potential $V(x)\ge 0$. The Dirichlet boundaries at $x=x_1,\,x_2,
\,x_3$ are embedded in $V(x)$. For convenience we shall in this section combine $V(x)$ and the mass term into a 
single potential
\be 
U(x)\equiv m^2+V(x)\, .
\ee 
In subsection 4.1 we compute the lattice vacuum stress tensor and vacuum energy for
the region $x_1\le x\le x_2$ between adjacent Dirichlet boundaries embedded in an
{\em arbitrary} background potential $V(x)$ which vanishes as 
$x\to \infty$. Then in subsection 4.2 we choose $V(x)$ to be the potential
\be\label{eq5.1}
V(x)\,=\,{\alpha^2-{1\over 4}\over (x-x_0)^2}\; ,\; \alpha\ge 1/2
\ee
and begin the principal numerical work of this paper.
We compute numerically $\langle T^{\mu\nu}\rangle$ in $x_1< x<x_3$
and the global vacuum energy in this same region. Then we compute the Casimir 
force on the middle Dirichlet boundary at $x=x_2=l$ first as the jump in
$\langle T^{11}\rangle$ at this point, and second using the global method. Very close 
agreement between these two forces (which should be the same, of course) is 
found. This is a crucial test of our lattice method, and it succeeds. A variety of other 
checks and tests will also be displayed in subsection 4.2.\hfil

We remind the reader that the potential (\ref{eq5.1}) has a specific physical 
interpretation. It defines what the authors call a \lq\lq Bessel 
boundary" \cite{ref2,ref5} -- i.e. semitransparent
surface texture represented by $V(x)$ attached to a core Dirichlet boundary at $x=x_0$.
This entire boundary was absent in the analysis of sec. 3. Parameter $\alpha\ge 1/2$
controls the amount of semitransparent material present and the potential $V(x)$ 
extends from $x_0$ into $x> x_0$ as far as we choose to let it. [We can also let $V(x)$
extend arbitrarily far into $x<x_0$ -- and in fact should do so if we wanted to 
calculate the Casimir force on the Bessel wall -- but so far as this section is
concerned, we need not bother about the region $x<x_0$.] Without the quantum field the
classical background material represented by $V(x)$ would not interact directly with 
the classical 
boundaries at $x=x_1,\, x_2,\, x_3$ nor would these objects interact with one another.
However, when the quantum field $\phi$ is present it is distorted 
by $V(x)$ and by each of the hard boundaries at $x=x_1,\, x_2,\, x_3$. Back forces are then 
exerted by $\phi$ on the extended structure $V(x)$ and on each of the hard
boundaries. Forces involving all parts of the spatial background (including the material 
represented by $V(x)$) come into existence. Thus we have a highly nontrivial Casimir 
system to study.\hfil\break
\subsection{Renormalized vacuum stress tensor for semihard $V(x)$}
\label{sec.semihard}
\subsubsection{$\langle T^{11}\rangle$}
\label{sec.T11}
For the region $x_1\le x\le x_2$ we have from eqs. (\ref{eq3.37}),
the lattice formula for arbitrary background potential $V(n)$
\bea
\T{11}_{reg}(x_1,x_2,x_1+na,a)&=&\sk\(\frac{(\varepsilon_{k}^2-U(x_1+na))(v^{k}_{n})^2}
{2a\,\varepsilon_k}\right.\\
& &+\left.\frac{(v^{k}_{n+1})^2-2(v^{k}_{n})^2+(v^{k}_{n-1})^2}
{8 a^3 \varepsilon_k}\).\nn
\eea
Here $x_2-x_1=N a \equiv l$. As we show in Appendix C this expression can be renormalized for 
arbitrary $V(n)$. The renormalized lattice $\langle T^{11}\rangle$ is (prior to
letting $a\to 0$)
\bea\label{eq5.2}
\T{11}(x_1,x_2,x_1+na,a,\mu)&=&\T{11}_{reg}(x_1,x_2,x_1+na,a)-\frac{2}{\pi a^2}-
f(l,n,a)\nn\\
& &\hspace{0cm}-\frac{U(x_1+na)}{4\pi}\,\ln(\mu a)-\frac{U(x_1+na)}{8}\,
h(\frac{l}{a},n)
\eea
where $h$ and $f$ are defined by eqs. (\ref{B8}), (\ref {B13}) respectively. As in the 
corresponding lattice formula (\ref{eq4.27}) for $V(n)=0$ the renormalization
process introduces dependence on an arbitrary renormalization mass parameter $\mu$
(as always because of the need to subtract a lattice sum diverging logarithmically as
the lattice constant $a\to 0$). Such an arbitrary renormalization parameter 
introduces
ambiguity, of course. Fortunately, because of the way $\langle T^{11}\rangle$ depends 
on $\mu$ in eq. (\ref{eq5.2}) this parameter does not contribute to Casimir forces on
e.g. the boundaries at $x=x_1$ and $x_2$. The latter are determined by the jump in
$\langle T^{11}\rangle$ across these points. If the background potential $V(x)$ is 
continuous across a boundary, then the term $U(x)\,\ln \mu a$ takes on the same value on 
both sides of this boundary: there is no jump in $\T{11}$ and therefore no Casimir 
force. Only if $V(x)$
makes a jump at a boundary will the Casimir force on this boundary depend on $\mu$.
Because Casimir forces are the principal experimental signature of the distortion of
the quantum field by its interaction with all
background structure the importance of $F_{Cas}$ {\em not} depending on $\mu$
for continuous background potential $V(x)$ is clear. 
This observation also highlights
the importance of not merely considering the internal region in studying a Casimir 
problem. To have a well-posed problem external regions have to be considered as well.
\hfil\break
Quite another question to be answered is this: Is the renormalized vacuum stress 
tensor itself well-defined? $\langle T^{\mu\nu}\rangle$ is after all our basic 
{\em local} mathematical instrument for representing/investigating the physical
properties of the system. If $\mu$ in eq. (\ref{eq5.2}) is truly arbitrary then
$\langle T^{\mu\nu}\rangle$ is not unique. What are the consequences of this?
Is the nonuniqueness associated with $\mu$ relatively trivial or is it symptomatic of
something deeper? Even if $\mu$ does not influence Casimir forces it does influence 
the energy density $\langle T^{00}\rangle$ and the total vacuum energy.\hfil\break
We see no compelling reason to view $\mu$ as being, in some sense, a universal 
constant. If it were, then why would it be absent for $U(x)=0$? 
Thus in sec. 3 where $V(x)\,=\,0$ we have fixed this ambiguity by an ad hoc
but reasonable condition
\be 
\lim_{l\to\infty}\,\langle T^{11}\rangle(m,l)\,=\,0\quad .
\ee 
Physically this means that there is no contribution to the Casimir force on the hard 
boundary at 
$x=l$ coming from the infinitely extended region $x>l$.\hfil\break
Here with $x$-dependent potential $V(x)>0$ we choose what we believe is the natural extension of the scheme
in sec. 3. We set
\be 
\mu^2\,=\,\tau^2 U(x)\,=\,\tau^2\,[m^2+V(x)]
\ee 
which of course reduces to the ansatz $\mu=\tau m$ for $V(x)=0$. Then 
eq. (\ref{eq5.2}) becomes
\bea
& &\hspace{-1.5cm}\T{11}(x_1,x_2,x_1+na,a)\,=\,\T{11}_{reg}(x_1,x_2,x_1+na,a)-\frac{2}
{\pi a^2}-f(l,n,a)\nn\\
& &-\frac{U(x_1+na)}{8\pi}\,\ln(\tau^2 U(x_1+na) a^2)-\frac{U(x_1+na)}{8}
\,h(\frac{l}{a},n)\, .\label{eq5.3}
\eea 
Note that $\tau$ here still does not influence any hard-boundary Casimir force as long
as $V(x)$ is continuous. Because $\tau$ is a constant we can fix it by going to 
spatial infinity where (we assume) $V(x)\to 0$ so that $U(x)\to m^2$:
\be 
\lim_{x\to\infty}\,\lim_{x_2\to\infty}\,\langle T^{11}\rangle(x_1,\,x_2,\,x,\,\mu)
\,=\,0\quad .
\ee 
Here $x_1< x<x_2$ prior to moving $x_2=l$ out to infinity and then following with $x$.
This condition is essentially the same as that posed in the $V(x)=0$ case in 
sec. 3 and, just as in
eq. (\ref{eq4.29}), we obtain the fixed $\tau$
\bea\label{eq5.4}
\ln\tau=-0.76066\ldots-\frac{\pi}{2}\,0.52125\ldots=-1.579438\ldots\,.
\eea
Eq. (\ref{eq5.3}) generalizes eq. (\ref{eq4.31}) in the sense $m^2\to U(x)$.
Now in the continuum limit with $x=x_1+ n a$ held constant, $l=x_2-x_1$ and 
$\hat{x}=x -x_1$
\bea
\T{11}(x_1,x_2,x)&=&\lim_{a\to 0}\(\T{11}_{reg}(x_1,x_2,x,a)-\frac{2}{\pi a^2}-
f(l,\frac{\hat{x}}{a},a)\right.\\ 
& &\left.\hspace{2.7cm}-\frac{U(x)}{8\pi}\,\ln(\tau^2 U(x)a^2)-\frac{U(x)}{8}
\,h(\frac{l}{a},\frac{\hat{x}}{a})\)\,.\nn
\eea
\subsubsection{$\langle T^{00}\rangle$}
\label{sec.T00b}
The lattice formula for $\langle T^{00}\rangle$ is
\bea
\T{00}_{reg}(x_1,x_2,x_1+na,a)=\sk\(\frac{\varepsilon_k}{2 a}(v^{k}_{n})^2+
\frac{(v^{k}_{n+1})^2-2(v^{k}_{n})^2+(v^{k}_{n-1})^2}{8 a^3 \varepsilon_k}\).
\eea
The renormalized lattice vacuum energy generalizing eq. (\ref{eq4.37}) is
\bea\label{eq5.5}
& &\T{00}(x_1,x_2,x_1+na,a,\mu)\,=\,\T{00}_{reg}(x_1,x_2,x_1+na,a)-\frac{2}{\pi a^2}-
f(l,n,a)\nn\\
& &\hspace{0cm}+\frac{U(x_1+na)}{8\pi}\,\ln(\mu^2 a^2)-\frac{U(x_1+na)}{8}\,
h(\frac{l}{a},n)\; .
\eea
With $\mu^2=\chi^2\,U(x)$
\bea
& &\hspace{-0.5cm}\T{00}(x_1,x_2,x_1+na,a)\,=\,\T{00}_{reg}(x_1,x_2,x_1+na,a)-\frac{2}{\pi a^2}-
f(l,n,a)\nn\\
& &+\frac{U(x_1+na)}{8\pi}\,\ln(\chi^2 U(x_1+na) a^2)-\frac{U(x_1+na)}{8}
\,h(\frac{l}{a},n)\; ,\label{eq5.6}
\eea
where
\bea\label{eq5.7}
\ln\chi=\ln\tau-1=-2.579438\ldots\,.
\eea
In the continuum limit
\bea 
\T{00}(x_1,x_2,x)&=&\lim_{a\to 0}\(\T{00}_{reg}(x_1,x_2,x,a)-\frac{2}{\pi a^2}-
f(l,\frac{\hat{x}}{a},a)\right.\\ 
& &\left.\hspace{1.5cm}+\frac{U(x)}{8\pi}\,\ln(\chi^2 U(x)a^2)-\frac{U(x)}{8}
\,h(\frac{l}{a},\frac{\hat{x}}{a})\).\nn
\eea
\subsubsection{Renormalized vacuum energy}
\label{sec.RenVacEn}
The lattice vacuum energy in the interval $x_1< x  < x_2$
\bea
E_{reg}(x_1,x_2,a)=\frac{1}{2}\,\sk\varepsilon_k
\eea
is renormalized in Appendix C with the result
\bea \label{eq5.8}
E(x_1,x_2,a,\mu)=E_{reg}(x_1,x_2,a) -\frac{2(x_2-x_1)}{\pi a^2}+
\frac{1}{2a}+\frac{1}{4\pi}\int_{x_1}^{x_2}U(x)\,dx\,\ln(\mu a)\,.
\eea
Assuming $E=\int \T{00}dx$ we find from eq. (\ref{eq5.6}) the same renormalized 
lattice vacuum energy
\bea\label{eq5.9}
E(x_1,x_2,a)&=&E_{reg}(x_1,x_2,a)-\frac{2(x_2-x_1)}{\pi a^2}+\frac{1}{2a}\nn\\
&+&\frac{1}{8\pi}\int_{x_1}^{x_2}U(x)\,\ln(\chi^2 U(x)a^2)\,dx\quad.
\eea
Note that in eqs. (\ref{eq5.8}), (\ref{eq5.9}) one can use in place of the
integral the lattice sum
\bea
\frac{1}{16\pi}\(\sum_{n=0}^{N-1}+\sum_{n=1}^{N}\)a\,U(x_1+na)\,
\ln(\chi^2 U(x_1+na)a^2)\quad .
\eea
The difference vanishes when $a\to 0$. In the continuum limit we define
\bea
E(x_1,x_2):=\lim_{a\to 0}E(x_1,x_2,a)\,.
\eea
\subsection{Numerical evaluation using the Bessel potential}
\label{sec.evaluation}
Now we proceed to a numerical investigation of the renormalized 
$\langle T^{\mu\nu}\rangle$ and $E(x_1,\,x_2)$ defined above for the background 
Bessel potential (\ref{eq5.1}). The core of the Bessel boundary is at $x=x_0$ and
 the surface texture represented by $V(x)$ extends to the right from $x=x_0$. We
 generally choose $x_0=-0.01$ and position the first Dirichlet boundary at
$x_1=0$. The other two Dirichlet boundaries are further to the right at 
$x=x_2,\, x_3>0$.
Before presenting our numerical results it is perhaps worth saying again how they are 
obtained. First the lattice mode equation (\ref{eq3.13}) has to be solved numerically
for the lattice modes $v^k_n$ and their energy eigenvalues $\varepsilon_k^2$.
Vanishing boundary conditions at each of the hard boundaries are additionally imposed
on the $v_n^k$. A given calculation is performed for specific values of lattice constant 
$a$, of parameter $\alpha$ in eq. (\ref{eq5.1}), of mass $m$ and of positions 
$x_1,x_2,x_3$. When any of these six values is changed the numerical calculation has to 
be redone. Given the numerical ingredients $\{v_n^k\}$ and $\{\varepsilon_k^2\}$
one can insert these ingredients into eqs. (\ref{eq5.2}), (\ref{eq5.6}) to obtain
the lattice $\langle T^{\mu\nu}\rangle$ numerically.\hfil\break
\subsubsection{Renormalized $\langle T^{00}\rangle$}
\begin{figure}
\label{sec.T00ren}
\centerline{%
\epsfxsize=4.4in
 \epsfbox{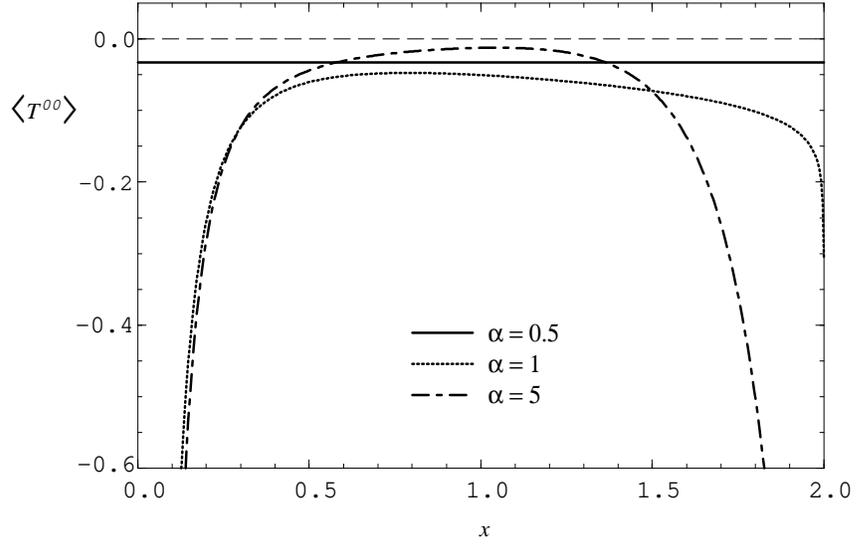}}
\caption[fig. 6] {\parbox[t]{12cm}{$\langle T^{00}\rangle$ from eq. (\ref{eq5.6}) with $m=0$, 
$x_0=-0.01$ for Dirichlet boundaries at $x_1=0$ and $x_2=2$. The lattice constant is
$a=1/400$.}}
\end{figure}
Fig. 4 shows the function $\langle T^{00}\rangle$ obtained from eq. (\ref{eq5.6}) in
$x_1< x<x_2$ with $x_0=-0.01,\,x_1=0,\,x_2=2$ for $m=0$ and three different values
of the parameter $\alpha$ in eq. (\ref{eq5.1}): these are $\alpha=1/2$ (or $V(x)=0$
everywhere between the Dirichlet boundaries), $\alpha=1$ and $\alpha=5$.
\begin{figure}
\centerline{%
\epsfxsize=4.25in
 \epsfbox{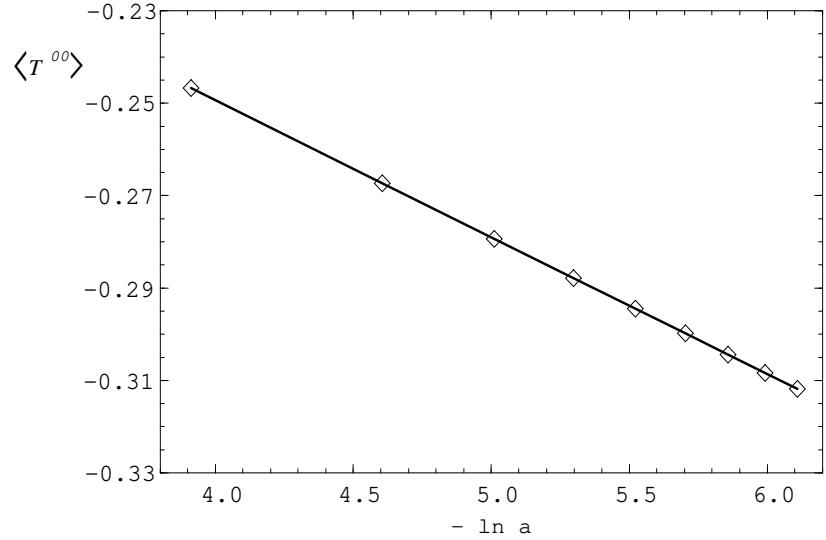}}
\caption[fig. 7] {\parbox[t]{12cm}{$\langle T^{00}\rangle$ from eq. (\ref{eq5.6}) with $\alpha=1$ and
other parameters as in fig. 4, evaluated on the right-hand boundary $x=x_2=2$.}}
\end{figure} 
Note again the arbitrary length unit along the horizontal axis. For
$\alpha=1/2$ we recover the well-known ($m=0$)-result (see e.g. ref.\cite{ref10})
$\langle T^{00}\rangle\,=\,\pi^2/(24\, l^2)$. For $\alpha>1/2$ we see that 
$\langle T^{00}\rangle$ is (i) asymmetric about the midpoint $x=1$ because of the
asymmetry of the potential (\ref{eq5.1}) and (ii) divergent as the boundaries are
approached. To investigate this divergence differently we display in fig. 5
$\langle T^{00}\rangle$ for $n=N$ or $x=x_2=2$ on the right hand boundary and 
$\alpha=1$ as a function of $(-\ln a)$. Clearly a logarithmic divergence is involved.

\subsubsection{Renormalized $\langle T^{11}\rangle$}
\label{sec.T11ren}
Here we use eq. (\ref{eq5.3}) with the potential (\ref{eq5.1}) to compute numerically 
the renormalized $\langle T^{11}\rangle$. First of all we wish to show what this 
function looks like when the Dirichlet boundaries are absent and the system consists 
of the scalar field coexisting with the Bessel boundary. Of course this system can only
approximately be realized on a finite lattice. Again we position a Dirichlet boundary 
at $x=x_1=0$ very close to the core $x_0=-0.01$ of the Bessel boundary. This is done
for computational reasons. We expect (and we may also conclude from our numerical 
results) that this hard boundary essentially disappears into the (Dirichlet like)
Bessel boundary and has no significant effect on $\langle T^{11}\rangle$ away from
$x\Buildrel{>}{\sim}0$. Also for computational reasons we introduce a second Dirichlet
boundary at $x_2=6$ \lq\lq far" from $x=0$. We obtain the numerical results shown in
fig. 6 for three different mass values $m=0,\,2,\,5$ (and $\alpha=1,\, a=1/300)$.
These need to be discussed in some detail.\hfil\break
\begin{figure}
\centerline{%
\epsfxsize=4in
 \epsfbox{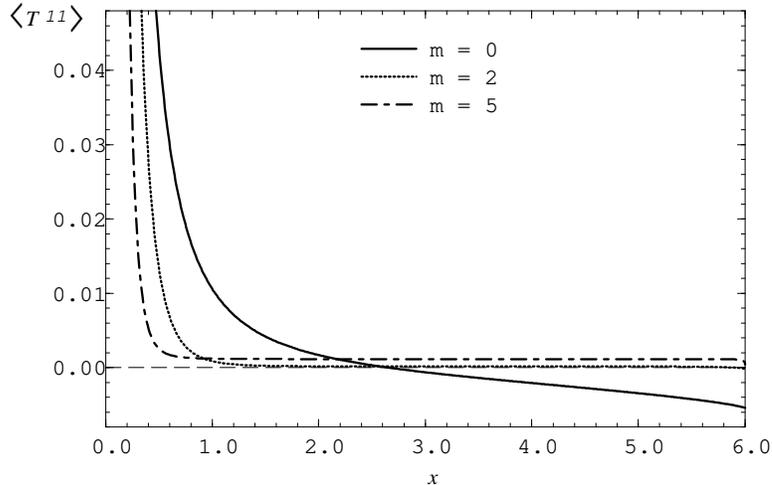}}
\caption[fig. 8] {\parbox[t]{12cm}{$\langle T^{11}\rangle$ from eq. (\ref{eq5.3}) with $x_0=-0.01,\,
\alpha=1$ for Dirichlet boundaries at $x_1=0$ and $x_2=6$. The lattice constant
 is $a=1/300$.}}
\end{figure}
\noindent For \lq\lq large" distance $x$ from the core of the Bessel boundary the potential (\ref{eq5.1})
is essentially constant (and small). Thus one expects $\langle T^{11}\rangle$
for large $x$ to approach a constant value equal to $\langle T^{11}\rangle$ in free
space for the same field mass $m$ (see sec. 4.2.1). This behavior is evident in
fig. 6 for masses $m=2,\,5$. For smaller mass $m=2$ the asymptotic value for large
$x$ is close to zero. For larger mass $m=5$ the asymptotic value of
$\langle T^{11}\rangle$ increases noticeably. We remind the reader of the free-space
vacuum stress tensor in one spatial dimension (see e.g. ref.\cite{ref10})
\be 
T^{\mu\nu}_{\mit free}\,=\,\eta_{\mu\nu}\,\Lambda\, ,\; \Lambda\,=\,-{m^2\over 4 \pi}\,
\ln{m\over \mu}\quad .
\ee 
This formula indicates that, if one may assume $m>\mu$, $T^{11}_{\mit free}$ is positive 
and increasing with increasing $m$ -- just what is observed in fig. 6. But remember
that we fixed $\tau$ in eq. (\ref{eq5.3}) by requiring $\langle T^{11}\rangle$
to vanish at large distance. (The same requirement applied to $T_{free}^{\mu\nu}$
would lead to $\mu=m$ and this tensor vanishing everywhere.) This is observed in 
fig. 6 for $m=2$ but not for $m=5$. The mathematical reason for the $m=5$ curve not 
vanishing is our imperfect knowledge of the value of parameter $\tau$. The numerical 
consequences of this grow with $m$, one consequence being that $\langle T^{11}\rangle$
does not quite vanish for large $x$ although it should. Note  also that with 
increasing $m$ the asymptotic value of $\langle T^{11}\rangle$ is reached in a shorter 
$x$ interval. This is entirely consistent with continuum results where boundary
effects are known to fall off exponentially in $m x$.\hfil\break
The $(m=0)$-curve in fig. 6 does not approach the value $\langle T^{11}\rangle=0$ 
asymptotically, nor should it. For $m=0$ there is no parameter $\mu$ to fix and no
 asymptotic condition on $\langle T^{11}\rangle$. For $m=0$ boundary effects in 
 $\langle T^{11}\rangle$ no longer diminish exponentially away from the $x=0$ 
boundary, but rather diminish as the inverse power $x^{-2}$. Consequently the boundary
at $x_3=6$ is no longer \lq\lq far away" and in fact is strongly influencing
$\langle T^{11}\rangle$ in fig. 6. To see $\langle T^{11}\rangle$ asymptotically
approaching zero one must choose a substantially  larger value of $x_2$. The $m=0$
continuum vacuum stress tensor between Dirichlet boundaries at $x=0,\,L$ is
\be 
\langle T^{\mu\nu}\rangle\,=\,-\delta_{\mu\nu}\,{\pi\over 24\, L^2}\quad .
\ee 
In fig. 6 the $(m=0)$-curve for $\langle T^{11}\rangle$ goes negative as it should.
\hfil\break
Now let us modify the system by introducing a third boundary at $x=x_2=1$, leaving
the other two boundaries in place at $x_1=0$ and $x_3=6$ (the previous $x_2$ is now
renamed $x_3$). Figs. 7, 8 display $\langle T^{11}\rangle$ throughout the interval
$0<x<6$. One sees in fig. 7 that $F_{Cas}$ on the intermediate boundary $x_2=1$
defined by
\begin{figure}
\centerline{%
\epsfxsize=4.3in
 \epsfbox{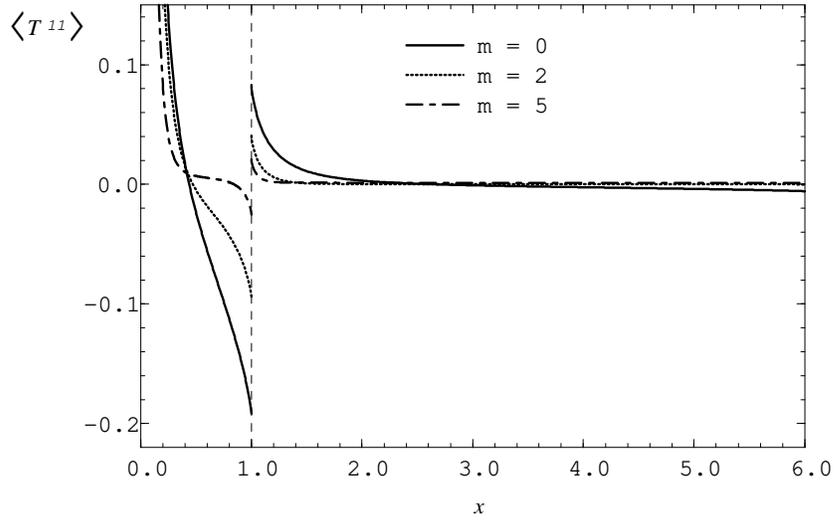}}
\caption[fig. 9] {\parbox[t]{12cm}{$\langle T^{11}\rangle$ from eq. (\ref{eq5.3}) with $x_0=-0.01,\, 
\alpha=1,\,a=1/300$ for Dirichlet boundaries at $x_1=0,\,x_2=1$ and $x_3=6$.}}
\end{figure}
\begin{figure}
\centerline{%
\epsfxsize=4.3in
 \epsfbox{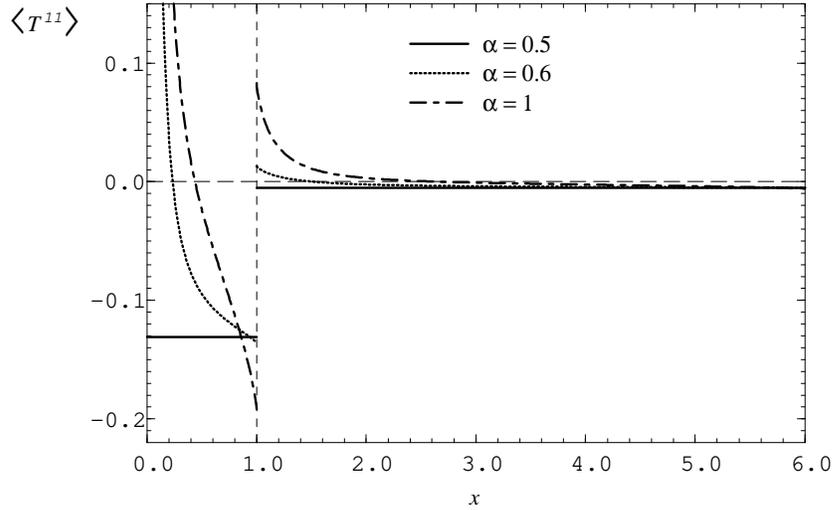}}
\caption[fig. 10] {\parbox[t]{12cm}{$\langle T^{11}\rangle$ from eq. (\ref{eq5.3}) with $x_0=-0.01,\, 
m=0,\,a=1/300$ for Dirichlet boundaries at $x_1=0,\,x_2=1$ and $x_3=6$.}}
\end{figure}
$$
F_{Cas}\,=\,\langle T^{11}\rangle_{x=1-\varepsilon}\,-\,
\langle T^{11}\rangle_{x=1+\varepsilon}\; ,\;\varepsilon\to 0_+
$$
 is directed to the left. This is predictable: the Bessel boundary's core $x_0=-0.01$ 
 and the Dirichlet point $x_1=0$ are much closer to $x_2=1$ than is the other Dirichlet 
boundary $x_3=6$. Dirichlet objects attract in Casimir theory and the Casimir force
between closer Dirichlet
objects dominates. Also predictable is the decrease shown in fig. 7 of $F_{Cas}$
with increasing mass $m$. For $V(x)=0$ between Dirichlet boundaries it is known that
Casimir forces weaken rapidly (exponentially) with increasing mass. There is no 
reason why any background potential $V(x)$ of the type considered here would qualitatively
alter this general rule.\hfil\break
Fig. 8 shows $\langle T^{11}\rangle$ for fixed $m=0$ and three values of the
strength parameter $\alpha=0.5,\,0.6$ and $1.0$ in potential V(x). Clearly $F_{Cas}$
increases with increasing $\alpha$. This too is predictable.
When $\alpha=1/2$ and $V(x)=0$ we know that 
$\langle T^{11}\rangle\,=\,-\pi/(24\,L^2)$ is constant (with $L=1$ to the left of 
$x_2=1$ and $L=5$ to the right of $x_2=1$).
Increasing $\alpha$ above $\alpha=1/2$ causes $V(x)=0$
to jump abruptly and nonuniformly to a positive value everywhere, and obviously
 to a much larger {\em average} value in $0< x< 1$ than in $1< x< 6$. The 
potential $V(x)$ represents \lq\lq Dirichlet material" added throughout the interval
$0< x< 6$. This nonuniformly added material pulls the $x_2=1$ boundary more strongly
to the left than to the right.\hfil\break
\subsubsection{Renormalized vacuum energy}
\label{sec.vacren}
These numerical calculations of the vacuum energy are based on eq. (\ref{eq5.9}),
where the integral can be done analytically for the potential (\ref{eq5.1}). With
$l=x_2-x_1$ and $b=\alpha^2-1/4$ we find
\bea  
\int_{x_1}^{x_2}U(x)\,\ln(\chi^2 U(x)a^2)\,dx&=&\ln\(\frac{\chi^2\,a^2}{l^2}\)
\int_{x_1}^{x_2} U(x)\,dx\\
& &+\frac{1}{l}\int_{0}^{1} U(x_1+x l)\,l^2\ln(U(x_1+x l)\,l^2)\,dx\, \nn
\eea
and
\bea
\bar{U}(x_1,x_2)&\equiv&\frac{1}{l}\int_{x_1}^{x_2}\(m^2+\frac{\alpha^2-\frac{1}{4}}
{(x-x_0)^2}\)\,dx=m^2+\frac{\alpha^2-\frac{1}{4}}{(x_1-x_0)(x_2-x_0)}\; ,\\
g(x_1,x_2)&\equiv&\int_{0}^{1}\((m l)^2+\frac{\alpha^2-\frac{1}{4}}
{(\frac{x_1-x_0}{l}+x)^2}\)\ln\((m l)^2+\frac{\alpha^2-
\frac{1}{4}}{(\frac{x_1-x_0}{l}+x)^2}\)\,dx\nn\\
&=&-\frac{2b l^2}{(x_1-x_0)(x_2-x_0)}\nn\\
& &+4m l\sqrt{b}\left[\arctan\(\frac{m\,(x_2-x_0)}{\sqrt{b}}\)-
\arctan\(\frac{m\,(x_1-x_0)}{\sqrt{b}}\)\right]\nn\\
& &+\frac{l\,(m^2(x_2-x_0)^2-b)}{x_2-x_0}\ln\(\frac{l^2(m^2(x_2-x_0)^2+b)}
{(x_2-x_0)^2}\)\nn\\
& &-\frac{l\,(m^2(x_1-x_0)^2-b)}{x_1-x_0}\ln\(\frac{l^2(m^2(x_1-x_0)^2+b)}
{(x_1-x_0)^2}\)\; .\nn
\eea
The renormalized vacuum energy in region $x_1< x< x_2$ is
\bea\label{eq5.10}
E(x_1,x_2,a)=\frac{1}{2}\,\sk\varepsilon_k-\frac{2 l}{\pi a^2}+\frac{1}{2a}+
\frac{l\,\bar{U}(x_1,x_2)}{4\pi}\ln(\frac{\chi \,a}{l})+\frac{g(x_1,x_2)}{8\pi l}\; .
\eea
Fig. 9 displays this energy as a function of boundary position $x_2$ for fixed
boundary position $x_1=0$. Nothing beyond the boundary at $x_2$ is considered here.
Fig. 9 qualitatively resembles fig. 1 giving the comparable vacuum energy for 
$V(x)=0$. As $x_2$ approaches $x_2=0$ here (the boundaries $x_1=0$ and $x_2$ move
close together) the function $E(0,x_2,a)$ approaches the function $-\pi/24\,l$ with
$l=x_2-x_1$. This is the $l\to 0$ limiting form of the vacuum energy for $V(x)=0$
 and $m\ge 0$ as it should be.\hfil\break
 \begin{figure}
 \centerline{%
\epsfxsize=4.4in
 \epsfbox{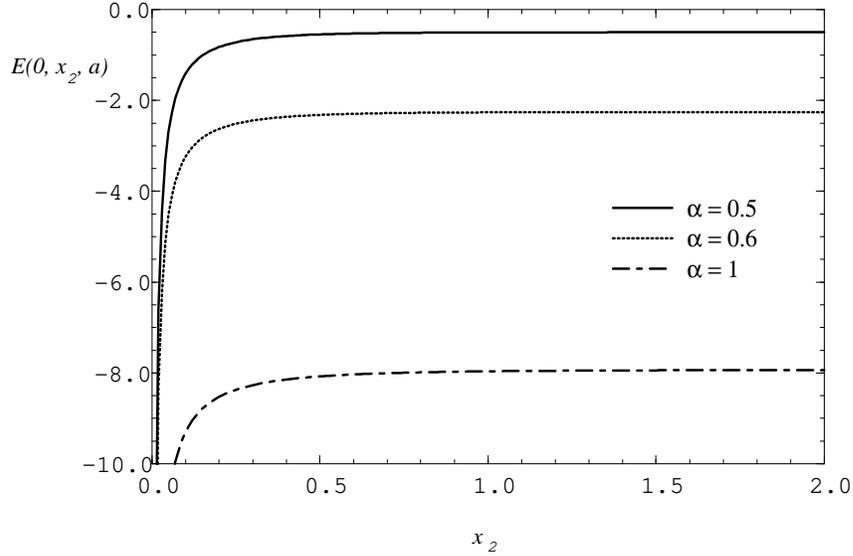}}
\caption[fig. 11] {\parbox[t]{12cm}{$E(x_1,\,x_2,\,a)$ from eq. (\ref{eq5.10}) with $x_0=-0.01,\,m=2,\,
a=1/500$ for Dirichlet boundaries at $x_1=0$ and variable $x_2$.}}
\end{figure}
 \subsubsection{Casimir forces}
 \label{sec.forces}
Now we proceed to the numerical evaluation of Casimir forces for various
configurations. It will be extremely important to verify that we obtain 
{\em locally} from $\langle T^{11}\rangle$ and {\em globally} from 
the vacuum energy the same Casimir forces. If this is what is found -- and it 
is --
then because the local and global mathematics differ quite substantially we can be 
confident both are working as they should. Indeed these calculations provide the 
crucial test of our lattice methods with nonzero background potential $V(x)$.\hfil\break
Global calculations of the vacuum energy will be done using the renormalized formula
(\ref{eq5.10}). Positioning three Dirichlet boundaries as before at $x_1< x_2< x_3$
we obtain the Casimir force on boundary $x_2$ (globally) as
\be\label{eq5.11}
F_{Cas}=\lim_{a\to 0}\(\frac{E(x_1,x_2,a)-E(x_1,x_2+a,a)+E(x_2,x_3,a)-
E(x_2-a,x_3,a)}{a}\).
\ee
Alternatively $F_{cas}$ can be calculated (locally) as
\be\label{eq5.12}
F_{Cas}=\lim_{a\to 0}(\T{11}_{reg}(x_1,x_2,x_2,a)-\T{11}_{reg}(x_2,x_3,x_2,a))\,. 
\ee
\begin{figure}
\centerline{%
\epsfxsize=4in
 \epsfbox{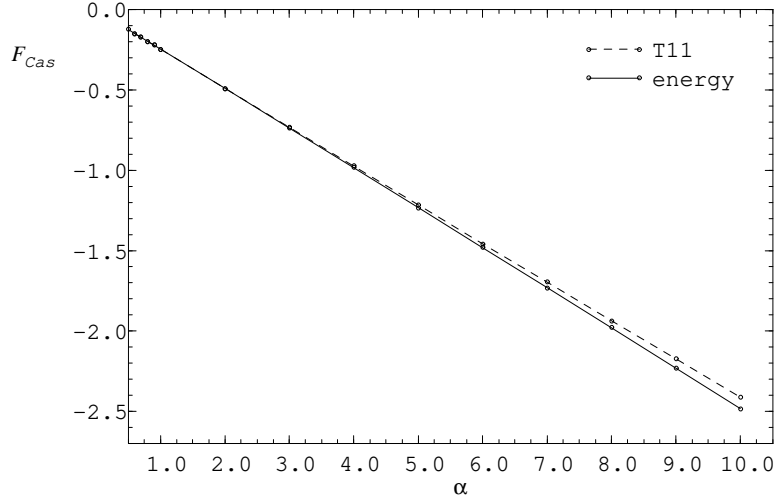}}
\caption[fig. 12] {\parbox[t]{12cm}{Casimir force on the middle boundary $x_2=1$ obtained globally and locally 
from eqs. (\ref{eq5.11}), (\ref{eq5.12}) with $x_0=-0.01,\,m=0,\,a=1/300$ and
$x_1=0,\,x_3=6$.}}
\end{figure}
Fig. 10 shows these two results for $F_{Cas}$ for fixed $x_1,x_2,x_3$ as a function of 
Bessel
parameter $\alpha$. The agreement is very good. Let us examine this in greater detail.
\noindent For small $\alpha$ we find very close agreement between the local and global
calculations even for relatively large lattice constant $a$. For example, in fig. 10 
at $\alpha=1$ the global and local values are -0.246007 and -0.246225 respectively.
The limiting value of $F_{Cas}$ for $\alpha\to 1/2$ is $-\pi/24\,+\,
\pi/600$.\hfil\break
As $\alpha$ increases the agreement between global and local becomes less
good. To improve the agreement one must go to smaller lattice constant $a$. This is 
demonstrated in fig. 11 where, for $\alpha=10$, we plot the globally and locally 
obtained results for $F_{Cas}$
versus $a$. Linear extrapolation of the two straight lines to $a=0$ yields the 
limiting values -2.45142 and -2.45105 respectively. This
 is extremely good agreement.\hfil\break
 \begin{figure}
 \centerline{%
\epsfxsize=4in
 \epsfbox{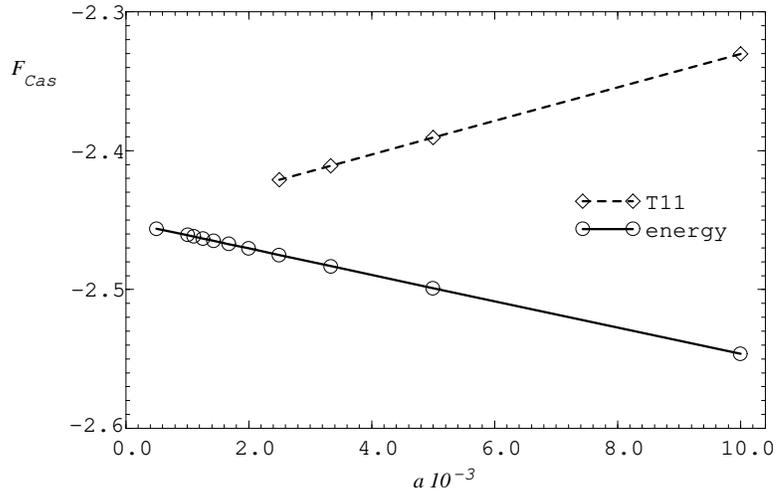}}
\caption[fig. 13] {\parbox[t]{12cm}{The global and local Casimir forces in fig. 10 for 
$\alpha=10$ plotted versus lattice constant $a$.}}
\end{figure}

\noindent Note that in fig. 10 the ($m=0$)-Casimir force is seen to depend linearly on $\alpha$.
\begin{figure}
\centerline{%
\epsfxsize=4in
 \epsfbox{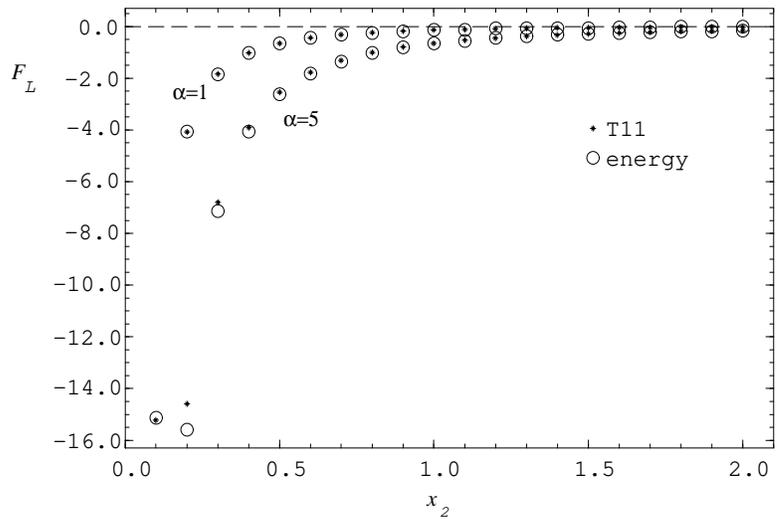}}
\caption[fig. 14] {\parbox[t]{12cm}{The left-hand Casimir forces (\ref{eq5.13}), (\ref{eq5.14}) on
the middle wall, the latter's position $x_2$ being variable. Here $x_0=-0.01,\,
m=1,\, a=1/400$ and $x_1=0$.}}
\end{figure}
\noindent For $m> 0$ this is not the case in general; however linear behavior does set in for 
sufficiently large $\alpha$.\hfil\break
\noindent To display the agreement between the global and local methods in even more detail let
us decompose $F_{Cas}$ on the $x_2$-boundary into left ($L$) and right ($R$) 
components
\be 
F_{Cas}\,=\,F_L\,+\,F_R\quad .
\ee 
$F_L(F_R)$ is the force exerted on the boundary $x_2$ from left
(right)
\bea\label{eq5.13}
F_{L}&\equiv&-\frac{\p E(x_1,x_2)}{\p x_2}=\T{11}(x_1,x_2,x=x_2)\,\, ,\\
F_{R}&\equiv&-\frac{\p E(x_2,x_3)}{\p x_2}=-\T{11}(x_2,x_3,x=x_2)\,\, .\nn
\eea 
In fig. 12 we plot $F_L$ computed locally using $\langle T^{11}\rangle(x_1,x_2,x_3,a)$
and globally using
\be\label{eq5.14}
F_L(x_1,x_2)\equiv-\frac{E(x_1,x_2+a,a)-E(x_1,x_2,a)}{a}\,.
\ee
\begin{figure}
\centerline{%
\epsfxsize=4in
 \epsfbox{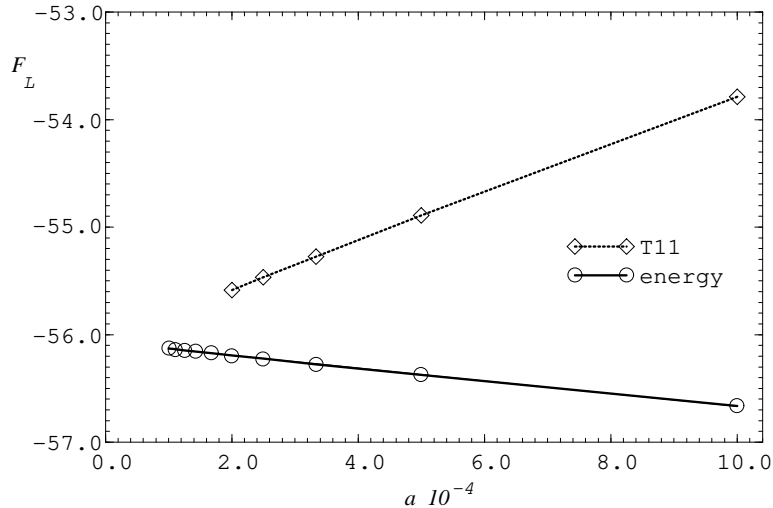}}
\caption[fig. 15] {\parbox[t]{12cm}{The left-hand Casimir forces in fig. 12 for $\alpha=5$ and
$x_2=0.1$ plotted versus lattice constant $a$.}}
\end{figure}
\noindent Again we find very good agreement as long as $\alpha$ is not too large and $x_2$ not
too small. For large $\alpha$ and small $x_2$ the influence of $V(x)$ is quite strong
and we have to go to smaller lattice constant to observe the local-global agreement.
We show in fig.~13 how this works for $\alpha=5$ and $x_2=0.1$. Linear
extrapolation to $a=0$ in fig.~13 yields the global and local values -56.073 and
-56.026 respectively for $F_L$; again the agreement is extremely close.
\hfil\break
\begin{figure}
\centerline{%
\epsfxsize=4.3in
 \epsfbox{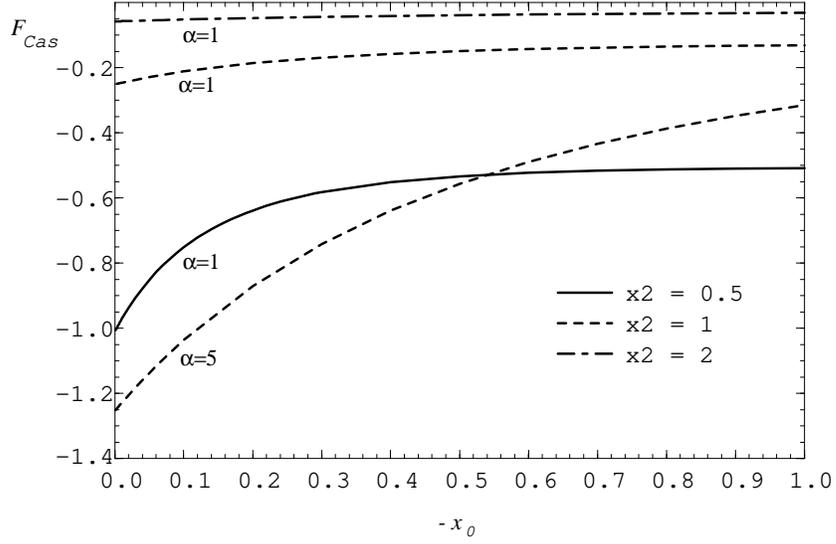}}
\caption[fig. 16] {\parbox[t]{12cm}{The Casimir force (\ref{eq5.11}) as a function of 
the position $x_0$
of the Bessel boundary for $x_2=0.5,1,2$. Here $a=1/500,\,m=0,\, x_1=0$ and $ x_3=6$.}}
\end{figure}
\noindent The final three figures in this section show Casimir forces computed globally using 
the vacuum energy.
\begin{figure}
\centerline{%
\epsfxsize=4.3in
 \epsfbox{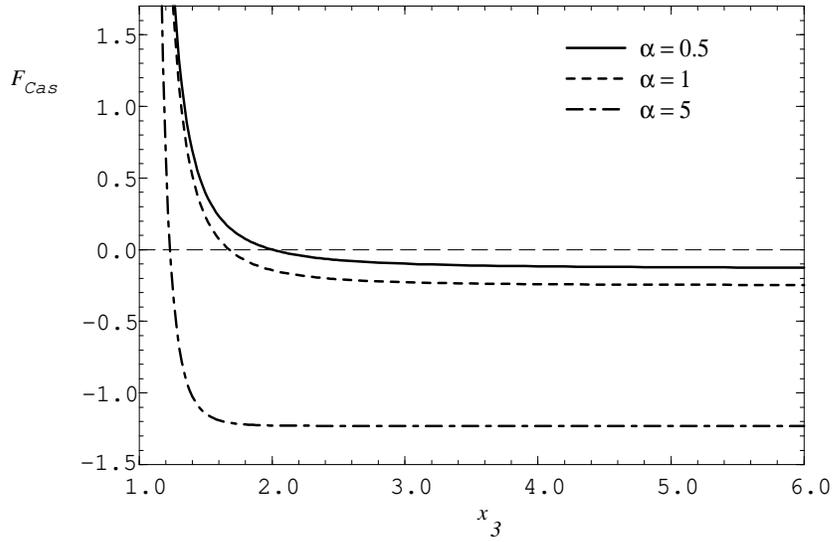}}
\caption[fig. 17] {\parbox[t]{12cm}{Global Casimir force (\ref{eq5.11}) for variable position $x_3$
of the rightmost Dirichlet boundary. Here $x_0=-0.01,\,m=0,\,a=1/500$ and $ x_1=0,\,
x_2=1$. The region $x> x_3$ is ignored.}}
\end{figure}
This is simpler than working with $\langle T^{11}\rangle$ which requires knowledge of
eigenvalues {\em and} eigenvectors. However we have verified that
 $\langle T^{11}\rangle$ provides the same results.\hfil\break
\noindent Fig. 14 shows how changing the position $x_0$ of the center of the Bessel boundary
 affects the Casimir force on boundary $x_2$ in the system with three Dirichlet 
boundaries at $x_1,x_2,x_3$, all embedded in the potential $V(x)\,=\,(\alpha^2-1/4)/
(x-x_0)^2$. $F_{Cas}$ in fig. 14 is the net Casimir force on boundary $x_2$ with 
$x_1,x_3$ fixed and variable $x_0$. Three different $x_2$-values are used. Essentially
in fig. 14 we are holding all three Dirichlet boundaries fixed and rigidly 
{\em translating} $V(x)$ by changing $x_0$. Clearly $F_{Cas}$ depends more 
sensitively on $x_0$ as $\alpha$ increases, which one would expect. Moreover the 
sensitivity of $F_{Cas}$ to $x_0$ increases as the distance $x_2-x_1$ between the
leftmost and middle Dirichlet boundaries decreases. This is also to be expected.
The conclusion is, that one has to choose $x_0$ very small in order to obtain relevant 
results for $F_{Cas}$. Our Choice $x_0=-0.01$ seems to be appropriate.
\hfil\break
\noindent Finally we investigate the dependence on $x_3$ of the Casimir force $F_{Cas}$ on 
boundary $x_2$. 
 For large $x_3$ this force
(fig. 15) is practically independent of $x_3$ as one would expect. At relatively
small $x_3$ a quite strong dependence on $x_3$ abruptly appears and $F_{Cas}$
reverses direction -- again quite predictable. The value of $x_3$ at which $F_{Cas}$
vanishes (i.e. changes sign) moves inward with increasing $\alpha$ as it should.\hfil
\break
\begin{figure}
\centerline{%
\epsfxsize=4.3in
 \epsfbox{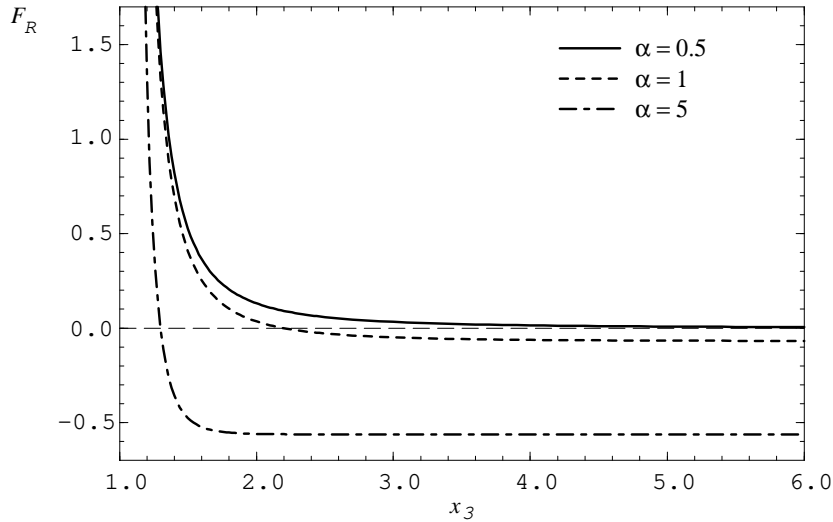}}
\caption[fig. 18] {\parbox[t]{12cm}{$F_R$ from eq. (\ref{eq5.13}) for variable $x_3$. 
Same parameters as fig. 15.}}
\end{figure}

\noindent Fig. 16 displays the part of the Casimir force in fig. 15 coming from $x_2< x< x_3$.
For $\alpha>1/2$ this force component does not vanish as $x_3\to \infty$ (as it
would if $V(x)$ were zero). The Casimir force on the Dirichlet boundary at $x_3$ 
obviously tends to zero as $x_3\to \infty$ because $V(x_3)\to 0$. Thus the asymptotic 
force for large $x_3$ in fig.~16 is entirely due to $V(x)$. Naturally this asymptotic 
force increases with increasing $\alpha$.\hfil\break
\section{Conclusion}
\label{sec.conclusion}
\setcounter{equation}{0}
Continuum QFT has always been used to formulate and study the diverse topics and 
systems of static and dynamical Casimir theory. We decided
several years ago to reformulate a broad range of these topics and systems in the 
language of lattice QFT, confident this would provide a tool powerful enough to solve
many Casimir problems not accessible to continuum methods. Here we have presented some
results on {\em static} finite lattice Casimir systems. A specific static 
spatial configuration was chosen: parallel Dirichlet boundaries with fixed separation
on a d-dimensional lattice. We formulated this problem first with no background
potential, then with a background lattice potential $V(n)$ depending on the coordinate
$x=n a$ perpendicular to the boundary planes at $x=0,\, l$. Even for vanishing
background potential $V(n)=0$ the presence of hard boundaries introduces technical 
features into the lattice formulation not present in lattice QFT with no boundaries.
We have examined many of these things in detail. With the introduction of a nonzero 
background potential $V(n)$ there appear additional technical complexities which
have also received careful attention.\hfil\break
Renormalization of lattice QFT in the continuum limit $a\to 0$ is a prerequisite to
success. One does lattice QFT primarily to be able to learn things about the
continuum theory which are difficult to find out using continuum methods. Certainly 
that is our attitude toward this work. Until one has understood how renormalization
works one's lattice results do not mean very much: the (informationless) 
$a\to 0$ divergences will swamp any physical information contained in the 
lattice mathematics.
For this reason we have paid close attention to renormalization, both for $V(n)=0$ and 
for $V(n)> 0$. Rather detailed results have been presented for one-dimensional 
lattices in secs. 3 and 4. We believe these results reveal a self-consistent and
 clear picture of the lattice QFT as it smoothly becomes (after renormalization) the
 continuum theory in the limit $a\to 0$, both for vanishing and for nonvanishing
 background potential $V(n)$.\hfil\break
 In unpublished work we have studied the extension of the $V(n)=0$ 1D-lattice
 analysis to $d\ge 2$ spatial dimensions. One can discern 
 in the mathematics of these systems how lattice Casimir QFT in
 higher dimensions including renormalization parallels continuum 
 theory. This is very
 important for us, because from the outset we planned to study 2D and 3D lattice
 configurations of {\em hard} boundaries which, because of 
 nonsymmetric or otherwise complicated boundary geometry, cannot be dealt with using
 continuum mathematics. Work on such systems is in progress.\hfil\break
 Another direction of research we are pursuing is the introduction of background 
 potentials $V(n)$ on $d\ge 1$ spatial lattices. The semihard Bessel potential 
(\ref{eq5.1}) is very useful in continuum theory because it leads to spatial modes
which are known explicitly. However on the lattice it matters little whether one
uses this potential or some other: The numerical work is made neither easier nor more
difficult by one's choice of potential. Thus we are trying out on the lattice various
semihard potentials to gain insights one would perhaps not be able to obtain 
differently. Another interesting type of enclosing boundary -- soft 
boundaries -- is also being studied on the lattice. We intend to report on different
aspects of this work elsewhere.\hfil\break
Beyond the static systems mentioned we have also been able to make progress in 
another important area: dynamical Casimir theory. One can, on a 
 lattice, simulate moving boundaries quite effectively. Time-dependent backgrounds 
excite the quantum field and one can see this clearly in one's numerical results.
We shall present lattice analysis of dynamical as well as static Casimir systems 
in subsequent articles.
\hfil\break
\vskip 1cm
\noindent {\bf Acknowledgements}\hfil\break
One of us (A. A) thanks the Institut f\"ur Theoretische Physik der Universit\"at
Heidelberg for its kind hospitality, and Penn State University and the 
LV Campus 
for financial support in various forms.\vfil\eject

\appendix
\section{Operators on the lattice}
\label{sec.AppA}
 \setcounter{equation}{0}
On a lattice one can reduce any differential operator to an ordinary numerical matrix 
once the fields $\phi_n$ have been labeled and ordered. Here we consider lattice
versions of the following operators
\appeqn
\be 
(\nabla\phi)^2\, ,\, \phi\triangle\phi\;\;\mbox{and}\;\nabla(\phi\nabla\phi)\quad .
\ee 
On a lattice there is, of course, more than one way to represent differentiation.
We shall not be concerned here with distinguishing among different possibilities;
the simplest and most direct definition will always be chosen.
\subsection{Laplace operator}
\label{sec.A1}
The lattice Laplace operator is relatively straightforward due to its symmetry.
However, the precise form of the matrix representing $\triangle$ will depend on the 
boundary conditions imposed on the field $\phi$. One can see this from the following
two simple examples.\hfil\break
Consider a 1D lattice having five points labeled by $i=0,\,\ldots,\,4$ and
Dirichlet boundary conditions at $i=0,\,4\,$; thus the field consists of
\be 
\phi_0,\phi_1,\phi_2,\phi_3,\phi_4,
\ee 
with
\be
\phi_0\,=\,\phi_4\,=\,0\, .
\ee
The second derivative is defined on the lattice in the usual way
\be 
\(\frac{\p^2\phi}{(\p x)^2}\)_i=\frac{\phi_{i+1}-2\phi_i+\phi_{i-1}}{a^2}\,.
\ee 
Because of the Dirichlet conditions we need -- e.g. in eq. (\ref{eq3.7}) -- this operator 
only for the {\em internal} points $n_i\, (i=1,2,3)$. Thus 
\bea\label{A1}
(\triangle\phi)_i&=&\sum_{j=1}^{3}\triangle_{i,j}\phi_j,\quad i=1,2,3\quad \mbox{with}
\nn\\
\triangle_{i,j}&=&\frac{1}{a^2}(\delta_{i+1,j}-2\delta_{i,j}+\delta_{i-1,j})\,,
\quad i,j=1,2,3\,\quad \mbox{or\, in\, matrix\, form} \nn\\
\triangle&=&\frac{1}{a^2}\(\begin{array}{rrr} -2&1&0 \\ 1&-2&1 \\ 0&1&-2 
\end{array}\)\,.
\eea
As a second example consider a 1D lattice having four points and a {\em periodic}
boundary condition; thus the field consists of
\be 
\phi_1,\phi_2,\phi_3,\phi_4\,.
\ee 
The lattice point $n=5$ is identified with the lattice point $n=1$, i.e. 
$\phi_1=\phi_5\,$.
With these boundary conditions we need -- e.g. in eq. (\ref{eq3.7}) -- the second
derivative also on the lattice \lq\lq boundary", i.e. on the lattice point $n=1$. 
One finds
\bea\label{A2}
(\triangle\phi)_i&=&\sum_{j=1}^{4}\triangle_{i,j}\phi_j,\quad i=1,2,3,4\quad\mbox{with}
\nn\\
\triangle&=&\frac{1}{a^2}\(\begin{array}{rrrr} -2&1&0&1 \\ 1&-2&1&0 \\ 0&1&-2&1 
\\ 1&0&1&-2 \end{array}\)\,.
\eea 
Using the periodic $\delta$ function we can also write
\be 
\triangle_{i,j}=\frac{1}{a^2}(\delta_{i+1,j}-2\delta_{i,j}+
\delta_{i-1,j})\,,\quad i,j=1,2,3,4\,.
\ee 
From these two simple examples one can immediately write down a general expression for 
the Laplace operator appearing in eq. (\ref{eq3.7})
\be\label{A3}
\triangle_{n,n'}=\sum_{j=1}^{d}\frac{1}{a^2}\(\delta_{n+\vec{e}_j,n'}-2\delta_{n,n'}+
\delta_{n-\vec{e}_j,n'}\)\,,\quad n,n'\in G\,.
\ee
Here the $\delta$-function $\delta_{n,n'}$ is appropriately periodic. This 
notation is, for several dimensions, neither very transparent nor very well suited for 
numerical work. One can however relabel the matrix elements according to e.g. the
scheme
\bea
& &(n_1,\ldots,\,n_d)\to m\, ,\\
 m&=&(n_1-1)\,(2N_\bot)^{d-1}+(n_2-1+N_\bot)\,(2N_\bot)^{d-2}\nn\\
 &+&\cdots+(n_{d-1}-1+N_\bot)\,(2N_\bot)+n_d+N_\bot\, ,\nn\\
 1&\le& n_1\le N_1-1\; ,\; -N_\bot+1\le n_i\le N_\bot\quad (i=2,\cdots, \,d)\; .\nn
 \eea
 This produces a symmetric matrix $\triangle_{m',m}$\, , whose dimension
 $(N_1-1)\,(2 N_\bot)^{d-1}$ equals the number of lattice points.
\subsection{The operators $(\nabla\phi)^2$ and $\nabla(\phi\nabla\phi)$}
\label{sec.a2}
The lattice representation of the differential operator $\nabla$ is not unique. Let us
again use the periodic $1D$ lattice with four lattice points to illustrate this.
We consider three definitions of $\nabla$:
\be 
(1)\quad(\nabla \phi)_i=\frac{\phi_{i+1}-\phi_i}{a}
\ee 
whose matrix representation $(\nabla 
\phi)_{i}=\sum_{j=1}^{4}\,D_{i,j}\,\phi_{j}$ is
\be 
D^{(1)}_{i,j}=\frac{1}{a}\(\begin{array}{rrrr} -1&1&0&0 \\ 0&-1&1&0 \\ 
0&0&-1&1 \\ 1&0&0&-1 \end{array}\)=\frac{1}{a}(\delta_{i+1,j}-
\delta_{i,j})\, ,
\ee 
\be 
(2)\quad(\nabla \phi)_i=\frac{\phi_{i}-\phi_{i-1}}{a}
\ee 
whose matrix form is
\be 
D^{(2)}_{i,j}=-(D^{(1)\,\top})_{i,j}=\frac{1}{a}\(\begin{array}{rrrr} 1&0&0&-1 \\ -1&1&0&0 \\
 0&-1&1&0 \\ 0&0&-1&1 \end{array}\)=\frac{1}{a}(\delta_{i,j}-
 \delta_{i-1,j})\, ,
\ee
\be 
(3)\quad(\nabla \phi)_i=\frac{\phi_{i+1}-\phi_{i-1}}{2a}
\ee 
whose matrix form is
\be 
D^{(3)}_{i,j}=\frac{1}{2a}\(\begin{array}{rrrr} 0&1&0&-1 \\ -1&0&1&0 \\ 0&-1&0&1 \\ 
1&0&-1&0 \end{array}\)=\frac{1}{2a}(\delta_{i+1,j}-\delta_{i-1,j})
\, .
\ee 
Definitions (1) and (2) have the disadvantage of producing unsymmetric matrices. 
However, it is not our purpose here to discretize $\nabla\phi$, but rather to define 
the lattice operator $(\nabla\phi)^2$.\hfil\break
For periodic boundary conditions one has in the continuum theory
\bea\label{A4}
\int dx\,(\nabla\phi(x))^2=-\int dx\,\phi(x)\triangle \phi(x)\,.
\eea
The analogous statement on the periodic lattice is (here $k=1,2,3$ labels the three
definitions)
\be 
 a\sum_{n}(\sum_{i}D^{(k)}_{n,i}\phi_i)(\sum_{j}D^{(k)}_{n,j}\phi_j)=
a\sum_{i}\sum_{j}
\phi_i(\sum_n (D^{(k)\,\top})_{i,n}D^{(k)}_{n,j})\phi_j)\,\equiv\,-a\sum_{i,j} \phi_i
\triangle_{i,j}\phi_j \, .
\ee 
It should be true that
\be\label{A5}
\sum_n (D^{(k)\,\top})_{i,n}D^{(k)}_{n,j}\,=\,-\triangle_{i,j}\,.
\ee
However, when we calculate the left-hand side of this equation we find
\be 
 \sum_{n}(D^{(1)\,\top})_{i,n}D^{(1)}_{n,j}=\sum_{n}(D^{(2)\,\top})_{i,n} 
D^{(2)}_{n,j}=\frac{1}
{a^2}\(\begin{array}{rrrr} 2&-1&0&-1 \\ -1&2&-1&0 \\ 0&-1&2&-1 \\ -1&0&-1&2 
\end{array}\)
\ee 
and
\be 
 \sum_{n}(D^{(3)\,\top})_{i,n}D^{(3)}_{n,j}=\frac{1}{a^2}\(\begin{array}{rrrr} 2&0&-2&0 \\ 
0&2&0&-2 \\ -2&0&2&0 \\ 0&-2&0&2 \end{array}\)\,.
\ee 
Since we use the lattice Laplace operator (\ref{A2}) for periodic conditions we see
that eq. (\ref{A5}) is fulfilled only for the unsymmetric discretizations $k=1,2$.
Thus we are led to our final (symmetric) choice of $(\nabla\phi)^2$
\bea
(\nabla\phi)_n^2&=&\frac{1}{2}\( \( \frac{\phi_{n+1}-\phi_n}{a}\)^2+\( 
\frac{\phi_n-\phi_{n-1}}{a}\)^2 \)\nn\\
&=&-\phi_n\frac{\phi_{n+1}-2\phi_n+\phi_{n-1}}{a^2}+\frac{\phi_{n+1}^2-2\phi_n^2+
\phi_{n-1}^2}{2a^2}\nn\\
&=&-\phi_n\sum_{n'}\triangle_{n,n'}\phi_{n'}+\frac{\phi_{n+1}^2-2\phi_n^2+
\phi_{n-1}^2}{2a^2}\,.\label{A6}
\eea
Using
\be 
(\nabla\phi)^2=\nabla(\phi\nabla\phi)-\phi\triangle\phi
\ee 
we obtain simultaneously from (\ref{A6}) a symmetrical definition for 
$\nabla(\phi\nabla\phi$)
\be\label{A7}
(\nabla(\phi\nabla\phi))_n=\frac{\phi_{n+1}^2-2\phi_n^2+\phi_{n-1}^2}{2a^2}\,.
\ee
An important  mathematical statement for periodic  boundary conditions 
is the ``spatial integral''
\be \label{A12}
\int_{0}^{l}dx\, \nabla(\phi\nabla\phi)\,\to\, \sum_{n=1}^{N}a\,
(\nabla(\phi\nabla\phi))_n =0\, .
\ee
With this lattice definition of $\int_{0}^{l}dx$ eq. (\ref{A4}) is satisfied on the periodic lattice.
Eqs. (\ref{A6}), (\ref{A7}) are also applicable for Dirichlet boundary conditions
as long as one stays away from the boundaries. Boundary points have to be dealt with
separately. For the left and right boundary points $n=0,N$ one can define right- and
left-handed derivatives as follows
\bea\label{A13}
(\nabla\phi)_0^2&=&(\nabla(\phi\nabla\phi))_0=\frac{1}{a^2}\phi_1^2 \,,\\
(\nabla\phi)_N^2&=&(\nabla(\phi\nabla\phi))_N=\frac{1}{a^2}\phi_{N-1}^2\,.
\label{A14}
\eea
Then we can also use eqs. (\ref{A6}), (\ref{A7}) at Dirichlet boundary points if we
adopt the definitions
\bea
\phi_{-1}^2&\equiv&\phi_{1}^2\,,\label{A8} \\
\phi_{N+1}^2&\equiv&\phi_{N-1}^2\,.\label{A9}
\eea
In the massive theory without potential we need more specifically
$\phi_{-1}=-\phi_{1}$ , $\phi_{N+1}=-\phi_{N-1}$.\hfil\break
Including eqs. (\ref{A8}), (\ref{A9}) we are led to a mathematical 
statement parallel to eq. (\ref{A12}), but for Dirichlet boundaries
\bea\label{A15}
\lefteqn{\frac{1}{2}\(\sum_{n=1}^{N}a\,(\nabla(\phi\nabla\phi))_n +
\sum_{n=0}^{N-1}a\,(\nabla(\phi\nabla\phi))_n\)=}\nn\\
& &=\sum_{n=1}^{N-1}a\,(\nabla(\phi\nabla\phi))_n+\frac{1}{2a^2}
(\nabla(\phi\nabla\phi))_N+\frac{1}{2a^2}(\nabla(\phi\nabla\phi))_0\nn\\
& &=-\frac{1}{2a^2}\phi_1^2-\frac{1}{2a^2}\phi_{N-1}^2+\frac{1}{2a^2}
\phi_{N-1}^2+\frac{1}{2a^2}\phi_{1}^2\\
& &=0\nn\, .
\eea
Thus the appropriate lattice definition of ``spatial integral'' for 
Dirichlet boundaries is
\be\label{A16}
\int_{0}^{l} dx\,\to\,{1\over 
2}\,a\,\left(\sum_{n=1}^{N}\,+\,\sum_{n=0}^{N-1}\right)\quad ,
\ee
and eq. (\ref{A4}) is again satisfied. \hfil\break
In summary: For the basic lattice of this paper with Dirichlet boundaries at $x_1=0,l$
and $d-1$ periodic directions we have, with eq. (\ref{A3})
\be\label{A10}
(\nabla\phi)_n^2=-\phi_n\sum_{n'\in G}\triangle_{n,n'}\phi_{n'}+\sum_{j=1}^{d}
\frac{\phi_{n+\vec{e}_j}^2-2\phi_n^2+\phi_{n-\vec{e}_j}^2}{2a^2}\,,
\ee
\be\label{A11}
(\nabla(\phi\nabla\phi))_n=\sum_{j=1}^{d}\frac{\phi_{n+\vec{e}_j}^2-2\phi_n^2+
\phi_{n-\vec{e}_j}^2}{2a^2}\,.
\ee
\section{A lattice artefact}
\label{sec.AppB}
\setcounter{equation}{0}
In calculating the ($d=1$)-vacuum stress tensor in sec. 3 with $V(x)=0$ we encountered a 
lattice artefact in the lattice formulation which (i) diverges in the
continuum limit $a\to 0$ and (ii) has no counterpart in the continuum theory. The 
origin of such a lattice quantity is to be found in the
noncommutativity of the mode sum over \lq\lq momentum" $k$ with the limit $a\to 0$.
Our task in this appendix will be to establish some needed properties of this lattice
artefact term
\be\label{B1}
A(m,l,n,a)=-\sk\frac{2 N^2\,\sin^4(\frac{\pi k}{2 N})\cos(\frac{2\pi k n}{N})}{l^2
\sqrt{(ml)^2+4 N^2\,\sin^2(\frac{\pi k}{2 N})}}\; \mbox{with}\;N={l\over a}
\ee
Most importantly we will show that $A$ contains no physical information.\hfil\break
To illustrate noncommutativity let us observe that
\be \lim_{a \to 0} A(m,l,n,a) \not=0\ee 
while in the reverse order
\be\label{B2} 
\sum_{k=1}^{\infty}\lim_{a\to 0}\(\frac{2 N^2\,\sin^4(\frac{\pi k}{2 N})
\cos(\frac{2\pi k n}{N})}{l^2\sqrt{(ml)^2+N^2\,\sin^2(\frac{\pi k}
{2 N})}}\)=0\,.
\ee
Note that $A$ has no global counterpart
\be
\label{B3}
\sum_{n=1}^{N}\,A(m,l,n,a)=0\,
\ee
because $\sum_{n=1}^{N}\cos(2 \pi k n/N)\,=\,0$ for $k\ne 0$.
Expanding $A$ in powers of $m^2$ we find
\bea\label{B4}
A(m,l,n,a)&=&-\frac{1}{al}\sk \sin^3(\frac{\pi k}{2N})\cos(\frac{2\pi k n}{N}) \nn\\
& & +\frac{m^2}{8}\sk\frac{1}{N}\sink \cos(\frac{2\pi k n}{N}) \\
& & -\frac{1}{l^2}\sum_{\nu=2}^{\infty}c_{\nu}\(\frac{ml}{2}\)^{2\nu}\sk\frac{
\cos(\frac{2\pi k n}{N})}{N^2\(N\sink\)^{2\nu-3}}\nn\\
&=&I\,+\,II\,+\,III\; .\nn
\eea
\underline{The term III}:\hfil\break
One can verify (numerically for example) that for 
Re\,$\mu\,>\,-1$
\bea\label{B5}
\lim_{N\to \infty}\(\sk\frac{\cos(\frac{2\pi k n}{N})}{N^2\(N\sink\)^{\mu}}\)=0,\quad 
\forall n \,,\,\mbox{Re}\,\mu>-1\,\, .
\eea
In the contribution III we need this statement for $\mu=2 \nu-3\ge 1$. To illustrate
 the limit (\ref{B5}) consider the $\mu=1$ case. Because
 \be 
 \sin(x)\ge\frac{2}{\pi} x,\quad 0\le x \le \frac{\pi}{2}\,,
 \ee 
 it follows that
\be 
\sk\frac{\cos(\frac{2\pi k n}{N})}{N^2N\sink}\le\frac{1}{N^2}\sk\frac{1}{k}\le 
 \frac{1}{N}={\cal O}(a)\quad \forall n\,.
 \ee 
For $\mu>1$
\be sk\frac{\cos(\frac{2\pi k n}{N})}{N^2(N\sink)^{\mu}}={\cal O}(a^2)\quad 
\forall n\,,\,\mu>1
\ee 
because the sum over $k$ now converges. In summary, the contribution III is at 
most ${\cal O}(a)$ and therefore this contribution to $A$ vanishes in the 
continuum limit. \hfil\break
\underline{The term II}:\hfil\break
We define
\bea\label{B6}
\sk\frac{1}{N}\sink \cos(\frac{2\pi k n}{N})=:h(\frac{l}{a},n)\,.
\eea
Clearly
\bea\label{B7}
\sum_{n=1}^{N} \,h(\frac{l}{a},n) =0 \,.
\eea
The finite sum (\ref{B6}) can be evaluated in closed form
\bea\label{B8}
h\(\frac{l}{a},n\)=\frac{a}{4l}\(\frac{1}{\tan(\frac{\pi a}{4l}+
\pi\frac{na}{l})}+
\frac{1}{\tan(\frac{\pi a}{4l}-\pi\frac{na}{l})}-2\)\,.
\eea
Writing $n=x/a$ and using the formula
\bea\label{B9}
\frac{1}{\tan(a+x)}+\frac{1}{\tan(a-x)}=-2\(1+\frac{1}{\tan^2x}\)a+{\cal O}(a^3)\quad 
\mbox{f\"ur}\quad  a\ll x
\eea
we obtain
\be 
 h\(\frac{l}{a},\frac{x}{a}\)\,= -\frac{a}{2l}+{\cal O}(a^2)\,,\quad \frac{a}{4}\ll x 
\ll (l-\frac{a}{4})\,.
\ee 
In the continuum limit
\bea
\lim_{a\to 0}h\(\frac{l}{a},\frac{x}{a}\)\,= 0\,,\quad 0<x<l 
\eea
and therefore
\bea\label{B10}
\lim_{a\to 0}h(\frac{l}{a},\frac{x}{a})=\sum_{k=1}^{\infty}\lim_{a\to 0}\(\frac{1}
{N}\sink \cos(\frac{2\pi k n}{N})\)=0\,,\quad 0<x<l \,.
\eea 
For $0< x< l$ it does not matter how one takes the limit $a\to 0$. However 
{\em on} the boundaries $x=0,l$ we have
\bea
h\(\frac{l}{a},0\)=h\(\frac{l}{a},N\)=\frac{a}{2l}\(\frac{1}{\tan(\frac{\pi a}{4 l})}-1\)
\stackrel{a \to 0}{\longrightarrow}\frac{2}{\pi}\, .
\eea
In other words, the ($N\to \infty$)-limit of the term II is {\em nonuniform}
in lattice position $n$. This phenomenon is quite familiar in infinite series. We 
see in eq. (\ref{B8}) that $h$ has poles at $n=1/4$ and $n=N-1/4$. These poles lie 
{\em between} the boundary and the first adjacent latti\-ce points -- i.e. they are
not on the lattice. They signal to us that the lattice function $h$ is nonuniform as
the boundaries are reached.\hfil\break
According to eq. (\ref{B10}) the continuum limit of II is zero between the Dirichlet 
endpoints. At these endpoints the value of II in the continuum limit jumps to
 $m^2/4 \pi$. As there is no $l$ dependence in this contribution to $A$ one sees
 that the Casimir force is unaffected.\hfil\break
\underline{The term I}:\hfil\break
We define
\be\label{B11} 
-\frac{1}{al}\sk \sin^3(\frac{\pi k}{2N})\cos(\frac{2\pi k n}{N})=:f(l,n,a)+\frac{1}
{2al}
\ee 
where the function $f$ can also be written
\bea\label{B12}
f(l,n,a)=-\frac{1}{l^2}\sk\(N\sin^3(\frac{\pi k}{2N})\cos(\frac{2\pi k n}{N})+
\frac{k}{N-1}\)\,.
\eea
The finite series (\ref{B11}) can be evaluated in closed form
\bea
&-&\frac{1}{al}\sk \sin^3(\frac{\pi k}{2N})\cos(\frac{2\pi k n}{N})=\frac{1}{2al}
\hspace{6cm}\\
& &\hspace{-1cm}+\frac{1}{16al}\(\,\frac{1}{\tan(\frac{3\pi}{4N}+\pi\frac{n}{N})}+
\frac{1}{\tan(\frac{3\pi}{4N}-\pi\frac{n}{N})}-\frac{3}{\tan(\frac{\pi}{4N}+
\pi\frac{n}{N})}-\frac{3}{\tan(\frac{\pi}{4N}-\pi\frac{n}{N})}\,\)  \nn
\eea
and
\bea \label{B13}
f(l,n,a)&=&\frac{1}{16al}\( \,\frac{1}{\tan(\frac{3\pi a}{4l}+\pi\frac{na}{l})}+
\frac{1}{\tan(\frac{3\pi a}{4l}-\pi\frac{na}{l})} \right.\nn \\ 
& &\hspace{1.3cm} \left.-\frac{3}{\tan(\frac{\pi a}{4l}+\pi\frac{na}{l})}-
\frac{3}{\tan(\frac{\pi a}{4l}-\pi\frac{na}{l})}\,\)\,.
\eea
We see from eq. (\ref{B13}) that $f(l,n,a)$ has poles at $n=1/4,\,3/4$ and 
$n=l-1/4,\,l-3/4$. These poles lie {\em between} the Dirichlet boundaries and
their first adjacent lattice points. From this we anticipate that the function $f$
will also exhibit nonuniform behavior as the boundaries are reached.\hfil\break
From eq. (\ref{B9}) it follows that
\be 
f(l,\frac{x}{a},a)={\cal O}(a^2)\,,\quad \frac{3a}{4}\ll x \ll l-\frac{3a}{4} 
\nn\, .
\ee 
This means that $f$ vanishes in the limit $a\to 0$ everywhere {\em between}
the boundaries
\be\label{B14}
\lim_{a\to 0}f(l,\frac{x}{a},a)=0\,,\quad 0<x<l\,,
\ee
and that for $0< x< l$ we can in eq. (\ref{B12}) freely interchange $\sum_k$ with
($a\to 0$). The behavior of $f$ on the boundary is obtained from the expansion
(for $n\ll N$)
\be 
f(l,n,a)=\frac{\pi}{4a^2}\(\frac{1}{3+4n}+\frac{1}{3-4n}-\frac{3}{1+4n}-
\frac{3}{1-4n}\) +{\cal O}(a^2)\,,\quad n\ll N \, .
\ee 
In the limit $a\to 0$ this expression diverges; however the value approached is 
independent of $l$. The Casimir force on the boundary is the discontinuity in
$\langle T^{11}\rangle$ across the boundary. An $l$-independent jump in
$\langle T^{11}\rangle$ makes no contribution to the Casimir force. On the boundary
$\sum_k$ and $a\to 0$ are not interchangeable. Note also that
\bea\label{B15}
\sum_{n=1}^{N}a\, f(l,n,a)=-\frac{1}{2a}\, .
\eea
In summary we have for the lattice artefact the following:
\bea
A(m,l,n,a)&=&\frac{1}{2al}+f(l,n,a)+\frac{m^2}{8}h(l,n,a)+{\cal O}(a)\,,
\label{B16}\\
\sum_{n=1}^{N}\,A(m,l,n,a)&=&0\,.\label{B18}
\eea
$A(m,n,l,a)-1/(2 a l)$ vanishes in the continuum limit $a\to 0$ for every point 
{\em between} the Dirichlet boundaries. This function on the boundaries $x=0,l$
jumps to a nonzero value which -- being independent of $l$ -- cannot influence the 
Casimir force.\hfil\break
The comment now to follow is speculative and mathematical. Let us define a lattice 
function
\bea 
\delta(l,n,a):=-2af(l,n,a)&=&\frac{1}{8l}\( \,-\frac{1}{\tan(\frac{3\pi a}{4l}+
\pi\frac{na}{l})}-\frac{1}{\tan(\frac{3\pi a}{4l}-\pi\frac{n a}{l})} \right. \\ 
& &\hspace{1.3cm} \left.+\frac{3}{\tan(\frac{\pi a}{4l}+\pi\frac{na}{l})}+
\frac{3}{\tan(\frac{\pi a}{4l}-\pi\frac{na}{l})}\,\)\nn
\eea
which seems to be rather like a $\delta$ function. Eq. (\ref{B15}) now reads
\be 
 \sum_{n=1}^{N}a\, \delta(l,n,a)=1\,.
 \ee 
The properties of $f$ discussed above show that
\be 
\delta(l,n,a)\,=\,\left\{\begin{array}{ll}
O(a) & 0<n<l\\
O(1/a) & n=0,\, l\end{array}\right.
\ee 
and the \lq\lq integration" (\ref{B18}) over the lattice only receives a contribution
from the boundaries. This suggests that
\be \lim_{a\to 0}\delta(l,n,a)=\frac{1}{2}\(\delta(x)+\delta(x-l)\)\, .
\ee 
To test this numerically
we applied a test-function polynomial $t(x)\,=\,A x^3+B x^2+C x+D$ and found that, with 
very high accuracy
\bea
\frac{1}{2}\(\sum_{n=1}^{N}a\,\delta(l,n,a)\,t(na)+\sum_{n=0}^{N-1}a\,\delta(l,n,a)\,
t(na)\)&\approx& \frac{1}{2}\(t(0)+t(l)\)\,,\\
\sum_{n=n_1(a)}^{n_2(a)}a\,\delta(l,n,a)\,t(na)&\approx& 0\,.\nn
\eea
In the latter sum, for each lattice constant $a$ used we required that
\be n_1(a)\,a=const=x_1>0\quad \mbox{and}\quad 
x_1<n_2(a)\,a=const=x_2<l\,.\ee 
Finally we mention that
\bea
\delta_1(l,n,a)&\equiv&\,\frac{1}{2l}\(\frac{1}{\tan(\frac{\pi a}{4l}+
\pi\frac{na}{l})}+\frac{1}{\tan(\frac{\pi a}{4l}-\pi\frac{na}{l})}\)\; , \\
\delta_2(l,n,a)&\equiv&\!\!\!\!-\frac{1}{2l}\(\frac{1}{\tan(\frac{3\pi a}{4l}+
\pi\frac{na}{l})}+\frac{1}{\tan(\frac{3\pi a}{4l}-\pi\frac{na}{l})}\)\nn
\eea
have the same properties as $\delta(l,n,a)$. Also
\be 
\delta(l,n,a)=\frac{3}{4}\delta_1(l,n,a)+\frac{1}{4}\delta_2(l,n,a).\ee 
\section{Perturbation theory on the one-dimensional lattice}
\label{sec.AppC}
\setcounter{equation}{0}
Here we calculate to first order in the lattice potential $U(n)=m^2+V(n)$ the lattice 
vacuum energy $E_{vac}$ and stress tensor component $\langle T^{11}\rangle$ for the
one-dimensional lattice with Dirichlet conditions at its endpoints. The 
\lq\lq unperturbed" system is the massless scalar field with vanishing background
potential $V(n)=0$. Our calculation for {\em arbitrary} $U(n)$ will determine 
the additional divergences in $E_{vac}$ and $\langle T^{11}\rangle$ arising from $U(n)$.
\hfil\break
For a given lattice potential one begins by first determining numerically the 
eigenvectors and spectrum of the matrix
\bea\label{C1}
{\cal O}_{n,n'}(\lambda)&=&\triangle_{n,n'}+\lambda \,U(n)\,\delta_{n,n'}\\
&=&-\frac{\delta_{n+1,n'}-2\delta_{n,n'}+\delta_{n-1,n'}}{a^2}+\lambda\, U(n)\,
\delta_{n,n'}\,,\quad 1\le n,n'\le N-1\nn\quad .
\eea
Then from these ingredients one can find the first-order shift in the vacuum energy and
in $\langle T^{11}\rangle$.\hfil\break
For $\lambda=0$ the spectrum $\omega_k^2$ and eigenvectors $\vec{w}^k$ 
of (\ref{C1})
are known
\bea
\omega_{k}^2&=&\frac{4}{a^2}\sin^2(\frac{\pi k}{2N})\,,\label{C2} \\
w^k_n&=&\sqrt{\frac{2}{N}}\sin(\frac{\pi k n}{N})\,, \label{C3}
\eea
where $1\le n\le N-1$ and $l=N a$. These eigenvalues are not degenerate, so we can
use nondegenarate perturbation theory.\hfil\break
For $\lambda\ne 0$ we name the eigenvalues and eigenvectors of (\ref{C1}) 
$\varepsilon_k^2(\lambda)$ and $\vec{v}^k(\lambda)$ respectively
\be \sum_{n'=1}^{N-1} {\cal O}_{n,n'}(\lambda)\,v^k_{n'}(\lambda)= 
\varepsilon_k^2(\lambda)\,v^k_n(\lambda)\,.\ee 
Expansion in powers of $\lambda$ yields (see e.g. ref. \cite{ref14})
\bea
\varepsilon_{k}^2(\lambda)&=&\omega_{k}^2+\lambda\,\varepsilon_k^{2\,(1)}+
\lambda^2\,\varepsilon_k^{2\,(2)}+\cdots \,\,,\label{C4}\\
\vec{v}^{\,k}(\lambda)&=&\vec{w}^{\,k}+\lambda\,\vec{v}^{\,k\,(1)}+\lambda^2\,
\vec{v}^{\,k\,(2)}+\cdots\label{C5}
\eea
where
\bea
\varepsilon_k^{2\,(1)}&=&(\vec{w}^{\,k} U\vec{w}^{\,k})=\sum_{n=1}^{N-1}\frac{2}{N} 
\sin^2(\frac{\pi k n}{N}) U(n)\label{C6}\\
& &\hspace{1.8cm}\stackrel{a\to 0}{\longrightarrow}\frac{2}{l}\int_0^l
\sin^2(\frac{\pi k x}{l}) U(x)\,dx \,,\nn\\
\vec{v}^{\,k\,(1)}&=&\sum_{p=1 \atop p\neq k}\frac{(\vec{w}^{\,p} U\vec{w}^{\,k})}
{\omega_k^2-\omega_p^2}\,\vec{w}^{\,p}\,.\label{C7}
\eea
\underline{Ground state energy}:\hfil\break
In the expansion of the lattice vacuum energy
\bea
E(\lambda)=\sum_{k=1}^{N-1}\frac{1}{2}\,\varepsilon_k(\lambda)=E^{\,(0)}+
\lambda\,E^{\,(1)}+\lambda^2\,E^{\,(2)}+\cdots\,\,.
\eea
the leading unperturbed term is
\bea\label{C8}
E^{\,(0)}=\frac{1}{2}\sk \omega_k=\frac{2l}{\pi a^2}-\frac{1}{2a}-\frac{\pi}{24l}+
{\cal O}(a^2)\,. 
\eea
To find the first correction to this we need the connection between the terms in
\be\label{C9}
\varepsilon_k(\lambda)=\sqrt{\varepsilon_k^2(\lambda)}=\omega_k+\lambda\,
\varepsilon_k^{\,(1)}+\lambda^2\,\varepsilon_k^{\,(2)}+\cdots\,\,
\ee
and the terms in eq. (\ref{C4}). From the identity
\bea
\(\sum_{\nu =0}^{\infty} a_{\nu}\,x^{\nu}\)^{\frac{1}{2}}&=&\sum_{\nu =0}^{\infty} 
b_{\nu}\,x^{\nu}\\
\Rightarrow \quad \sum_{\nu =0}^{\infty} a_{\nu}\,x^{\nu}&=&\sum_{\nu =0}^{\infty} 
b_{\nu}\,x^{\nu}\sum_{\mu =0}^{\infty} b_{\mu}\,x^{\mu}=\sum_{\nu =0}^{\infty}
\(\sum_{\mu =0}^{\nu} b_{\nu-\mu}\,b_{\mu}\)\,x^{\nu}\nn
\eea 
one finds
\bea
a_{\nu}=\sum_{\mu =0}^{\nu} b_{\nu-\mu}\,b_{\mu}\,.
\eea
Thus
\bea
\nu =0:\quad & & b_0=\sqrt{a_0}\,\,.\\
\nu =1:\quad & & b_1=\frac{a_1}{2\sqrt{a_0}}\,\,.\nn\\
\nu =2:\quad & & b_2=\frac{1}{2\sqrt{a_0}}\(a_2-\frac{a_1^2}{4\,a_0}\)\,.\nn
\eea
Then from
\bea\label{C10}
\varepsilon_k^{\,(1)}=\frac{\varepsilon_k^{2\,(1)}}{2\omega_k}=
\frac{(\vec{w}^{\, k} U\vec{w}^{\, k})}{2\omega_k}
\eea 
we find
\bea\label{C11}
E^{\,(1)}=\sk\frac{1}{2}\varepsilon_k^{\,(1)}=\frac{l}{8}\sk\frac{1}
{N\sin(\frac{\pi k}{2N})}\sum_{n=1}^{N-1}\frac{2}{N} \sin^2(\frac{\pi k n}{N})\,U(n)\,.
\eea
This is the first-order correction to the lattice vacuum energy for arbitrary lattice 
potential $V(n)$.\hfil\break
Let us define
\bea
U(x)=\bar{U}+\widehat{U}(x)\,,
\eea
with
\bea
\bar{U}&\equiv&\frac{1}{l}\int_{0}^{l} U(x)\,dx\,,\\
& &\hspace{-1cm}\int_{0}^{l}\widehat{U}(x)\,dx=0\quad .\nn
\eea
Substitution in eq. (\ref{C11}) yields with the help of (\ref{eq4.10})
\bea
E^{\,(1)}&=&\frac{1}{8}\int_{0}^{l}U(x)\,dx\,\sk\frac{1}{N\sin(\frac{\pi k}{2N})}
+\frac{l}{8}\sk\frac{1}{N\sin(\frac{\pi k}{2N})}\sum_{n=1}^{N-1}\frac{2}{N} 
\sin^2(\frac{\pi k n}{N})\widehat{U}(n)\nn\\
&=&\frac{1}{8}\int_{0}^{l}U(x)\,dx\,\(\frac{2}{\pi}\ln N+c(N)\)+\frac{l}{8}
\sk\frac{d_k}{N\sin(\frac{\pi k}{2N})}\,, \label{C12}
\eea
where
\bea
d_k=\sum_{n=1}^{N-1}\frac{2}{N} \sin^2(\frac{\pi k n}{N})\widehat{U}(n)
\stackrel{a\to 0}{\longrightarrow}\frac{2}{l}\int_0^l\sin^2(\frac{\pi k x}{l})
\widehat{U}(x)\,dx\,,
\eea
\[ \lim_{k\to \infty} d_k =0\,.\] 
The second term in the second equality of eq. (\ref{C12}) is finite for $a\to 0$, 
while the first term is logarithmically divergent.\hfil\break
Now we have what we need write down the divergent terms in $E(\lambda)$
\be\label{C13}
E(\lambda)\,=\,{2 l\over \pi a^2}\,-\,{1\over 2 a}\,-\,{\lambda\over 4 \pi}\,
\ln a\,\int_0^l dx\,U(x)\,+\,\mbox{terms finite for } a\to 0\quad .
\ee
Terms of ${\cal O}(\lambda^2)$ and higher do not contain $a\to 0$ divergences. This can be
checked numerically. For $U(x)=m^2$ we recover eq. (\ref{eq4.11}). Note that on the
lattice one can replace
\be \int_{0}^{l}U(x)\,dx \longrightarrow \frac{1}{2}\(\sum_{n=0}^{N-1}+\sum_{n=1}^{N}\)
U(n)\,a\quad .\ee 
The difference is at most of $O(a)$ and, because $a\,\ln a\to 0$ in the continuum 
limit, this difference is irrelevant.\hfil\break
\underline{$\langle T^{11} \rangle$}:\hfil\break
For the lattice $\langle T^{11} \rangle$ we have from eq. (\ref{eq3.37})
\bea\label{C14}
\tn(\lambda)&=&\sk\frac{(\varepsilon_{k}^2(\lambda)-\lambda\,U(n'))v^k_{n'}
(\lambda)^2}{2a\varepsilon_k(\lambda)}\nn\\
&+&\sk\frac{v^{k}_{n'+1}(\lambda)^2-2v^{k}_{n'}(\lambda)^2+v^{k}_{n'-1}(\lambda)^2}
{8 a^3 \varepsilon_k (\lambda)}\\
&=&\tn^{\,(0)}_{n'}+\lambda\,\tn^{\,(1)}_{n'}+\lambda^2\,\tn^{\,(2)}_{n'}+\cdots\,\,.
\nn\eea
From eq. (\ref{eq4.25}) the unperturbed tensor is
\bea\label{C15}
\tn^{\,(0)}=\frac{2\pi}{a^2}-\frac{\pi}{24l^2}+f(l,n',a)+{\cal O}(a^2)
\eea
where $f$ is the function (\ref{B12}). The first order correction to this is found
by sub\-sti\-tuting eqs. (\ref{C4}), (\ref{C5}) and (\ref{C9}) into (\ref{C14})
\bea
& &\tn^{\,(1)}=\sk\left[\frac{(w^k_{n'})^2}{2a}\(\frac{\varepsilon_k^{2\,(1)}
-U(n')}{\omega_k}-\varepsilon_k^{\,(1)}\)+\frac{\omega_k}{a}w^k_{n'}v_{n'}^{k\,(1)}
\right]\hspace{2cm}\\
& &\hspace{2.5cm}+\sk\(-\frac{(w^{k}_{n'+1})^2-2(w^{k}_{n'})^2+(w^{k}_{n'-1})^2}{8 a^3 \omega_k^2}
\varepsilon_k^{\,(1)}\right.\nn\\
& &\hspace{2.5cm}\left. +\, \frac{w^{k}_{n'+1}v^{k\,(1)}_{n'+1}-2w^{k}_{n'}v_{n'}^{k\,(1)}
+w^{k}_{n'-1}v^{k\,(1)}_{n'-1}}{4 a^3 \omega_k}\)\,.\hspace{-1cm}\nn
\eea
Next we introduce the potential
\be  \widetilde{U}(n)\,\equiv\,U(n)-U(n')\,,\quad\mbox{i.e.}
\quad\widetilde{U}(n')=0 \ee 
which leads to
\bea
\varepsilon_k^{2\,(1)}&=&\tilde{\varepsilon}_k^{2\,(1)}+U(n')\,,\\
\varepsilon_k^{\,(1)}&=&\tilde{\varepsilon}_k^{\,(1)}+\frac{U(n')}{2\omega_k}\,,\nn\\
v_{n'}^{k\,(1)}&=&\tilde{v}_{n'}^{k\,(1)}\,.\nn
\eea
Here the quantities with tildes are defined with $\widetilde{V}$ in place of $V$.
Continuing with this tilde notation we find for the first-order correction to
$\langle T^{11} \rangle^{(0)}$
\bea\label{C16}
\tn^{\,(1)}&=&\widetilde{\tn}^{\,(1)}-\frac{U(n')}{8}\sk\frac{1}{N\sink}+
\frac{U(n')}{8}\sk\frac{1}{N}\sink \cos(\frac{2\pi k n'}{N})\nn\\
&=&\widetilde{\tn}^{\,(1)}-\frac{U(n')}{8}\sk\frac{1}{N\sink}+\frac{U(n')}{8}
\,h(N,n')\,,
\eea
where $h(N,n')$ is the function eq. (\ref{B6}).\hfil\break
The lattice quantity $\langle \widetilde{T}_n'^{11} \rangle^{(1)}$ remains finite in 
the limit $a\to 0$ and is the first-order correction to $\langle T^{11} \rangle^{(0)}$
for a lattice potential $\widetilde{V}(n)$ which vanishes at the point $n=n'$. It is
therefore  relatively obvious that $\langle \widetilde{T}_{n'}^{11} \rangle^{(1)}$
contains no $a\to 0$ divergent terms because the lattice should be (in first order)
proportional to $\widetilde{V}(n')$. For $V(x)=m^2$ eq. (\ref{C16}) reduces to the system 
discussed in sec. 4.2. From (\ref{C15}), (\ref{C16}) we can read off the divergent 
terms in eq. (\ref{C14})
\bea\label{C17}
\langle T_n^{11}\rangle (\lambda)\,=\,\frac{2\pi}{a^2}+f(l,n,a)+\frac{U(n)}{4\pi}\,
\ln a\,+\,\cdots \, .  
\eea
Second-order corrections need not be discussed. Numerical probes show that all
divergent terms are displayed in eq. (\ref{C17}).\hfil\vfil\eject  

\vfill\eject
\end{document}